\documentclass[aos,preprint]{imsart}

\RequirePackage[OT1]{fontenc}
\usepackage{amsthm,amsmath,amsfonts,amssymb,multirow,graphicx,subfigure,booktabs,url,algorithm,mathtools,wrapfig,mathrsfs,bm,relsize,caption,color,soul,makecell}

\RequirePackage[authoryear]{natbib}
\RequirePackage[psamsfonts]{amssymb}
\usepackage[noend]{algpseudocode}
\usepackage[colorlinks=true,
linkcolor=blue,
urlcolor=blue,
citecolor=blue]{hyperref}

\startlocaldefs
\numberwithin{equation}{section}
\theoremstyle{remark}\newtheorem{rem}{Remark}
\newtheorem{conj}{Conjecture}
\theoremstyle{default}\newtheorem{thm}[conj]{Theorem}
\newtheorem{lemma}[conj]{Lemma}

\newcommand{\sgn}{\mathop{\mathrm{sign}}}

\DeclareFontFamily{U}{mathx}{\hyphenchar\font45}
\DeclareFontShape{U}{mathx}{m}{n}{
	<5> <6> <7> <8> <9> <10>
	<10.95> <12> <14.4> <17.28> <20.74> <24.88>
	mathx10
}{}
\DeclareSymbolFont{mathx}{U}{mathx}{m}{n}
\DeclareFontSubstitution{U}{mathx}{m}{n}
\DeclareMathAccent{\widecheck}{0}{mathx}{"71}

\def\wh{\widehat}
\def\wt{\widetilde}
\def\To{\longrightarrow}
\def\KL{\mathcal{KL}}
\def\to{\rightarrow}

\def\EE{\mathbb{E} }
\def\PP{\mathbb{P} }
\def\QQ{\mathbb{Q} }

\def\G{\mathcal{G}}

\def\M{\mathcal{M}}

\def\A{\mathcal{A}}
\def\RR{\mathbb{R} }
\def\supp{\textrm{supp}}

\def\i{\infty}
\def\H{\mathcal{H}}
\def\bl{\setminus}
\def\E{\mathcal{E}}
\def\I{\mathcal{I}}
\def\1{\bm{1}}
\def\r{{\infty,1}}
\def\c{{1,\infty}}
\def\O{\Omega}

\def\H{\mathcal{H}}
\def\ll{\Sigma_{11}}
\def\lr{\Sigma_{12}}
\def\rr{\Sigma_{22}}
\def\rl{\Sigma_{21}}
\def\rrl{\Sigma_{22\cdot1}}
\def\tr{\textrm{tr}}
\def\q{{\infty,q}}
\def\0{\bm{0}}
\def\bI{\bm I}

\def\rm{\textrm}
\def\wc{\widecheck}
\let\emptyset\varnothing
\makeatletter
\def\BState{\State\hskip-\ALG@thistlm}
\makeatother

\endlocaldefs

\begin{document}
	
		\begin{frontmatter}
		\title{ Adaptive Estimation in  Structured  Factor Models
			with Applications to Overlapping Clustering}
		\runtitle{Adaptive estimation in structured factor models}
		
		\begin{aug}
			\author{\fnms{Xin} \snm{Bing} \ead[label=e1]{xb43@cornell.edu}}
			\and
			\author{\fnms{Florentina} \snm{Bunea} \ead[label=e2]{fb238@cornell.edu}}
			\and
			\author{\fnms{Yang} \snm{Ning} \ead[label=e3]{yn265@cornell.edu}}
			\and
			\author{\fnms{Marten} \snm{Wegkamp}	\ead[label=e4]{mhw73@cornell.edu}}
			
			\runauthor{X. Bing, F. Bunea, Y. Ning and M. Wegkamp}
			
			\affiliation{Cornell University}
			
			\address{
				X. Bing\\
				Department of Statistical Science\\ 
				Cornell University\\
				Ithaca, New York 14853-3801\\
				USA\\
				\printead{e1}}
			\address{
				F. Bunea\\
				Department of Statistical Science\\ 
				Cornell University\\
				Ithaca, New York 14853-3801\\
				USA\\
				\printead{e2}}
			\address{
				Y. Ning\\
				Department of Statistical Science\\ 
				Cornell University\\
				Ithaca, New York 14853-3801\\
				USA\\
				\printead{e3}}
			\address{
				M. Wegkamp\\
				Department of Mathematics \&\\
				Department of Statistical Science\\
				Cornell University\\
				Ithaca, New York 14853-3801\\
				USA\\
				\printead{e4}}
		\end{aug}
		
		\begin{abstract}
			
			This work  introduces a novel  estimation method, called {LOVE}, of the entries and structure of a loading matrix $A$ in a latent factor model $X=AZ+E$, 
			for an observable random vector $X \in \RR^p$, with correlated unobservable  factors $Z\in\RR^K$, with $K$ unknown, and   uncorrelated 
			noise $E$. 
			Each row  of $A$ is   scaled, and allowed to be sparse. In order to identify the loading matrix $A$ we require the existence of pure variables, which are 
			components of $X$   that are  associated, via $A$, with one and only one latent factor. Despite the fact that  the number of factors $K$, the number of  the pure variables,  and their location are all unknown, we only require a  mild condition on the covariance matrix of $Z$, and a minimum of only two pure variables per latent factor to show  that  $A$  is uniquely defined, up to  signed permutations. Our proofs for model  identifiability are constructive,   and lead to our novel  estimation method of  the number of factors and of the set of pure variables, 
			from a sample of size $n$ of observations on $X$.
			This is the first step of our LOVE algorithm, which is optimization-free,  and   has low computational complexity of order $p^2$.
			The second step of LOVE is an easily implementable linear program that estimates $A$. We prove that  the resulting estimator is near minimax rate optimal for $A$, with respect to the $\| \ \|_{\infty, q}$ loss, for $q \geq 1$, up to logarithmic factors in $p$, and that it can be minimax-rate optimal in many cases of interest.
			
			The model structure  is motivated by the problem of overlapping variable clustering, ubiquitous in data science.
			We define the population level clusters as groups of those components of $X$ that are associated, via the matrix $A$, with the same unobservable latent factor, and multi-factor association is allowed.  Clusters are respectively anchored by  the pure variables, and form overlapping sub-groups of the $p$-dimensional random vector $X$.  The \textbf{L}atent model approach to \textbf{OVE}rlapping clustering
			is reflected in the name of our algorithm, LOVE.
			
			The third step of LOVE estimates the clusters from the support of the columns of the estimated $A$. 
			We guarantee cluster recovery with zero false positive proportion,  and with false negative proportion control.  The practical relevance of LOVE is illustrated through the analysis of an RNA-seq data set, devoted to determining the functional annotation of genes with unknown function.	
		\end{abstract}
	
		
	\end{frontmatter}

\section{Introduction}

In this work we consider the problem of estimating the $p \times K$,  possibly  sparse, loading matrix $A$  that parametrizes  the factorization  
of a  zero-mean observable random vector, $X \in \RR^p$ as 
\begin{equation}\label{mod}   X = AZ + E \end{equation}
from  $n$ i.i.d. realizations of $X$.  The zero mean random vector $Z   \in \RR^K$ is unobservable, and can be viewed as a latent factor vector.  $E \in \RR^p $  is a zero-mean, unobservable random noise vector, with   uncorrelated  entries. The number of factors $K$ is not known, and both $p$ and $K$ are allowed to grow, and be larger, than $n$.   
Factor models have been  used as dimension reduction devices  in virtually any scientific discipline for nearly a century, and generated an enormous amount of literature.  We refer to the classical monographs of \cite{bollen1989} and \cite{anderson_book} for earlier work, and to   \cite{izenman}   for a more recent survey and applications. 

In this work, we revisit some of the open problems in factor model definition and estimation, and also consider one of their  much less explored  applications, to overlapping  clustering. For the latter, we deem two components $X_i$ and $X_j$ of $X$  similar  if they  have non-zero association, via the matrix $A$,  with the same latent factor $Z_a$. Similar variables are placed in the same cluster, $G_a$:
\begin{equation}\label{groups1} 
G_a := \bigl\{j \in \{1, \ldots, p\}:\ A_{ja}\ne0\bigr\},\quad \text{for each } a \in \{1, \ldots, K\}.
\end{equation}
Since each $X_j$ can be associated with multiple latent factors, the clusters will overlap. The  problem  of overlapping clustering is of wide-spread interest in virtually any scientific  area,  for instance in 
neuroscience \citep{Craddock2, Craddock} and genetics \citep{GeneExpressionCluster,wiwie2015comparing}, to give  a  very limited number of examples. The solutions are typically algorithmic in nature, and their quality is assessed against a ground scientific truth or via extensive simulation studies, for instance   \cite{krishnapuram2001low,bezdek2013pattern}, among many others. These  problems  have not received a systematic analysis in the statistical literature and, in particular, the problem of estimating overlapping  clusters of variables, with theoretical guarantees, remains largely unexplored.  

In this work, we propose model-based clustering via $A$. However, $A$ cannot be uniquely defined in (\ref{mod}),  without further restrictions, a phenomenon well understood over six decades ago.  Most notably, \cite{anderson1956} provided an in-depth analysis of this problem, and proved that in the absence of  conditions on $A$ and $C :=\text{Cov}(Z)$, $A$ is not identifiable in model (\ref{mod}). We revisit some of these conditions here, with a view towards our application 
to overlapping clustering.
We defer a detailed literature review of related identifiability conditions for model (\ref{mod}) to  Section \ref{sec_related_works}.

Using overlapping clustering as  motivation, we formalize our first modeling assumption on $A$.  We  consider models (\ref{mod}) in which each row of $A$ is scaled, to avoid scale ambiguities. Specifically, we assume that: 
\begin{itemize}
	\item[(i)] $\sum_{a=1}^{K}|A_{ja}| \le 1$.
\end{itemize}
The inequality in (i)  allows for $\sum_{a} |A_{ja} | =0$, which renders more flexibility to model (\ref{mod}),  relative to the more commonly used equality conditions.  If   $\sum_{a} |A_{ja} | =0$, then  $X_j=E_j$, and $X_j$ is  not associated with any of the latent factors, via this model.  The interpretation to  clustering is that the  corresponding $X_j=E_j$ does not belong to any cluster given by this model, which is a desired feature in many practical applications, including the one presented in this paper in Section \ref{genes}. 
Furthermore, in order to use the model for clustering, we need to  avoid the trivial situation in which each component $X_j$ is associated with all latent factors. From this perspective, we allow the rows  $A_{j\cdot}:= (A_{j1},\ldots,A_{jK})$ to be  sparse, for  $j \in \{1, \ldots, p\}$, but this property is not required for the identifiability of $A$. 

Condition (i) alone cannot ensure that $A$ in model (\ref{mod}) is uniquely defined, as one can still construct an invertible  matrix $Q$ such that $AZ = AQQ^{-1}Z$, 
with both $A$ and $AQ$ satisfying (i). Moreover,  when $A$ is sparse, $A$ and $AQ$  may not have the same sparsity pattern, creating ambiguity in the cluster definition.  We introduce below  two additional requirements  that allow us to show, in Section \ref{id} below, that $A$ is  identifiable.

We call (ii) given below  {\it the pure variable assumption}. Informally, it postulates the existence of at least two {\it pure variables} $X_j$, which are components of $X$ associated with one and only one latent factor.   In Section \ref{id}  we  provide examples that show that if pure variables do no exist, $A$ in (\ref{mod}) is not uniquely defined.   
\begin{itemize}
	\item[(ii)] For every $a\in\{1,\ldots,K\}$, there exist at least two indices  $j \in \{1, \ldots, p \}$ such that $|A_{ja}|=1$ and $A_{jb}=0$ for all $b\ne a$. 
\end{itemize}
We note that in the very particular case of known  $\Gamma:=\text{Cov}(E)$,   only {\em one} pure variable per group is required for identifiability, which follows 
from the proof of Theorem \ref{ident} in Section A.1.  	
The pure variable assumption has an immediate practical implication to variable clustering. Since clusters $G_a$  given by (\ref{groups1}) are  defined relative to the unobservable factor  $Z_a$, a pure variable $X_j$ is an  observable proxy of $Z_a$, and that helps explain the otherwise unclear nature of $G_a$.

For future reference, we let $I$ denote the index set corresponding to pure variables. 
In psychology, these variables are called factorially simple items (\cite{mcdonaldbook}).  A similar condition 
can be traced back to the econometrics literature,  and an early reference is \cite{koopmans1950}, further discussed in \cite{anderson1956}, who called it ``zero elements in {\it specified} positions". These works  prove that (ii) corresponding to a {\it known} set $I$  is a sufficient condition for identifying $A$, for latent factors with arbitrary correlations.   However, full generality  on the positive definite covariance matrix $C$ of the latent factors comes at the  steep price of knowing $I$ a priori, which  is often unrealistic in practice.
Appropriate conditions on $C$ that guarantee identifiability of  $I$ in (ii),  in the general case when $I$ is not known and, moreover, $K$ is unknown,  have not been investigated for the general model (\ref{mod}), to the best of our knowledge.  To this end, we introduce the following condition on the covariance matrix $C$. 

\begin{itemize}
	\item[(iii)] 
	$\Delta(C) := \min_{a\ne b}\left(C_{aa}\wedge C_{bb} - |C_{ab}|\right) > 0$ and $C$ positive definite, 
\end{itemize}
where $a\wedge b:= \min(a,b)$. If (iii) holds, then $ \text{Cov}(Z_a \pm Z_b) = \text{Var}(Z_a) +\text{Var}(Z_b) \pm 2\cdot\text{Cov}(Z_a,Z_b) \ge C_{aa}+C_{bb} - 2|C_{ab}| > 0,$ which implies  that the latent factors are different, up to signs, that is $|Z_a| \ne |Z_b|$ a.s. for any $a\ne b$.

Condition (iii)  holds trivially under the much stronger assumption that  the latent factors are independent, or have a slight departure from independence, corresponding to  diagonal dominance in $C$. These type of assumptions are commonly made in latent factor models, but may often be unrealistic, see, for instance, \cite{anderson1956, anderson_book, bollen1989, everitt1984, izenman} and our discussion in Section \ref{sec_related_works}. Condition (iii) therefore relaxes  the independent factor assumption, and we comment further on it below. 

Condition (iii) is a companion of our Conditions (i) and (ii). When the last two are being made, Condition (iii) admits relaxations, which have   been established only in special set-ups. 

Under the pure variable assumption  (ii), if $I$  {\it is known} in advance,  the arguments employed in the proof of  our Theorem \ref{ident}  of Section \ref{id} show that  (iii) is not required, and the assumption that $C$ is a positive definite covariance matrix suffices. This  is  consistent with the classical literature on general latent models, see, for instance, \cite{anderson1956}. 

Identifiability results corresponding to the realistic situation when $I$ is not known are scarce, and correspond to particular instances of the model we consider in this work. In the limit case of our model, when all $p$ variables are pure variables, which corresponds to non-overlapping clustering, \cite{bunea_cord_pecok} showed that, once again, $C$ being positive definite suffices for identifiability.

The problem of identifying  $A$ under (ii), with $I$ unknown, has been  revived more recently, in the particular case 
of modeling random vectors  $X$  with only non-negative  values,  when $A$ and $Z$ also have only non-negative entries. This  set-up corresponds  to the area known as non-negative matrix factorization (NMF), in which one studies  positive matrix factorizations of the type ${\bm X} = A{\bm Z} +{\bm E}$, 
where the observed data ${\bm X}$ is a $p \times n$ matrix, ${\bm Z}$ is the $K \times n$ unobservable matrix of the latent vectors, and ${\bm E}$ is the $p \times n$ noise matrix.  In this context, when ${\bm E} = 0$, and conditioning on ${\bm Z}$,    \cite{donohoNMF} was among the first works to propose a condition similar to (ii),  with $I$ unknown, coupled with appropriate conditions on ${\bm Z}$, 
leading to an  NMF decomposition  with unique factors.  Moreover, the  unique determination of $I$  under (ii), for $\bm E \neq 0$, but with very small component-wise variances,  was solved in \cite{bittorfNMF},  for known $K$, and for scaled NMF models, in which the columns of ${\bm X}, {\bm Z}$ and $A$ sum up to 1. These  results were proved  under their  that no row  of  a scaled version of ${\bm Z}$ is a  convex combination of the other rows. Conditioning on $Z$, this requirement is weaker than our Condition (iii), should we impose it on $n^{-1}{\bm Z}{\bm Z}^T$, but it is not readily generalizable outside the NMF framework.

In light of this discussion,  our Condition (iii) on $C$ is a key ingredient in the identification of $I$, in the context of the more general model (\ref{mod}), when $E$ is not negligible, and $K$ is not known. The details are given  in  Section \ref{id} below.  If all the latent variables have the same variance, then Condition (iii) becomes  the very mild  requirement that the correlations between pairs of  latent variables are strictly less than 1, $\text{Cor}(Z_i, Z_j)<1$,  for $1\le i< j\le K$. When the factors have unequal variances, Condition (iii)  may still hold, but  it becomes stronger. We view this as the price to pay for the identifiability of $I$, and consequently of $A$ in the general model (\ref{mod}).

Summarizing, this work is devoted to estimation in model  (\ref{mod}) with $A, C$  satisfying (i) - (iii). The number of factors $K$ is not known, and both $K$ and $p$ are allowed to grow  and be larger than $n$.
In Section \ref{sec_contri} below we present our contributions and the structure of this paper. A detailed contrast with existing literature  is presented in Section \ref{sec_related_works}.

\subsection{Our contributions}\label{sec_contri}

\paragraph{\bf 1. Identifiability of the allocation matrix $A$ in  sparse latent models with pure variables} We show, in Proposition \ref{ident} of Section \ref{id},  that the allocation matrix $A$, which is allowed to have entries of arbitrary signs, is uniquely defined, up to trivial orthogonal transformations, namely signed permutation matrices.  This is a consequence  of one of our main results, Theorem \ref{I} of Section \ref{id}.  In this result we highlight and resolve  the main difficulty in this problem,  that of distinguishing between the pure variables and the non-pure variables.    Both proofs are constructive, and show that the pure variable set $I$ and allocation matrix $A$ can be determined uniquely from $\Sigma:=\textrm{Cov}(X)$.  Moreover, the number of factors $K$ is not assumed to be known, and its determination is also a consequence of  Theorem \ref{I}. To the best of our knowledge, these are new results in both the latent factors literature and other related matrix factorization literature. We comment on connections to related results in Section \ref{sec_related_works}. 

\paragraph{\bf 2. Estimation of the allocation matrix $A$ and of the overlapping clusters. The LOVE algorithm}  We provide an estimator $\wh A$ of the sparse and structured matrix $A$ that is tailored to our model  specifications. Our approach follows the constructive techniques used in our identifiability proofs. We first construct $\wh I$, an estimator of the pure variable set $I$, and  $\wh K$, an estimator of the number of clusters, $K$. These are used to estimate the rows in $A$ corresponding to pure variables. The remaining rows of $A$ are estimated via an easily implementable linear program that is 
tailored to this  problem.  As part of our procedure, we also develop a novel estimator (\ref{decomp1}) and (\ref{decomp2}) of a precision matrix, $C^{-1}$.  Our procedure  is presented in Sections \ref{estpure}, \ref{sec_est_AI} and \ref{sec_est_AJ}, respectively. To the best of our knowledge,  our estimation strategy is new, and complements the large body of literature in factor models.  In particular, we do not resort to optimizing a complicated quasi likelihood function via  computationally demanding EM algorithms. These  algorithms  require, in addition, a notoriously delicate initialization, especially in high dimensions, and typically  only convergence to a stationary point can be  guaranteed, see \cite{Rubin1982}. 
Moreover, as our procedure is not Bayesian, we do not employ distributional assumptions to construct our estimator.  In Section \ref{sec_est_group}, we  build  a collection of overlapping clusters $\wh {\mathcal{G}}$, using the estimated allocation matrix $\wh A$. The combined procedure is summarized in a new 
algorithm, LOVE, highlighting our \textbf{L}atent model approach to \textbf{OVE}rlapping clustering.

\paragraph{\bf 3. Statistical guarantees}   
Our estimation procedure does not depend on distributional assumptions, but for the purpose of our statistical analysis,  and in particular our minimax analysis,
we assume that $X\in\RR^p$ has a sub-Gaussian distribution with $\log p= o(n)$ as $n\to\infty$.
LOVE, for appropriate choices of tuning parameters,  recovers the population level clusters with  a zero false positive proportion  and generally low false negative proportion,  with high probability, and under a  mild condition on the cluster separation as measured by the quantity $\Delta(C)$.  
This is a direct consequence of  a number of results regarding estimation of identifiable loading matrices in factor models satisfying (i) - (iii) and, to the best of our knowledge, they are all new.
\begin{enumerate}
	\item[(1)] Consistent estimation of the number of factors  $K$; 
	\item[(2)] Control of the relationship between $\wh I$ and  $I$ for $A$ with entries of arbitrary strength.  In particular, we show $I\subseteq\wh I\subseteq I\cup J_1$,  where we carefully define and  characterize $J_1$ as the set of quasi-pure variables.  
	\item [(3)] Minimax lower bounds on the norms $L_q(\wh A,A)$, defined below,  for all $q\ge 1$, in particular for $q=+\infty$, 
	for $A$ given by  model (\ref{mod}) under (i) - (iii). 
	\item[(4)]  Attainment of these bounds, showing that our procedure is minimax optimal and adaptive. 
	\item[(5)] Control of the relationship between the support of $A$ and the support of $\wh A$. 
	\item[(6)] Control of cluster recovery. 
\end{enumerate}
The details  are given in  Sections  \ref{stat_guarantee_A} and \ref{stat_guarantee_group}.  
In particular, we emphasize that  (2) above, proved in Theorem \ref{consistI}
of Section \ref{whI},  guarantees recovery of $I$ with minimal mistakes.  This result does not require the necessary, yet unpleasant, signal strength restrictions encountered in the typical 
exact support recovery literature. However, under such restrictions, we also obtain $\widehat{I} = I$, with high probability, in Remark \ref{rem1} of Section \ref{whI}.  Since placing restrictions on the entries in $A$ reduces the number of configurations of interest, the more general result (2) is a new and practically relevant  result for pure variable recovery.

Results (3) and (4) are  given in Theorems \ref{prop_rates}, \ref{thm:Dantzig} and  \ref{thm_minimax}  of Section \ref{stat_guarantee_A}.
We consider the loss function  
\begin{equation}\nonumber 
L_q(\wh A, A):= \min_{P }\|\wh AP-A\|_{\i,q},\quad 1\le q\le \infty,
\end{equation}
with the minimum  taken over all 
$K\times K$ signed permutation matrices $P$ and 
$$\|A\|_{\i,q}:=\max_{1\le i\le p}\left\|A_{i\cdot}\right\|_q = \max_{1\le i\le p}\left( \sum_{j=1}^K|A_{ij}|^q\right)^{1/q},$$ is the maximum $\ell_q$ norm of the rows of $A$. We let  $s= \max_{i\in [p]} \|A_{i\cdot}\|_0$ be the   row-sparsity index.

We show that the error of estimation with respect to the $L_{q}$ loss function, for each $q$,   is proportional to  $ s^{1/q} n^{-1/2} $, multiplied by $\| C^{-1} \|_\r$. This is consistent with the most recent results regarding  error rates expressed in terms of the  $\ell_q$-sensitivity of $C$ in \cite{gt2011} and \cite{belloni2016},  as discussed in Section \ref{stat_guarantee_A}. The results hold up to  logarithmic factors in $p$ and $s$.  

Results (5) and (6) are presented in Theorem \ref{propgroup} of Section \ref{stat_guarantee_group}. Moreover, we can further partition the variables in each cluster into two signed sub-groups consistently. In our model formulation,  $A$ is allowed to have positive and negative entries.  
Since  $A$ can only be identified up to signed permutations,   one cannot expect sign consistency for $\wh A$. However,  we can identify consistently the two sub-groups of each 
cluster that contain variables that are associated with the common latent factor in the same direction, although the direction itself is not identifiable. These results are presented in Section \ref{stat_guarantee_group}.

We conduct an extensive simulation study in Section \ref{sec_sim} to assess the numerical performance of our proposed strategy. The study confirms our theoretical findings. We conclude the validation of our approach with a data analysis, devoted to determining the functional annotation of genes with unknown function. Our analysis confirms existing biological ground truths,  as our procedure   tends to cluster together genes with the same Gene Ontology (GO) biological process, molecular function, or cellular component terms. 


We summarize our contributions in the table below, restricting attention to estimation in general latent models (\ref{mod}) under (i) - (iii), without any further restrictions  on the signs or scales of $X$ and $Z$.

\begin{table}[ht]
	\centering
	\resizebox{\textwidth}{!}{
		\begin{tabular}{|c|c|c|}
			\hline
			Model (\ref{mod}) under (i) - (iii) 	& Our results  & Existing results in comparable factor models\\\hline
			
			Identifiability Conditions&  \thead{Existence of $I$ with $I$ and $K$ {\it unknown}.\\ $C$ is positive definite and satisfies (iii).} &  \thead{Existence of {\it known} $I$ and $K$.\\ $C$ is positive definite.}
			\\ \hline 
			Estimation: $I$ &  \footnotesize{Runs in $O(p^2)$ time; optimization-free.} &  $\times$ \\ \hline 
			Estimation: $A$ &  \thead{Not MLE-based approach.\\
				Unique solution.\\ Linear program; runs in $O(p^2 + pK)$.} & \thead{MLE-based approach. \\
				Multiple solutions.\\ EM algorithm; computationally involved.}\\ \hline 
			Guarantees: $I$ & \small{Recovered}   &  $\times$ 	\\ \hline 
			Guarantees: $A$  & \thead{Finite sample $\| \ \|_{\infty, q}$ lower bounds.\\ 
				Adaptive finite sample upper bounds.\\ 
				Both $p$ and $K$ can grow with n.}
			& \thead{
				Row-wise asymptotic normality of MLE.   \\Only $p$ can grow with $n$ and $K$ is {\it fixed}.}
			\\\hline 
			Cluster recovery & \small{Guaranteed}   & $\times$ \\ \hline 
		\end{tabular}
	}
\end{table}

In Section \ref{sec_related_works} we discuss our results further, and provide a detailed comparison between our work and related contributions. 
All proofs are deferred to Section A of \ref{suppA}.

\subsection{Notation} 
We use the following notation throughout this paper.
For the $n$ consecutive integer set starting from $1$, we write $[n] = \{1,\ldots,n\}$. The sign of any generic number $N$ is denoted by $\textrm{sign}(N)$. For any $m\times d$ matrix $M$ and index sets $I\subseteq\{1,\ldots,m\}$ and $J\subseteq\{1,\ldots,d\}$,  we write $M_I$ to denote the $|I|\times d$ submatrix $(M_{ij})_{i\in I, 1\le j\le d}$ of $M$ consisting of the rows in the index set $I$, 
while we denote by  $M_{IJ}$ the $|I|\times |J|$ submatrix with entries $M_{ij}$, $i\in I$ and $j\in J$.
The $i$th row of $M$ is denoted by $M_{i\cdot}$, and the $j$th column of $M$ is denoted by $M_{\cdot j}$. Let $\|M\|_{\infty}=\max_{1\leq j\leq m ,1\leq k\leq d}|M_{jk}|$, $\|M\|_1 = \sum_{1\leq j\le m,1\le k\leq d}|M_{jk}|$, $\|M\|_F=(\sum_{j=1}^{m }\sum_{k=1}^{d}M_{jk}^2)^{1/2}$, $\|M\|_{{\r}}=\max_{1\leq j\leq m }\sum_{k=1}^{d}|M_{jk}|$ and $\|M\|_{{\c}}=\max_{1\leq k\leq d }\sum_{j=1}^{m}|M_{jk}|$ denote the matrix max norm, matrix $\ell_1$ norm, matrix Frobenius norm, matrix $1$ norm and matrix $\infty$ norm. We denote by $\langle\ \cdot\ \rangle$ the Frobenius scalar product. For a vector $v\in\RR^d$, define $\|v\|_q=(\sum_{i=1}^d|v_j|^q)^{1/q}$ for $1\leq q<\infty$, $\|v\|_\infty=\max_{1\leq j\leq d}|v_{j}|$ and $\|v\|_0=|\textrm{supp}(v)|$, where $\textrm{supp}(v)=\{j: v_j\neq 0\}$ and $|S|$ is the cardinality of the set $S$. 
For a vector $v \in \RR^d$, we denote by $v_S$ the vector $w\in \RR^d$ that has    the same coordinates $w_i=v_i$  as $v$ on the index set $S\subseteq\{ 1,\ldots,d\}$ and zero coordinates otherwise ($w_i=0$ for all $i \in \bar{S} := [d] \setminus S$). 
We write  $M^T$ for the transpose of $M$ and $\text{diag}(m_1,\ldots, m_{d})$ for the $d\times d$ diagonal matrix with elements $m_1,\ldots,m_d$ on its diagonal, while $\text{diag}(M)$ is the diagonal matrix obtained from the diagonal elements of a square matrix $M$. The identity matrix in $\RR^{d\times d}$ is denoted by $\bm{I}_d$, the vector in $\RR^{d}$ with all entries equal to one is denoted by $\bm{1}_d$ and a vector/matrix with all zero entries is denoted by $\0$ whose dimension might vary line by line. We use $c_0, c_1, \ldots$ to denote generic constants.  Finally, a signed permutation matrix is an orthogonal matrix that permutes the index and switches the sign within each column. We write $\H_K$ as the hyperoctahedral group of   $K\times K$ signed permutation matrices. 

\section{Identifiability} \label{id}  

In this section we show that the allocation matrix $A$  given by Model (\ref{mod})  and (i) - (iii) is identifiable, up to multiplication with a signed permutation matrix.

For any $A\in \RR^{p\times K}$ which satisfies Model (\ref{mod}), we can partition the set $[p]=\{1,\ldots,p\}$ into two disjoint parts: $I$ and its complement $J := [p] \setminus I$ such that for each row $A_{i\cdot}$ of $A_I$, there exists only one $a\in[K]$ such that $|A_{ia}|=1$.  We name $I$ the pure variable set and $J$ the non-pure variable set. Specifically, for any given $A$, the pure variable set $I$ is defined as 
\begin{equation}\label{defI}
I(A) := \bigcup_{a=1}^KI_a,\quad
I_a := \left\{i\in [p]: |A_{ia}| = 1, A_{ib} = 0, \text{ for any } b\ne a \right\}.
\end{equation}
We write $I(A)$ in (\ref{defI}) to emphasize that the pure variable set is defined relative to $A$. In the following, we will not write this explicitly when there is no confusion.  We also note that the sets $\{ I_a\}_{1 \leq a \leq K}$ form a partition of $I$.

To show the identifiability of $A$, it suffices to show that $A_I$ and $A_J$ are identifiable, respectively, up to signed permutation matrices. By the definition of $A_{I}$,  this matrix is identifiable provided the partition of the pure variable set $I$ is. The identifiability of $I$, and thus the problem of distinguishing between  the sets $I$ and $J$,  on the basis of the distribution of $X$ alone, is  the central challenge  in this problem.  We meet this  challenge in Theorem \ref{I} below:   part ({\bf a}) offers  a necessary and sufficient  characterization of $I$;   part ({\bf b})  shows that, as a consequence, $I$ and its partition $\mathcal{I}:= \{I_a\}_{1 \leq a \leq K}$ are identifiable.  Let
\begin{equation}\label{mi}
M_i := \max_{j\in [p]\bl\{i\}}|\Sigma_{ij}|
\end{equation}	
be the largest absolute value of the entries of row $i$ of $\Sigma$ excluding $|\Sigma_{ii}|$. Let $S_i$ be the set of indices for which $M_i$ is attained:
\begin{equation}\label{si}
S_i := \bigl\{j\in[p]\bl\{i\}: |\Sigma_{ij}|= M_i\bigr\}.
\end{equation}

\begin{thm}\label{I}  Assume that model (\ref{mod}) and (i) - (iii) hold. Then: 
	\begin{itemize}
		\item[({\bf a})]
		$
		i \in I\quad  \Longleftrightarrow\quad  M_i = M_j ~~\text{for all }j\in S_i.$
		\item[({\bf b})]  The pure variable set $I$ can be determined uniquely from $\Sigma:=\text{Cov}(X)$. Moreover, its partition $\mathcal{I} := \{I_a\}_{1 \leq a \leq K}$ is unique and can be determined from $\Sigma$ up to label permutations.
	\end{itemize}
\end{thm}

\noindent  The identifiability of the allocation matrix  $A$ and  that of the collection of clusters $\mathcal{G}=\{ G_1,\ldots,G_k\}$ in (\ref{groups1}) use the results from Theorem \ref{I} in crucial ways.  We state the result in Theorem  \ref{ident} below.  

\begin{thm}\label{ident} 
	Assume that Model (\ref{mod}) with (i) - (iii) holds. Then, there exists a unique matrix $A$, up to a signed permutation, such that $X=AZ+E$. This implies that the associated overlapping clusters  $G_a$, for  $1 \leq a \leq K$,  are identifiable, up to label switching.  
\end{thm} 

\begin{rem} We show below that the pure variable assumption (ii) is needed for the  identifiability of $A$, up to a signed permutation. Assume that $X=AZ+E$ satisfies (i) and (iii),  but not (ii). We construct an example in which  $X$ can also be written as $X=\tilde A\tilde Z+E$, where $\tilde A$ and $\tilde Z$ satisfy the same conditions (i) and (iii), respectively, but  $\tilde A\neq AP$ for any $K\times K$ signed permutation matrix $P$ and $\tilde A$ may have a sparsity pattern different from $A$. To this end, we construct  $\tilde A$ and $\tilde Z$ such that $\tilde A\tilde Z=AZ$. Let $\tilde A=AQ$ and $\tilde Z=Q^{-1}Z$,  for some $K\times K$ invertible matrix $Q$ to be chosen such that Cov$(\wt Z) = Q^{-1}C(Q^{-1})^T$ satisfies (iii). In addition, we need to guarantee that $\tilde A=AQ$ satisfies (i). For simplicity, we set $K=3$. The following example satisfies all our requirements: 
	\[
	C = \begin{bmatrix}
	1 & 0 & 0\\ 0 & 2 & 0\\
	0 & 0 & 3
	\end{bmatrix},\qquad Q = \begin{bmatrix}
	1 & 1/3 & 0\\
	1/3 & 2 & 1/2\\
	0 & 1/2 & 2
	\end{bmatrix}.
	\]
	It is easy to verify that Cov$(\wt Z) = Q^{-1}C(Q^{-1})^T$ satisfies (iii). For any $1\le j\le p$, consider 
	\[
	A_{j\cdot}^T = (1/8, -3/8, 0)
	\]
	then 
	\[
	\widetilde{A}_{j\cdot}^T = A_{j\cdot}^TQ = (0, -17/24, -3/16)
	\]
	which also satisfies condition (i). However, $A_{j\cdot}$ and $\widetilde{A}_{j\cdot}$ have different sparsity patterns. Thus, if the matrix $A$ does not satisfy (ii),  $A$ is generally not identifiable. 
\end{rem}

\section{Estimation}\label{estimation} 
We develop  estimators from the observed data,  which is assumed to be  a sample of $n$ i.i.d. copies $X^{(1)},\ldots,X^{(n)}$ of $X \in \RR^p$, where $p$ is allowed to be larger than $n$.  Our estimation procedure consists of the following four steps: \\
(1) Estimate the pure variable set $I$, the number of clusters $K$ and the partition $\I$;\\ (2) Estimate $A_I$, the submatrix of $A$ with rows $A_{i\cdot}$ that correspond to $i\in I$;\\ (3) Estimate $A_{J}$, the submatrix of $A$ with rows  $A_{j\cdot}$  that correspond to $j\in J$;\\ (4) Estimate the overlapping clusters $\G = \{G_1,\ldots,G_K\}$.

\subsection{Estimation of $I$ and $\I$} \label{estpure} 
Given the different nature of their entries, we estimate the submatrices $A_I$ and $A_J$ separately. For the former, we first  estimate $I$ and its  partition $\mathcal{I} =\{ I_1, \ldots, I_K\}$, which can be both  uniquely constructed from $\Sigma$, as shown by Theorem  \ref{I}.   We   use the  constructive proof of Theorem \ref{I} for this step,  replacing the unknown   $\Sigma$ by the sample covariance matrix $$\wh {\Sigma}=  \frac{1}{n}\sum_{i=1}^nX^{(i)} (X^{(i)} )^T.$$ Specifically, we iterate through  the index set $\{1, 2, \ldots, p\}$, and use the sample version of part $({\bf a})$ of Theorem \ref{I} to decide 
whether an index $i$ is pure. If it is not deemed to be pure, we add it to the  set that estimates $J$. Otherwise,  we retain the estimated index set $\wh S_i$ of $S_i$ defined in (\ref{si}), which corresponds to an estimator of $M_i$ given by (\ref{mi}).  We then use the constructive  proof of part $({\bf b})$ of Theorem \ref{I}  to declare $\wh S_i \cup \{i\} := \wh {I}^{(i)} $ as an estimator of one of the partition sets of $\mathcal{I}$.  The resulting procedure has complexity $O(p^2)$, and we give all the specifics in Algorithm 1 of Section \ref{love}. The algorithm requires the specification of a tuning parameter $\delta$, which will be discussed in Section \ref{choice}.

\subsection{Estimation of the allocation submatrix $A_I$}\label{sec_est_AI}
Given the estimators $\wh I$, $\wh K$  and $\wh \I = \{\wh I_1, \ldots, \wh I_{\wh K}\}$ from Algorithm 1, we estimate the  matrix $A_I$ by a  $|\wh I|\times {\wh K}$  matrix  
with rows $i\in\wh I$ consisting of ${\wh K}-1$ zeros and one entry equal to either $+1$ or $-1$ as follows. For each $a\in [\wh K]$,  
\begin{enumerate}
	\item[(1)] Pick an element $i\in \wh I_a$ at random, and set $\wh A_{ia}=1$. 
	Note that $\wh A_{ia}$ can only be $+1$ or $-1$ by the definition of a pure variable. 	
	\item[(2)] For the remaining $j\in \wh I_a\bl\{i\}$, we set $\wh A_{ja} = \sgn(\wh \Sigma_{ij})$.
\end{enumerate} 
This procedure induces a partition of $\wh I_a = \wh I_a^1 \cup \wh I_a^2$, where $\wh I_a^1$ and $\wh I_a^2$ are defined below: 
\begin{equation}\label{subgroup}
\left\{\begin{array}{ll}
\wh A_{ka} = \wh A_{la}, & \text{ for } k,l\in \wh I_a^1 \text{ or }k,l\in \wh I_a^2\\
\wh A_{ka} \ne \wh A_{la}, &  \text{ for } k\in \wh I_a^1 \text{ and }l\in \wh I_a^2
\end{array}\right..
\end{equation}

\subsection{Estimation of the allocation submatrix $A_J$}\label{sec_est_AJ}

We continue by estimating the matrix $A_J$,  row by row.  To motivate our procedure, we begin by highlighting the 
structure of each row $A_{j\cdot}$ of $A_{J}$, for  $j \in J$. 
We recall that $A_{j\cdot}$ is sparse,  with $\|A_{j\cdot}\|_1 \leq 1$,  for each $j \in J$, as specified by  assumption 
(i).  In addition, model (\ref{mod}) subsumes a further constraint on each row $A_{j\cdot}$ of $A$, 
as explained below.  To facilitate notation, we rearrange $\Sigma$, $A$ and $\Gamma$ as follows:
$$\Sigma = \begin{bmatrix}
\Sigma_{II} & \Sigma_{IJ}\\
\Sigma_{JI} & \Sigma_{JJ} 
\end{bmatrix},\quad A= \begin{bmatrix}
A_I \\
A_J \end{bmatrix}
\quad \text{and}\quad  \Gamma= \begin{bmatrix}
\Gamma_{II} & 0\\
0 & \Gamma_{JJ} \end{bmatrix}.
$$ 
Model (\ref{mod}) implies the following decomposition of the covariance matrix $\Sigma$ of $X$: 
$$
\Sigma = \begin{bmatrix}
\Sigma_{II} & \Sigma_{IJ}\\
\Sigma_{JI} & \Sigma_{JJ} 
\end{bmatrix}  =
\begin{bmatrix}
A_I CA_I^T & A_ICA_J^T\\
A_JCA_I^T& A_J CA_J^T
\end{bmatrix} + \begin{bmatrix}
\Gamma_{II} & 0\\
0 & \Gamma_{JJ}
\end{bmatrix}.
$$
In particular, $\Sigma_{IJ} =  A_ICA_J^T$.
Thus, for each $i\in I_a$ with some $a\in[K]$ and $j\in J$, we have 
\begin{equation}\label{display4}
A_{ia} \Sigma_{ij} = A_{ia}^2 \sum_{b=1}^{K}   A_{jb}C_{ab}  =  \sum_{b=1}^{K}A_{jb}C_{ab}=C_{a\cdot}^TA_{j\cdot}.
\end{equation}
Averaging display (\ref{display4}) over all $i\in I_a$ yields
\begin{equation}\label{forbeta}
\frac{1}{|I_a|}\sum_{i\in I_a}A_{ia}\Sigma_{ij} = C_{a\cdot}^TA_{j\cdot}, \quad\text{for each }a\in[K]. 
\end{equation}
For each $j \in J$, we let   $$\beta^j := A_{j\cdot}$$ and
\begin{equation}\theta^j = \left(\frac{1}{|I_1|}\sum_{i\in I_1}A_{i1}\Sigma_{ij},\ldots,\frac{1}{|I_K|}\sum_{i\in I_K}A_{iK}\Sigma_{ij} \right)^T.
\end{equation}
Since $A_{ia} \in \{-1, 1\}$, for   each $i \in I_a$ and $a \in [K]$,  the entries of $\theta^j$  are respective averages 
of the sign corrected entries of $\Sigma$ corresponding to the partition of the  pure variable set.   Summarizing,   modeling assumption (i) and equation (\ref{forbeta}) above show that the estimation of $A_J$ reduces to estimating, for each $j \in J$, a $K$-dimensional vector  $\beta^j$ that  is sparse,  with norm $\| \beta^j\|_1\le1$,  and that satisfies the equation $$ \theta^j = C\beta^j.$$
Both $C$ and $\theta^j$, for each $j \in J$,  can be estimated directly from the data as follows.
For each $j \in  \wh J$,  we  estimate the $a$-th entry of $\theta^j$ by 
\begin{equation}\label{theta}
\wh \theta_{a}^j = \frac{1}{|\wh I_a|}\sum_{i\in \wh I_a}\wh A_{ia}\wh \Sigma_{ij},\ a\in[\wh K],
\end{equation}
and compute
\begin{equation}\label{Chat}
\wh C_{aa} = 
\frac{1}{|\wh I_a|(|\wh I_a|-1)}\!\sum_{i, j\in \wh I_a, i\ne j}\!\!\!|\wh \Sigma_{ij}|,\quad \wh C_{ab} = 
\frac{1}{|\wh I_a||\wh I_b|}\!\sum_{i\in \wh I_a, j\in \wh I_b}\!\!\!\wh A_{ia}\wh A_{ib}\wh \Sigma_{ij},
\end{equation}
for each $a\in[\wh K]$ and $b\in [\wh K]\bl\{a\}$ to form the estimator  $\wh C$ of $C$.  
The estimates (\ref{theta}) and (\ref{Chat}) rely crucially on having first estimated the  pure variables and their partition, according to  the steps  described  in Sections \ref{estpure} and \ref{sec_est_AI} above. 
\\ 

We have developed a   computationally efficient method to estimate $\beta^j$.
	We  exploit  the fact that the square matrix $C$ is  invertible  
	and take the   equation $\beta^j=C^{-1} \theta^j$ as our starting point. The idea is to
	first construct  a pre-estimator $\bar{\beta}^j = \wh {\Omega}\wh \theta^j$, based on an appropriate estimator $\wh {\Omega}$ of the precision matrix $\Omega:=C^{-1}$, followed by  a sparse projection of $\bar{\beta}^j$. Alternatively, and recommended to speed up the computation, we could use a simple hard threshold operation in the second step   as described in Remark 5.4.
	We first motivate our proposed estimator of $\Omega$. From the decomposition
	\begin{eqnarray}\label{decomp1}
	\bar\beta^j - \beta^j &=& \wh \Omega (\wh\theta^j-\theta^j) + (\wh\Omega-\Omega) \theta^j\nonumber\\
	&=&  \wh \Omega (\wh\theta^j-\theta^j) + (\wh\Omega C- I) \beta^j,
	\end{eqnarray}
	we immediately have
	\begin{eqnarray}\label{decomp2}
	\|\bar\beta^j - \beta^j \|_\infty &\le& \|  \wh \Omega \|_{\infty,1} \|\wh\theta^j-\theta^j\|_\infty + \| \wh\Omega C- I\|_\infty \| \beta^j\|_1.
	\end{eqnarray}
	Since we  can show, in Lemma 12 of the supplementary material,  that $\|\wh\theta^j-\theta^j\|_\infty$ has optimal convergence rate,  and  since $\|\beta^j\|_1 \leq 1$ under our model, our estimator  $\wh \Omega$ should ideally render  values for  $\|  \wh \Omega \|_{\infty,1}$ and $\| \wh\Omega C- I\|_\infty $ that are as small as possible.
	With this in mind, we propose 
	the linear program
	\begin{equation}\label{obj_omega}
	(\wh\Omega,\wh  t\ )=\arg\min\limits_{t\in\RR^{+},\ \Omega \in \RR^{\wh K \times \wh K}}t
	\end{equation}
	subject to
	\begin{equation}\label{constr_omega}
	\Omega = \Omega^T, \quad \|\Omega \wh C-I\|_\i\le  \lambda t, \quad \|\Omega\|_\r\le t,
	\end{equation}
	with   tuning parameter $\lambda$.
	This linear programming problem is clearly  tailored to our purpose, and its optimal solution 
	$\wh \Omega$  adds
	a novel estimator for $C^{-1}$ to the rich literature on  precision matrix estimation (\cite{meinshausen2006high,yuan2007model,friedman2008sparse,cai_CLIME,cai2016}, to name a few).
	Its novelty consists in (a) the usage  of the  matrix $\| \cdot\|_{\infty,1}$ norm, instead  of  the  commonly used matrix $\|\cdot\|_1$ norm, 
	and (b) the fact that  this norm appears in the upper bound of the restriction   (\ref{constr_omega}). 
	After  we compute $\bar \beta^j= \wh \Omega\wh\theta^j$, for each $j\in\wh J$, we  solve the following optimization problem
	\begin{equation}\label{obj_beta}
	\wh\beta^j=\arg\min_{\beta\in \RR^{\wh K}} \|\beta\|_1
	\end{equation}
	subject to
	\begin{equation}\label{constr_beta}
	\|\beta - \bar \beta^j\|_\i \le \mu,
	\end{equation}
	for some tuning parameter $\mu$ that is   proportional to $\| C^{-1}\|_{\infty,1}$,  to obtain our final  estimate $\wh\beta^j$ as the optimal solution of this linear program.
	This solution  is also sparse and properly scaled, in accordance to our model specification (i).
	Then,  $\wh A_{\wh J}$ is the matrix with rows $\wh\beta^j$, for  $j\in\wh J$. Our final  estimator $\wh A$ of $A$ is obtained by concatenating $\wh A_{\wh I}$ and $\wh A_{\wh J}$.   Its  statistical property is analyzed in Section \ref{stat_guarantee},  along with  precise forms of the tuning parameters needed for its construction. \\
	
	An alternative way to estimate $\beta^{j}$ is by the following Dantzig-type estimator. 	 Starting with the equation $\theta^j = C\beta^j$, we can consider, for each $j \in \wh J$,  the linear program 
	\begin{equation} \label{Dantzig}
	\min_{\beta\in\RR^{\wh{K}}} \| \beta\|_1 
	\end{equation}
	subject to
	\begin{equation}
	\label{Dantzig_cond}
	\| \wh C \beta - \wh\theta^j \|_\infty \le \lambda',
	\end{equation}
	with tuning parameter $\lambda'$. 
	The solution  is sparse and properly scaled, in accordance to our model specification (i).
	Our final goal of support  recovery  of $\beta^j$ still requires an additional hard thresholding step of the solution of this linear program. 
	In this case, the  appropriate threshold $\mu$ is   proportional to the  $\ell_\infty$-sensitivity of the matrix $C$, introduced by \cite{gt2011}. The latter quantity depends on the unknown support  of 
	the different rows $\theta^j$, but can be upper bounded by $\| C^{-1} \|_{\infty,1}$. The statistical properties of this procedure are  analyzed in Section \ref{stat_guarantee} as well.\\
	
	Both procedures require, in practice, the  estimation of the quantity $\| C^{-1}\|_{\infty,1}$. 
	The procedure in (\ref{obj_beta}) - (\ref{constr_beta}) recovers the support of $\beta$ automatically while the procedure in (\ref{Dantzig}) - (\ref{Dantzig_cond}), even though it renders a sparse solution, requires a further hard-thresholding step for the support recovery.

\subsection{Estimation of the overlapping groups}\label{sec_est_group}

Recalling the definition of groups in (\ref{groups1}), the overlapping groups are estimated by
\begin{equation}\label{estG}
\wh \G = \bigl\{\wh G_1,\ldots,\wh G_{\wh K}\bigr\},\quad \wh G_a= \bigl\{i\in[p]: \wh A_{ia} \ne 0\bigr\}, \text{ for each $a\in[\wh K]$}.
\end{equation}
Variables $X_i$ that are  associated (via $\wh A$)  with the same latent factor $Z_a$ are  therefore  placed  in the same group $\wh G_a$.  
To accommodate  potential  pure noise variables, we further define
\begin{equation}\label{noise_group}
G_0 := \bigl\{j \in \{1, \ldots, p\}:\ A_{ja}=0, \text{ for all } a \in \{1, \ldots, K\}\bigr\}
\end{equation}
as the pure noise cluster. 
We can estimate $G_0$ in (\ref{noise_group}) by
\begin{equation}\label{estG_noise}
\wh G_0= \bigl\{i\in[p]: \wh A_{ia} = 0, \text{ for all $a\in[\wh K]$}\bigr\}.
\end{equation}
However, our main focus is on $\G$ because it completely determines $G_0$.

In many applications, it may be of interest to identify the sub-groups of variables that are all either positively  or negatively associated with the same latent factor. To this end, we define 
\begin{align}\label{groups2}
\G^s &:=  \bigl\{ G^s_1,\ldots, G_{ K}^s\bigr\}, \\ \nonumber G_a^s &:= \bigl\{ G_a^{1},  G_a^{2}\bigr\} := \Bigl\{\bigl\{i\in G_a:  A_{ia} > 0 \bigr\}, \bigl\{i\in G_a:  A_{ia} < 0 \bigr\}\Bigr\},
\end{align} 
for each $a\in[K]$, and they are estimated by 
\begin{align}\label{estG_s}
\wh \G^s &= \bigl\{\wh G^s_1,\ldots,\wh G_{\wh K}^s\bigr\},\\
\wh G_a^s&= \Bigl\{\bigl\{i\in\wh G_a: \wh A_{ia} > 0 \bigr\}, \bigl\{i\in\wh G_a: \wh A_{ia} < 0 \bigr\}\Bigr\},\nonumber
\end{align}
for each $a\in[\wh K]$.
The fact that $A$ is only identifiable up to a signed permutation matrix, has  the repercussion that the labels of the two sub-groups in $G_a^s$ are not identifiable. Thus, variables placed in  the subgroups $G_a^1$ and $G_a^2$  are, respectively,  associated with $Z_a$ in the same direction.
The directions between two sub-groups, henceforth called direction sub-groups,  are opposite. This can be identified, although the direction itself cannot. We show in Section \ref{stat_guarantee}  that the direction sub-groups can be  identified, and well estimated.

\subsection{LOVE: A {\bf L}atent variable model approach for {\bf OVE}rlapping clustering.} \label{love}

We give  below the specifics of Algorithm \ref{alg1}, motivated in Section \ref{estpure},  and summarize our final algorithm, LOVE in Algorithm \ref{alg3}.  

{\begin{algorithm}[ht]\small
		\caption{Estimate the partition of the pure variables $\I$ by $\wh \I$}\label{alg1}
		\begin{algorithmic}[1]
			\Procedure {PureVar}{$\wh \Sigma$, $\delta$}
			\State $\wh \I \gets \emptyset$.
			\ForAll {$i\in [p]$} 
			\State $\wh I^{(i)} \gets \bigl\{l\in [p]\setminus\{i\}: \max_{j\in [p]\setminus\{i\}}|\wh \Sigma_{ij}| \le |\wh \Sigma_{il}|+2\delta\bigr\}$
			\State $Pure(i) \gets True$.
			\ForAll {$j \in \wh I^{(i)}$}
			\If {$\bigl||\wh \Sigma_{ij}|- \max_{k\in [p]\bl\{j\}}|\wh\Sigma_{jk}|\bigr| > 2\delta$}   
			\State $Pure(i) \gets False$,
			\State \textbf{break}
			\EndIf	
			\EndFor
			\If {$Pure(i)$}
			\State $\wh I^{(i)} \gets \wh I^{(i)}\cup \{i\}$
			\State $\wh\I \gets$ \textsc{Merge($\wh I^{(i)},\ \wh \I$)}
			\EndIf
			\EndFor
			\State \Return $\wh \I$ and $\wh K$ as the number of sets in $\wh \I$
			\EndProcedure
			\Statex
			\Function {Merge}{$\wh I^{(i)}$, $\wh \I$}\label{alg2}
			\ForAll {$G \in \wh \I$}
			\Comment $\wh \I$ is a collection of sets
			\If {$G \cap \wh I^{(i)}\ne \emptyset$} 
			\State  $G\gets G\cap \wh I^{(i)}$
			\Comment Replace $G\in \wh \I$ by $G\cap \wh I^{(i)}$
			\State \Return $\wh \I$
			\EndIf
			\EndFor
			\State $\wh I^{(i)} \in \wh \I$
			\Comment add $\wh I^{(i)}$  in $\wh \I$
			\State \Return $\wh \I$
			\EndFunction
		\end{algorithmic}
\end{algorithm}}

{\begin{algorithm}[ht]\small
		\caption{ The LOVE procedure  for overlapping clustering.}\label{alg3}
		\begin{algorithmic}[1]
			\Require $\wh \Sigma$ from I.I.D. data $(X^{(1)},...,X^{(n)})$, the tuning parameters $\delta$, $\lambda$ and $\mu$. 	
			\State Apply Algorithm \ref{alg1} to obtain the number of clusters $\wh K$, the estimated set of pure variables $\wh I$ and its  partition of $\wh \I$. 
			\State Estimate $A_I$ by $\wh A_{\wh I}$ from  (\ref{subgroup}). 
			\State Estimate $C^{-1}$ by $\wh \O$ from (\ref{obj_omega}) and $\bar{\beta}^j$ for each $j\in \wh J$.
			\State Estimate $A_{J}$ by $\wh A_{\wh J}$ from (\ref{obj_beta}).  Combine $\wh A_{\wh I}$ with $\wh A_{\wh J}$ to obtain $\wh A$.
			\State Estimate overlapping groups $\wh \G=\{\wh G_1,...,\wh G_{\hat K}\}$ and its direction subgroups $\wh \G^s=\{\wh G^s_1,...,\wh G^s_{\hat K}\}$ from (\ref{estG}) - (\ref{estG_s}) by using  $\wh A$.\\
			Output $\wh A$, $\wh \G$ and $\wh \G^s$.
		\end{algorithmic}
\end{algorithm} }

\section{Statistical guarantees}\label{stat_guarantee}
We provide in this section statistical guarantees for: 
\begin{enumerate}
	\item[(1a)] The estimated number of clusters $\wh K$;
	\item[(1b)] The estimated pure variable set $\wh I$ and its estimated  partition $\wh \I$; 
	\item[(2)]  The estimated allocation matrix $\wh A$ and its adaptation to the unknown  row sparsity of $A$.  
	\item[(3)] The individual Group False Positive Proportion (GFPP), the individual Group False Negative Proportion(GFNP), the Total False Positive Proportion (TFPP) and the Total False Negative Proportion (TFNP) for the estimated overlapping groups.
\end{enumerate}
We make the blanket assumption for the remainder of this paper that $X$ is {\em sub-Gaussian}, that is,  the {Orlicz} norm $\|X_j\|_{\psi_2}$ of each $X_j$ is bounded by a common constant $\sigma_*$.\footnote{
	The Orlicz norm of $X_j$ is defined as $\Vert X_j\Vert_{\psi_2}=\inf\left\{ c>0 :
	\EE\left[ \psi_2\left( {\vert X_j\vert} / {c}\right)\right] <1 \right\},$
	based on the {Young} function $\psi_2(x)=\exp(x^2) -1$.} The sub-Gaussian condition implies $\max_{j\in[p]}\Sigma_{jj} \le 2\sigma_*^2$ and $  \| C \|_\infty \le 2\sigma_*^2$. Let 
\begin{equation}\label{delta}  
\E =\E(\delta) := \left\{  \max_{1\le i < j \le p} \left| \wh \Sigma _{ij} - \Sigma_{ij} \right| \le \delta \right\}. \end{equation} 
We assume throughout that
$\delta = c_0\|\Sigma\|_\i\sqrt{\log (p\vee n) /n}$, for some absolute constant $c_0$, 
and $\log p = o(n)$, so that $\delta = o(1)$, for $n$ large enough, where $a\vee b=\max(a,b)$. Taking $c_0>0$ large enough, Lemma 2 in \cite{bien2016convex} guarantees that $\E$ holds with high probability: 
\begin{equation}\label{max}
\PP(\E) \ge 1-c_1(p\vee n)^{-c_2}
\end{equation}
for some  positive, finite constants $c_1$ and $c_2$.  Apart from $\delta$, the quantity 
\begin{equation}\label{deltanu}   
\Delta(C) := \nu > 0, 
\end{equation}
plays an important role in our analysis. 
Indeed, assumption (iii)  requires that $\nu>0$
in order to guarantee  that the latent factors are  distinguishable from one another. We can view $\nu$ 
as a measure of their separation, and naturally therefore, the size of $\nu$
impacts
the quality of all our estimators, in addition to the magnitude  of $\delta$.

\begin{rem}
	It is   common practice to standardize the data in a pre-processing step, and perform  statistical analyses  on the standardized data. Our model can be easily adapted to this case
	by assuming 
	that the latent variable model holds for a standardized version of $X$, specifically for $\wt X:= (\text{diag}(\Sigma) )^{-1/2}(X-\EE(X))$, leading to 
	\begin{equation}\label{eqmodel2}
	\wt X = AZ+E
	\end{equation}
	with $A$, $Z$ and $E$ satisfying the same conditions (i), (ii) and (iii). Recall that in model (\ref{mod}) we have already assumed that $X$ has mean zero. Transforming model (\ref{eqmodel2}) back to the original scale, we have $X-\EE[X] = [ (\text{diag}(\Sigma) )^{1/2} A ] Z + [(\text{diag}(\Sigma)) ^{1/2} E]$.  We note that the new allocation  matrix $\wt A:= [ (\text{diag}(\Sigma)) ^{1/2} A ]$ has the same support as $A$. Moreover, a pure variable $j$ in cluster $a$ satisfies $|\wt A_{ja}|=\Sigma^{1/2}_{jj}$. Therefore,  pure variables are given different weights, proportional to their respective standard deviations, which relaxes the equal weight restriction in Condition (ii). The caveat is that, under (\ref{eqmodel2}), we have   $1=\text{Cov}(\wt X_j)=A_{j\cdot}^TCA_{j\cdot}+\Gamma_{jj}$ for any $1\leq j\leq p$. This further implies that $\Gamma_{jj}=\Gamma_{j'j'}$ for any $j, j'\in I_a$, that is, model (\ref{eqmodel2}) subsumes 
	that the random noise has the same variance for all pure variables in each cluster.
	Depending on what modeling assumptions best fit a particular problem, either (\ref{mod}) or (\ref{eqmodel2}) can be considered.  The identifiability of   model (\ref{eqmodel2}) follows directly from the proof of Theorem \ref{ident}. The LOVE algorithm, presented in the next subsection, is also applicable, provided we replace the sample covariance matrix $\wh\Sigma$ with the sample correlation matrix $\wh R$ with entries
	$$
	\wh R_{jk}=\frac{1}{n}\sum_{i=1}^n (X^{(i)}_{j}-\bar X_{j})(X^{(i)}_{k}-\bar X_{k})/(\textrm{sd}(X_{j})\textrm{sd}(X_{k})),
	$$
	with $\bar X_{j}=n^{-1}\sum_{i=1}^n X^{(i)}_j$ and $\textrm{sd}(X_{j})=\{{n}^{-1}\sum_{i=1}^n (X^{(i)}_{j}-\bar X_{j})^2\}^{1/2}$. Then, all our theoretical guarantees hold unchanged on the new event 
	\begin{equation*}
		\E =\E(\delta) := \left\{  \max_{1\le i < j \le p} \left| \wh R _{ij} - R_{ij} \right| \le \delta \right\}. 
	\end{equation*} 
	Since	\cite{bunea2015minimax}	 showed that $\E$ holds with high probability by choosing 
	$\delta = c_0\sqrt{\log (p\vee n) /n}$, for some constant $c_0$, we can obtain the same statistical guarantees under the  model (\ref{eqmodel2}).
\end{rem}

\subsection{Statistical guarantees for $\wh K$, $\wh I$  and $\wh \I$}\label{whI}

We first analyze   the performance of our estimator $\wh I$  of $I$, and its corresponding partition.  This problem 
belongs to the general class of pattern recovery problems, and it is well understood that under strong enough signal conditions one can expect   $\wh I = I$, with high probability. This turns out to be  indeed the case for our problem, but we obtain this as a corollary of a more general result. We set out to quantify when our estimated set contains the least  taxing type of errors, under minimal assumptions.  To make this precise, we introduce the concept of  {\it quasi-pure} variables.
A quasi-pure variable $X_i$ has very strong association with only one latent factor, say $Z_a$, in that $|A_{ia}| \approx1$, 
and very low association with the rest: $|A_{ib}| \approx 0$, for all $b \neq a$.  Formally, we define the set of quasi-pure variables as: 
\begin{equation}\label{J1}
J_1 := \{ j\in J: \text{there exists }a\in [K],\text{ such that }|A_{ja}| \ge 1-4\delta/\nu\}.
\end{equation}
For each $a\in[K]$ we further define the set of quasi-pure variables associated with the same factor:
\begin{equation}\label{Ja}
J_1^a := \{j\in J_1: |A_{ja}| \ge 1-4\delta/\nu \}.
\end{equation}
When $\nu$ is a strictly positive constant, $\epsilon:=4\delta/\nu =  o(1)$.  The lower bound $|A_{ja}| \ge 1-\epsilon$ in 
$(\ref{Ja})$ implies, under condition (ii), that  $|A_{jb}| \le \epsilon$,  for any $b\ne a$ and $j\in J_1^a$,  justifying  the name quasi-pure variables
for those components of $X$ with indices in $J_1$.  We observe, for future reference, that  $\{J_1^1,\ldots, J_1^K\}$ forms a partition of $J_1$. 

We show in Theorem \ref{consistI}  that, with very high probability, the estimated $\wh I$ contains the pure variable set $I$, and is in turn contained in a set that includes all pure variables and quasi-pure variables.
Importantly, $\wh I$ will not include indices of variables $X_j$ that are associated  with multiple latent factors at a level 
higher than $\epsilon$.  Equally importantly, if a quasi-pure variable $X_i$ is  included in $\wh I$,  then this variable will 
have the corresponding $|A_{ia}| \approx 1$, and it will be placed together with the pure variables associated with the same factor $Z_a$, for some $a$, and not in a new cluster. This is crucial for ensuring that the number of clusters $K$ is consistently estimated, and also for establishing the cluster misclassification proportion in Section \ref{stat_guarantee_group} below.

\begin{thm}\label{consistI}
	Assume Model (\ref{mod}) with (i) - (iii), and
	\begin{equation}\label{nu}
	\nu>2\max\left( 2\delta,\sqrt{2\| C\|_\infty \delta} \right).
	\end{equation}   
	Then: 	\begin{enumerate}
		\item[(a)] $\wh K = K$;
		\item[(b)] $I\subseteq \wh I \subseteq I\cup J_1$.
	\end{enumerate}
	Moreover, there exists a label permutation $\pi$ of the set $ \{1,\ldots,K\}$, such that the output $\wh \I=\bigl\{\wh I_a\bigr\}_{a\in [K]}$ from Algorithm \ref{alg1} satisfies:
	\begin{enumerate}
		\item[(c)] $I_{\pi(a)} \subseteq \wh I_{a} \subseteq I_{\pi(a)}\cup J_1^{\pi(a)}$.
	\end{enumerate}
	All results hold with probability larger than $1 - c_1(n\vee p)^{-c_{2}}$, for $c_1, c_2$ positive constants defined in (\ref{max}).
\end{thm}

\noindent
The conclusion of Theorem \ref{consistI} holds only under condition (\ref{nu}), which stipulates that the separation between the latent factors, as measured by $\nu$,  is not only strictly positive,  which was needed  for identifiability, but slightly above a quantity that depends on the estimation error $\delta$, and which becomes $o(1)$ for $n$ large enough. From the inspection of the proof, condition (\ref{nu}) can be relaxed to $\nu > 4\delta$ when $J_1 = \emptyset$.

\begin{rem}\label{rem1} Let $e_1 = (1,0,\ldots, 0)^T$ and $\H_K$ be the hyperoctahedral group of signed permutation matrices.
	If   $A_I$ and $A_J$ are well separated in the sense that
	\[	\min_{j\in J,\ P\in \H_K} \bigl \|A_{j\cdot} - P e_1\bigr \|_1 > 8\delta/\nu,
	\]
	then $J_1 = \emptyset$, and  Theorem \ref{consistI}  yields exact recovery of the pure variable set and of its partition: $\wh I = I$ and $\wh \I = \I$, with high probability.  However, we  expect $J_1 \neq \emptyset$, as we  expect quasi-pure variables to be present in a  high dimensional model, which is the context for  which  Theorem \ref{consistI} has been established. 
\end{rem}

\subsection{Statistical guarantee for $\wh A$}\label{stat_guarantee_A}

In this section we state, and comment on, the statistical properties of the estimate $\wh A$ obtained in Sections \ref{sec_est_AI} and 
\ref{sec_est_AJ}.  Recall that $\delta = O(\sqrt{\log (p\vee n)/n})$ was given in (\ref{delta}) above, and the estimation 
of $A_J$ made use of two tuning parameters:  $\lambda$, in (\ref{constr_omega}),   and $\mu$,   in (\ref{constr_beta}). 
Theorem \ref{prop_rates} 
establishes the properties of our estimates relative to the theoretically optimal values of these tuning parameters, both of which are functions of $\delta$, while their data adaptive calibration is discussed in Section \ref{choice} below. 
We let   
$\lambda= 2\delta'$ and $ \mu = 5\| C^{-1} \|_\r\delta'$, 
with
\begin{equation}\label{delta'}
\delta'  = \left( \frac{8}{\nu} \|C\|_\i -3 \right) \delta, 
\end{equation}
for $\nu$ defined in (\ref{deltanu}) above.  When  $\nu$ and $\|C\|_{\infty}$ are strictly positive constants  we thus have 
$\lambda = O(\sqrt{\log (p\vee n)/n})$ and $\mu = O(\| C^{-1} \|_\r\sqrt{\log (p\vee n)/n})$. 
We consider the loss function for two $p\times K$ matrices $A, A'$ as 
\begin{equation}\label{loss}
L_q(A, A'):= \min_{P\in \H_K}\|AP-A'\|_{\i,q},\quad 1\le q\le \infty.
\end{equation}
Here $\H_K$ is the  hyperoctahedral group of all $K\times K$ signed permutation matrices and 
$$\|A\|_{\i,q}:=\max_{1\le i\le p}\left\|A_{i\cdot}\right\|_q = \max_{1\le i\le p}\left( \sum_{j=1}^K|A_{ij}|^q\right)^{1/q},$$ 
for a generic matrix $A\in \RR^{p\times K}$.

\begin{thm}\label{prop_rates}
	Assume the conditions in Theorem \ref{consistI} hold.  Let $\lambda$ and $\mu$ be as defined above, and set $s= \max_{i\in [p]} \|A_{i\cdot}\|_0$. Then,  
	\begin{equation*}
		L_q( \wh A, A)   \le  10 s^{1/q}\| C^{-1} \|_\r\delta',\quad 1\le q\le \infty,
	\end{equation*}
	with probability larger than $1 - c_1(n\vee p)^{-c_2}$, for $c_1, c_2$ positive constants defined in 	(\ref{max}),  provided that
	$(2\mu+ 4\delta/\nu )< 1$.	We use the convention that $s^{1/q}=1$ for $q=+\infty$.
\end{thm}	

\begin{rem}
	\mbox{}
	\begin{enumerate}
		\item In fact, we prove the stronger result 
		\begin{equation*}
			\min_{P\in \H_K}  \bigl\| \wh A_{i\cdot} - (AP)_{i\cdot} \bigr\|_q \le 10(s_i)^{1/q}\| C^{-1} \|_\r\delta',\quad 1\le q\le \infty,
		\end{equation*} with sparsity index $s_i=\| A_{i\cdot}\|_0$ 
		for each row $A_{i\cdot}$, $i\in[p]$ of $A$.  The signed permutation matrix $P$ that achieves the minimum is determined by the alignment of the pure variables and is the same for each $i\in [p]$.

		\item
		Inspection of the proof of this result quickly reveals that
		$\|\wh A_{i\cdot}\|_1 \leq 1$,  for each $i\in[p]$, with high probability, in accordance with our model requirement (i). 
		\item
		The size of $\|C^{-1}\|_\r $ ranges 
		from the constant $\|C^{-1}\|_\infty$, when all latent factors are independent, to   the fully general case of $\|C^{-1}\|_\r = O(K)$. In the latter case the bounds become meaningful when $K < O( \sqrt{n/\log p})$. However, if $C^{-1}$ is sparse,  then $\|C^{-1}\|_\r$ may be considerably smaller than $K$. 
		In particular, if $Z$ has a multivariate normal distribution and many factors $Z_i$ are   conditionally independent,   then $\|C^{-1}\|_\r$ is small. We do not make any of these assumptions here, and regardless of the situation, Theorem \ref{prop_rates} shows that our estimation procedure  adapts automatically to it.
	\end{enumerate}
\end{rem}

Our primary focus is the bound for $q=+\infty$, as this leads to inference on support recovery of 
$A$. 
More generally, for any $q \geq 1$, it is well understood  that the quality of estimating a sparse vector in high-dimensional regression-type models   depends on the interplay between  its sparsity and  the behavior of the appropriate Gram matrix associated with the model, which reduces to $C = \EE [ZZ^T]$ in our case.  The concept of  $\ell_q$-sensitivity,  introduced by  \cite{gt2011}, is  the most general characterization of this interplay to date. It  offers a link between  the $\ell_q$-norm of sparse vectors $\beta $ and  the $\ell_\infty$-norm of the   product between the Gram matrix  and $\beta$,
uniformly over vectors $\beta$ of sparsity $s$, ranging over a collection of  cones.
Formally,  the $\ell_q$-sensitivity of the matrix $C$ is defined as
\begin{equation}\label{sensi}
\kappa_q(C,s):= \inf_{|S|\le s} \inf_{v \in \mathcal {C}_S } \frac{\| C v \|_\infty}{\| v\|_q},
\end{equation}
with $\mathcal {C}_S:=\left\{v\in \RR^{K}:\ \| v_{\bar S}\|_1\le \| v_{S}\|_1 \right\}$ and $S\subseteq [K]$ with $|S|\le s$.
In our context,  that of a square, {\em invertible}  matrix $C$, the reciprocal of the $\ell_\infty$-sensitivity $\kappa_\infty(C,s)$ becomes essentially $\| C^{-1} \|_{\infty,1}$ with $[\kappa_\infty(C,K)]^{-1} = \| C^{-1}\|_\r$, which indeed links $\| \beta\|_\infty$ to $\| C\beta\|_\infty$.     Similarly,  the quantities 
$(2s)^{1/q} \| C^{-1}\|_{\infty,1}$ provide   concrete substitutes of the reciprocals of the $\ell_q$-sensitivities of $C$, and  all of our   rates in Theorem \ref{prop_rates} 
match the lower bounds in Theorem \ref{thm_minimax}, up to a logarithmic factor, and the quantities $\| C^{-1} \|_\r$ and $\lambda_1(C)$.\\

Another possible  estimation procedure is   the linear program (\ref{Dantzig}) - (\ref{Dantzig_cond}) with tuning parameter
$\lambda'=3 \delta'$. We denote its solution by $\widehat{A}_{D}$. 

\begin{thm}\label{thm:Dantzig}
	Assume the conditions in Theorem \ref{consistI} hold.  Let $\lambda'=3\delta'$   and set $s= \max_{i\in [p]} \|A_{i\cdot}\|_0$. Then,  
	\begin{eqnarray} \label{uno}
	L_q( \widehat{A}_D, A)   &\le& 6 [ \kappa_q(C,s)] ^{-1} \delta',\\
	&\le& 6 \| C^{-1}\|_\r(2s)^{1/q}  \delta', \qquad 1\le q\le \infty,\label{duo}
	\end{eqnarray}
	with probability larger than $1 - c_1(n\vee p)^{-c_2}$, for $c_1, c_2$ positive constants defined in 	(\ref{max}). 
	We use the convention that $s^{1/q}=1$ for $q=+\infty$.
\end{thm}	

As discussed in Section \ref{sec_est_AJ}, we would need to further threshold $\wh A_{D}$ in order to build the desired clusters. The thresholding level is   proportional  to  $\|\wh A_{D} -A\|_{\i}$, and its practical implementation would require an estimator of  $[\kappa_\infty(C,s)]^{-1}$, which  cannot be computed. One can however bound $[\kappa_\infty(C,s)]^{-1}$ by $\| C^{-1}\|_{\infty,1}$ as in (\ref{duo}), which becomes identical to the rate of convergence of $\wh A$ in Theorem \ref{prop_rates}.

We now show that the rates of convergence in Theorems \ref{prop_rates} and \ref{thm:Dantzig} are optimal (up to a logarithmic factor in $p$) in a minimax sense for all estimators over
the parameter space 
\begin{align*}
\A_s& := \left\{  A\in [-1,1]^{p\times K}: A \text{ satisfies   (i) and (ii) and }  \max_{1\le i\le p}   \| A_{i\cdot} \| _0 \le s  \right\}.
\end{align*}
For our purpose of establishing a minimax lower bound,  it suffices to consider a particular sub-Gaussian distribution of $X$ and a particular covariance matrix $C$.
We choose to take the multivariate Gaussian  
$N_p(\0, ACA^T + \sigma^2{\bI_p})$ with $A\in \A_s$, any positive definite $C$ and  
some constant $\sigma^2>0$, satisfying (\ref{ass_rate}) below.

\begin{thm}\label{thm_minimax}
	Assume  $X\sim N_p(\0,ACA^T+\sigma^2\bI_p)$. Let $K \ge 2$, $p \ge 2K+1$, $1\le s \le 4K/5$ and 
	\begin{equation}\label{ass_rate}
	s\sqrt{\sigma^2\over \lambda_1(C)}
	\sqrt{\log (K/s) \over n} \le c_1,
	\end{equation}
	for some constant $c_1>0$.
	Then, for all $1\le q\le \infty$,
	\begin{equation}\label{minimax}
	\inf_{\wh A}\sup_{A\in \A_s }\PP_{A }\left\{
	L_q(\wh A, A) 
	\ge  c_2 s^{1/q} \sqrt{\frac{\sigma^2}{\lambda_1(C)}}  \sqrt{\frac{\log (K/s)}{n}}\right\} \ge c_3,
	\end{equation}
	for some positive constants $c_2, c_3$ depending solely on 
	$c_1$.
	The infimum is taken  over all estimators $\wh A$ of $A$ and we use the convention $s^{1/q}=1$ for $q=+\infty$. 
\end{thm}
We attain this bound, up to logarithmic factors, even when $I$ and its partition are  not known,  for suitable covariance matrices $C$. Indeed,  
Theorems \ref{prop_rates}, \ref{thm:Dantzig} and \ref{thm_minimax} immediately imply   that our procedures are not only adaptive in $s$, but minimax optimal over $A\in \A_s$,  
up to a logarithmic  $\log(K/s)$ and $\log ( p\vee n)$,   for any covariance matrix $C$ with bounded (constant) $\nu$,  
$  \lambda_1(C)$ and  $ \| C^{-1}\|_{\infty,1}$.
We note that if $Z$ were observed, then an $\ell_0$ penalized least squares estimator of $A$ would have an error upper bound containing the factor $\log(K/s)$.  From this perspective, the factor  $\log(K/s)$ in the lower bound (\ref{minimax}), derived for unobservable $Z$, is sharp. 
The $\log(p)$-term in the upper bound of our estimator stems directly from our choice of $\delta$ in (\ref{delta}) that controls $\| \widehat\Sigma -\Sigma\|_\infty$, for sub-Gaussian distributions, and cannot be dispensed with in our  estimation procedure of $I$ and $A$. 
Finally, our bounds are established over large classes $\A_s$, without additional assumptions on $A$, at the expense of placing conditions on $C$. Even in the classical linear regression model, there is a mismatch -- for instance, in terms of largest and smallest eigenvalues of the Gram matrix -- between minimax lower bounds for estimating the vector of regression coefficients and achievable upper bounds. Our rates coincide with the minimax rates obtained by \cite{belloni2016} in the errors in variables context, where, just like in our case, the design is not observed.

\subsection{Statistical guarantee for $\wh \G$ and $\wh \G^s$.}\label{stat_guarantee_group}
For easy of presentation, and without loss of generality, throughout this section, we continue to write $A$ for its  orthonormal transformation  $AP$ that uses the optimal signed permutation matrix $P\in\H_K$ from Theorem \ref{prop_rates} to align the  columns and signs of $A$ with that of $\wh A$. 

We define two criteria  to evaluate the estimated clusters $\wh \G$ on the event $\wh K = K$.  The latter holds   with high probability by Theorem \ref{consistI}.
We first define the individual Group False Positive Proportion (GFPP) and the individual Group False Negative Proportion (GFNP) as,
\begin{equation}\label{g_fpr_fnr}
\rm{GFPP}(\wh G_a) := \frac{|(G_a)^c\cap \wh G_a|}{|(G_a)^c|},\quad  \rm{GFNP}(\wh G_a) := \frac{|G_a\cap (\wh G_a)^c|}{|G_a|},
\end{equation}
for each $a\in [K]$,
where $(G_a)^c := [p]\setminus G_a$ and $(\wh G_a)^c := [p]\setminus \wh G_a$, with the convention $\rm{GFPP}(\wh G_a) = 0$ if $|(G_a)^c| = \emptyset$.
GFPP and GFNP quantify the misclassification proportion within each group $\wh G_a$. Furthermore, with the same convention, we can define the Total False Positive Proportion (TFPP) and Total False Negative Proportion (TFNP) to quantify the overal misclassification proportion of $\wh \G$.
\begin{equation}\label{t_fpr_fnr}
\textrm{TFPP}(\wh \G) := \frac{\sum_{a=1}^{K}|(G_a)^c\cap \wh G_a|}{\sum_{a=1}^{K}|(G_a)^c|},\quad  \textrm{TFNP}(\wh \G) := \frac{\sum_{a=1}^{K}|G_a\cap (\wh G_a)^c|}{\sum_{a=1}^{K}|G_a|}.
\end{equation}
Finally, given $\mu = 5\|\O\|_\r\delta'$ with $\delta'$ specified in (\ref{delta'}), we define 
\begin{equation}\label{goodJ}
J_2 := \{ i\in J:  \text{for any }a \text{ with } A_{ia}\ne0,\ |A_{ia}| > (2\mu)\vee (4\delta/\nu)\}.
\end{equation}
and $J_3 := J\bl (J_1\cup J_2)$. $J_2$ can be viewed as the set where every non-zero entry of $A_{j\cdot}$ is separated away from $0$ for each $j\in J_2$. The following theorem shows that $J_2$ plays a critical role in quantifying both the support recovery of $\wh A$ and the misclassification proportion of $\wh \G$. Let 
$\wh S := \supp(\wh A)$. 

\begin{thm} \label{propgroup}
	Under the conditions of Theorem \ref{prop_rates},  with probability greater than $1-c_1(n\vee p)^{-c_2}$ for some positve constant $c_1$ and $c_2$ defined in (\ref{max}), we have:
\begin{enumerate}
	\item[(a)] 
	$\supp\bigl(A_{J_2}\bigr)\subseteq \supp(\wh A) \subseteq \supp\left(A\right),\quad\sgn(\wh A_{\wh S}) = \sgn\bigl(A_{\wh{S}}\bigr).$
	\vspace*{2mm}
	\item[(b)] Let $s_j^a = 1\{|A_{ja}|\ne 0\}$ and $t_j^a =1\{|A_{ja}|\le (2\mu)\vee (4\delta/\nu)\}$, for each $j\in J$ and $a\in [K]$. 
	\begin{equation}\label{gfnr}
	\textrm{GFPP}(\wh G_a) = 0 ;\quad
	\textrm{GFNP}(\wh G_a) \le \frac{\sum_{j\in J_1 \cup J_3\setminus J_1^a}t_j^a}{\sum_{j\in J}s_j^a+|I_a|}.
	\end{equation}
	\item[(c)]  Let $s_j = \sum_{a=1}^{K}1\{|A_{ja}|\ne 0\}$ and $t_j = \sum_{a=1}^{K}1\{|A_{ja}|\le (2\mu)\vee (4\delta/\nu)\}$, for each $j\in J$.
	\begin{equation}\label{fnr}
	\textrm{TFPP}(\wh \G) = 0 ;\qquad
	\textrm{TFNP}(\wh \G) \le \frac{\sum_{j\in J_1\cup J_3}t_j}{ \sum_{j\in J}s_j + |I|}.
	\end{equation}	
\end{enumerate}
\end{thm}		

\begin{rem}\label{rem3} \mbox{}
	\begin{enumerate}
		\item
		From our proof of Theorem \ref{propgroup}, it is easy to verify that the expression of $\textrm{TFNP}$ in (\ref{fnr}) continues to hold for the Direction False Positive Proportion (DFPP) and the Direction False Negative Proportion (DFNP) defined in (\ref{eqsfpr}) below with $s_j$ replaced by $\sum_{a=1}^{K}\1\{ A_{ja} < 0\}$ or $\sum_{a=1}^{K}\1\{ A_{ja} > 0\}$, $t_j$ replaced by $\sum_{a=1}^{K}1\{ -(2\mu)\vee (4\delta/\nu) \le A_{ja} < 0 \}$ or $\sum_{a=1}^{K}1\{0 <  A_{ja} \le  (2\mu)\vee (4\delta/\nu)\}$ and $I$ replaced by $I^{+}$ or $I^{-}$, where $I^{\pm} := \cup_{a\in [K]}\{i\in I_a: A_{ia}=\pm1\}$.
		\item According to display (\ref{gfnr}), it is easy to see that $\rm{GFNP}(\wh G_a)$ will be small if either $t_j^a$ is small for $j\in J_1\cup J_3$ or $|J_1| + |J_3|-|J_1^a|$ is dominated by $|I_a| + \sum_{j \in J}s_j^a$. Moreover, from display (\ref{fnr}), TFNP will be small in the following two cases:
		\begin{itemize}
			\item[-]  $|J_1|+| J_3|$ is dominated by  $|I| + |J_2|$;
			\item[-]  $t_j$ is small relative to $s_j$, for $j\in J_1 \cup  J_3$.
		\end{itemize}
		To illustrate this, consider  $t_j \equiv t$ and $s_j \equiv s$, for each $j\in J$, to simplify the expressions a bit, and assume $|J_1|+|J_3| = \alpha(|I|+|J_2|)$, for some $\alpha \ge 0$. We show in the supplementary material that
		\[
		\textrm{TFNP}(\wh \G) \le   {t} \left/  \left\{s + \frac{1}{\alpha} \left(1+\frac{(s-1)|J_2|}{|I|+|J_2|} \right)\right\},\right. 
		\]
		Thus, when either $t$ or $\alpha$ is small, that is,  when $|J_1|+| J_3|$ is dominated by  $|I| + |J_2|$, then TFNP will be small. Note that even when $t$ itself is large but bounded by some constant, TFNP might also be small since $s$ can be close to $K$ which is allowed to grow as $O(\sqrt{n/\log p})$.

		\item If $J_2 = J$ with $\mu = 3\|C^{-1} \|_\r\delta$, from noting that $J_2 \subseteq J\setminus J_1$, Remark \ref{rem1} in Section \ref{whI} yields $\wh I = I$. We can choose
		$\lambda = \delta$ in (\ref{constr_omega}) and $\mu = 3\|C^{-1} \|_\r\delta$ in (\ref{constr_beta}), and follow the proof of Theorems \ref{prop_rates} and \ref{propgroup} to arrive at the following conclusions:
		$$
		\supp(\wh A) = \supp\left(A\right),\quad  \sgn(\wh A) = \sgn(A).$$
		Moreover, we get exact cluster recovery:  
		\begin{enumerate}
			\item[(a)]  $\textrm{GFPP}(\wh G_a) = \textrm{GFNP}(\wh G_a) = 0$, for each $a\in [K]$.
			\item[(b)] $\textrm{TFPP}(\wh \G) = \textrm{TFNP}(\wh \G)=0$.
		\end{enumerate}
		This immediately yields 
		$\wh G_0 = G_0$.
		Again, all statements hold with probability greater than $1-c_1(n\vee p)^{-c_2}$.
		
		\item We prove that
		Theorem \ref{propgroup} also holds for the hard threshold estimator $\wt A$ in  which we combine $\wh A_{\wh I}$ with $\wt A_{\wh J}$. Each row of $\wt A_{\wh J}$ is estimated by $ \wt \beta_a^j =  \bar \beta_a^j 1\{ |\bar \beta^j_a|>\mu \}$ of $\beta_a^j= A_{ja}$,
		$a\in[\wh K]$,  using the same $\mu=5\| C^{-1}\|_\r\delta' $ as before for the threshold $\mu$. However, we cannot guarantee that the  scaling  restriction of condition (i) holds for this estimator.
		
		\item
		Theorem \ref{propgroup}  holds for the Dantzig type procedure $\widehat A_D$, followed by the  hard-threshold procedure described in the above item, using this time the threshold
		$\mu=6\| C^{-1}\|_\r\delta' $. In this case, the scaling restriction of condition (i) continues to hold as it holds for $\widehat A_D$, with high probability.
	\end{enumerate}
\end{rem}

\subsection{Discussion and related work}\label{sec_related_works}

To the best of our knowledge,  optimal estimation of  identifiable sparse loading matrices $A$ in  model (\ref{mod}) satisfying (i) - (iii),  when both $I$  and $K$ are unknown,  and when the entries in $X$, $Z$ and $A$  are allowed to have arbitrary signs, has not been considered elsewhere and our results bridge this gap. There exists, however, a very large body of literature on related problems. We review the most closely related results below, and explain the differences with our work. 


Results regarding the identifiability of $A$ in general  latent models, typically  not  sparse, are scattered throughout over  more than six decades of  literature. They all involve conditions on both $A$ and $C$, and there is typically a trade-off between the restrictions on $A$ versus those on $C$, as  first summarized and proved in \cite{anderson1956},  reviewed in \cite{lawyley1971} and  later in  \cite{anderson1988}. We recall them briefly here for the convenience of the reader. 

By far the most commonly used assumption  is that the latent factors are uncorrelated, so that  $C$ is either the identity or a diagonal matrix. In this case, it is typically further assumed  that  the scaled columns of $A$ are orthogonal, see, for instance, the literature review in \cite{izenman}. An alternative requirement is that  $A$ contain a $K \times K$  lower diagonal matrix, see, e.g.,  \cite{geweke1996} and, moreover, that the placement of  this  matrix within $A$ is known, which requires careful  justification  \citep{Carvalho2008},  and may be problematic from a practical perspective \citep{dunson2011}.


In general, latent factors are correlated,  which is our point of  view   in this work.  Then,  starting with \cite{anderson1956}, one places on the structure of $A$ constraints that are different than those made when $C$ is diagonal.   The most common of those assumptions involves the existence of a  pure variable set $I$, similar to our assumption   (ii).  If $I$ is  known,  classical results in \cite{anderson1956} and   the proof of  our Theorem \ref{ident}   show that $C$ can be an arbitrary positive definite matrix.  When $I$ is  unknown, conditions on the latent factors also need to be imposed. Sufficient conditions on $Z$, with provable  guarantees for the identification of $I$,  are only known, to the best of our knowledge, in the NMF literature: the uniqueness of $I$  follows from the uniqueness of the solution of an appropriate linear program, applied to population quantities, and tailored to matrices with non-negative entries, see \cite{bittorfNMF}.
In contrast, the arguments of Section \ref{id}  above are optimization-free and can be used for matrices that have entries of arbitrary sign. Therefore, we  provide a  new addition to the literature on pure-variable and loading matrix identification, in general latent models, and also in the particular case of  NMF. We continue this line of reasoning  in \cite{topic18}, that adapts the LOVE procedure to search for the anchor words in the topic model.

A related, but different,  identifiability question  regards the covariance matrix  $\Sigma$ of $X$ which,  under (\ref{mod}),   can be written as the  sum between a rank $K$ matrix and a diagonal matrix:  
\begin{equation}\label{sig} \Sigma = ACA^{T} + \Gamma, \end{equation} and $\Gamma = \text{Cov}(E) $ is a diagonal matrix with possibly different entries. In these models, the identifiability question is whether $\Sigma$ can be decomposed uniquely as the sum  between $ACA^{T}$  and  $\Gamma$.  Answers to this question generated a large amount of  literature.  We refer the reader to  \cite{Ledermann1937,anderson1956,Shapiro1982, Shapiro1985, Bekker1997} for earlier results, and to 
\cite{BaiNg2002, ch11, ch12a, 2011robust, Hsu11},  \cite{fan2013large}, \cite{wegkamp2016}  for more recent works,  that also address the problems of  rank estimation and optimal estimation of high dimensional  covariance matrices.  It is noteworthy that these works, relative to one another,  give different types of  sufficient conditions under which one can separate the low rank matrix $ACA^T$ from $\Gamma$. However, since we always have $ACA^{T}= (AQ)(Q^{T}CQ)(Q^{T}A^{T})$, for any orthonormal $Q$, they do not guarantee the identifiability of $A$ itself. Conversely, 
we show in Theorem \ref{ident} in Section \ref{id}  that  under  conditions (i) - (iii),    $C$ and $\Gamma$ are identified, and   $A$ is identified up to signed permutations. Therefore, we also identify uniquely the decomposition of $\Sigma$. Our conditions are not always comparable to those employed for the unique decomposition of $\Sigma$, but in special cases they imply them. 
Although the uniqueness of the decomposition of $\Sigma$ is a by-product of our results, we do not pursue the covariance estimation problem in this work, but   we included the above discussion for completeness. 

Furthermore, 
we do not view the problem of  estimating the number of factors $K$ as that of estimating the rank of a matrix. This approach is taken in \cite{BaiNg2002}, via  penalized least squares, but provided that either  $C=I$ or $ AA^T=I$ and that  $K$ is bounded by a fixed integer.  Alternatively, we could adapt the criteria in \cite{bunea2011, bing2018, wegkamp2016} to (\ref{mod}) to allow for $K\to\infty$ in the rank estimation problem. However, proving that such an estimator is consistent would ultimately require an unnecessary lower bound restriction  on the $K$-th largest eigenvalue of $ACA^T$.  In contrast, our Theorem \ref{consistI} shows that such conditions can indeed be avoided. We estimate directly the set $I$ and its partition  via LOVE,  and as a byproduct $K$,  at a low computational cost of order $p^2$.

Estimation of $A$ in identifiable  factor models  is typically  based on  iterative alternating least squares procedures or the EM algorithm, see for instance  \cite{Rubin1982, bai2012} and the references therein. As discussed in these works, the resulting algorithms are not suitable for large data sets due to their notoriously slow convergence to a solution that is typically  not the global  optimum.  Bayesian estimation, see,  e.g. \cite{Carvalho2008} and the references therein, offers an alternative approach which may become computationally very demanding in high dimensions, requires a likelihood framework, and careful prior specification.  Moreover, existing procedures do not estimate $A$ under our model specifications 
(i) - (iii), and any adaptation would still require the challenging estimation of $I$.   Our procedure  offers a  solution to the computational problem,  as LOVE does not require a likelihood or other prior distributional specifications,  is tailored to our model with unknown $I$, and has  provable low computational complexity.

The statistical properties  of estimators of $A$  in model (\ref{mod}) (i) - (iii)  have not been studied, and even particular cases of the model have received a very limited amount of attention, from a theoretical perspective.  When $I$ is known and $K$ is fixed, \cite{bai2012} established the asymptotic  normality of the MLE in a model  similar to ours, although  the estimator they ultimately construct is not necessarily the MLE under this  model, but rather an appropriate  transformation of  the stationary point of a quasi-likelihood 
for a different factor model. We give the specific details of their construction in Section C.1 of the supplementary material.
If $I$ is unknown, but  $K$ is known, and moreover,  the columns of ${\bm X}$, $A$ and ${\bm Z}$ have non-negative entries that sum up to 1,   \cite{arora2013practical} provide a practical algorithm for the estimation of $A$ and offer bounds on the  $\ell_1$ matrix norm loss of their estimator. 
The extra restrictions on this model are motivated by a specific model, the topic model, appropriate for vectors with discrete distributions, for instance multinomial. The construction and analysis of these estimates  are not transferable to our general framework, as they depend heavily on these restrictions. Our results of Section \ref{stat_guarantee_A} bridge this gap in the literature and offer lower and upper bounds for the performance of estimators of $A$  in model (\ref{mod}) (i) - (iii).

Finally, to the best of our knowledge, overlapping clustering based on  model (\ref{mod}) has not been analyzed. A  particular case of this model, corresponding 
to a matrix $A$ with binary entries, has been considered in \cite{bunea2015minimax,bunea2016pecok} for non-overlapping clustering. According to their model, all $p$ variables are pure variables, as the model assume that  $X_j = Z_k + E_j$, for all $j \in G_k$ and $ k \in \{1, \ldots, K\}$, $\{G_k\}_{1 \leq k \leq K}$ form a partition of $\{1, \ldots, p\}$. 
When $C$ is positive definite, the non-overlapping clusters are shown to be identifiable, and the work of 
\cite{bunea2015minimax,bunea2016pecok} is  devoted to exact recovery of clusters with minimax optimal cluster separation, a very different problem than the one considered here.

\section{Simulation Studies}\label{sec_sim}

In this section, we first discuss our procedure for selecting the tuning parameters, then evaluate the performance of LOVE based on estimation error and overall clustering misclassification proportion. 
In the supplementary materials, 
we compare LOVE with existing overlapping clustering algorithms and  study the performance of LOVE for the non-overlapping clustering problem.

\subsection{Data driven choice of the tuning parameters}\label{choice}
\paragraph{Tuning parameter $\delta$}
Proposition \ref{consistI} specifies the theoretical  rate of $\delta$, but only up to constants that   depend on the underlying data generating mechanism. We propose below a data-dependent way to select $\delta$, based on data splitting.  Specifically, we split the data set into two independent parts, of equal sizes. On the first set, we calculate the sample covariance matrix $\wh \Sigma^{(1)}$. On the second set, we choose a  fine  grid of values $\delta_\ell = c_\ell\sqrt{\log p/n}$, with $1\le \ell\le M$, for $\delta$, by varying the proportionality constants $c_\ell$. For each $\delta_\ell$, we obtain the estimated number of clusters $\wh K(\ell)$ and the pure variable set $\wh I(\ell)$ with its partition $\wh \I(\ell)$. Then we construct the $|\wh I(\ell)|\times \wh K(\ell)$ submatrix $\wh A_{\wh I(\ell)}$ of $\wh A$, and estimate $\wh C(\ell)$ via formula (\ref{Chat}). Finally, we calculate the $|\wh I(\ell)|\times |\wh I(\ell)|$ matrix $W_\ell = \wh A_{\wh I(\ell)}\wh C(\ell)\wh A_{\wh I(\ell)}^T$. In the end, we have constructed a family $\mathcal{F} = \{ W_1, \ldots,  W_M\}$ of the fitted matrices  $W_\ell$, each corresponding to different $\wh \I(\ell)$ that depend in turn on $\delta_\ell$, for $\ell\in \{1,\ldots,M\}$. Define
\begin{equation}\label{cvcrit}
CV(\wh \I(\ell)) := \frac{1}{\sqrt{| \wh I(\ell)|\bigl(| \wh I(\ell)|-1\bigr)}}\left\|\wh \Sigma_{\wh I(\ell)\wh I(\ell)}^{(1)}-W_\ell\right\|_{\textrm{F-off}},
\end{equation}
where $\|B\|_{\textrm{F-off}}:= \| B -\text{diag}(B)\|_F$ denotes the Frobenius norm over the off-diagonal elements of a square matrix $B$. We choose $\delta^{cv}$ as the value $\delta_\ell$ that minimizes $CV(\wh \I(\ell))$ over the grid $\ell\in [M]$.
To illustrate how the selection procedure works, we provide an example in Section B of the  supplementary material.

\paragraph{Tuning parameters $\lambda$ and $\mu$}  The tuning parameter  $\lambda$ in the linear program   (\ref{constr_omega}) for estimating $\O = C^{-1}$ is specified by $\lambda = 2\delta'$ with $\delta'$ defined in (\ref{delta'}). Since $\delta'$ is proportional to $\delta$, we use $\lambda = c_0\delta^{cv}$ where $c_0$ is some constant and could be tuned by a cross-validation strategy used in the related work on the precision matrix estimation, for instance \cite{cai_CLIME}. More precisely, we randomly split the data into two parts. For a given grid of $\lambda$, we compute $\wh \O$ on the first dataset for each value in the grid. Then we choose the one which gives the smallest likelihood loss from the second dataset, where the likelihood loss is defined by
\[
L(\Omega,C) = \langle \O, C\rangle - \log\det(\O).
\]
From  Remark \ref{rem3} (3) in Section \ref{stat_guarantee_group}, when $J_2 = J$, we can choose $\lambda = \delta$ which is the smallest $\lambda$ we should consider. Therefore, we set the grid of $\lambda$ equal to $[\delta^{cv}, 3\delta^{cv}]$. From our simulation, the selected $\lambda$ is $\delta^{cv}$ in most cases. Hence we recommend to use $\lambda = \delta^{cv}$ and our simulations are based on this choice.

Recall that $\mu = c_1\|C^{-1}\|_\r \delta$ for some constant $c_1$, and that $\wh \Omega$ estimates $C^{-1}$.  Our extensive simulations show that the choice of $\mu = \|\wh \O\|_\r \delta^{cv}$ yields stable performance, with $\wh\O$ solved from (\ref{obj_omega}) and $\delta^{cv}$ selected via cross-validation.

\subsection{Estimation error and cluster recovery with LOVE}\label{sec_sim_love}

In this section, we study the numerical performance of LOVE  in terms of clustering and estimation accuracy. To the best of our knowledge, there is no comparable algorithm with provable guarantees developed for  our framework, especially if the set $I$ is unknown, as explained in detail in  Section \ref{sec_related_works} above, and further re-visited in Section C.1 of the supplementary material.

We generate the data in the following way. We set the number of clusters $K$ to be $20$ and simulate the latent variables $Z=(Z_1, \ldots, Z_{K})$ from $N(0,C)$.  The diagonal elements of $C$ is given by $C_{ii} = 2+({i-1})/{19}$ for $i=1,\ldots,20$, and the off-diagonal elements are generated as  $C_{ij} = (-1)^{(i+j)}0.3^{|i-j|}\left(C_{ii}\wedge C_{jj}\right)$ for any $i\ne j$. In addition, the error terms $E_1,...,E_p$ are independently sampled from $N(0, \sigma_p^2)$, where $\sigma_p^2$ itself is sampled from a uniform distribution on $[1,3]$. Since the rows of $A$ corresponding to pure variables in the same cluster are allowed to have different signs, we consider the following configuration of signs for pure variables in each cluster:  $(3,2)$, $(4,1)$, $(2,3)$, $(1,4)$ and $(5,0)$, with the convention that the first number denotes the number of positive pure variables in that group and the second one denotes the number of negative pure variables. Among the 20 groups, each sign pattern is repeated 4 times. To generate $A_J$, for any $j\in J$, we randomly assign the cardinality $s_j$ of the support of $A_{j\cdot}$ to a number in  $\{2, 3, 4, 5\}$, with equal probability. Then, we randomly select the support from $\{1,2,\ldots, K\}$ with cardinality equal to $s_j$. For $A_{jk}$ which is nonzero, we set it as $A_{jk}=\sgn\cdot(1/s_j)$ with $\sgn$ randomly sampled from $\{-1, 1\}$. Thus, we can generate $X$ according to the model $X=AZ+E$. In the simulation studies, we vary $p$ from $200$ to $1000$ and $n$ from $300$ to $1000$.  Each simulation is repeated $50$ times.

Recall that the true allocation matrix $A$ and our estimator $\wh A$ are not directly comparable, since they may differ by a permutation matrix. To evaluate the performance of our method, we consider the following mapping approach \citep{wiwie2015comparing}.  If $A$ and $\wh A$ have the same dimension, we first find the mapping (i.e., the signed permutation matrix $P\in \H_K$) such that $\|A-\wh AP\|_F$ is minimized. Thus, we can compare the permuted estimator $\wt A=\wh AP$ with $A$ to evaluate the estimation and recovery error. Under this mapping approach, we can evaluate TFPP and TFNP defined in (\ref{t_fpr_fnr}). Moreover, in order to  account for the direction sub-groups defined in (\ref{groups2}), we can define Direction False Positive Proportion (DFPP) and Direction False Negative Proportion (DFNP) as follows:
\begin{equation}\label{eqsfpr}
\textrm{DFPP}=\frac{\sum_{a=1}^{K} |G_a^1\cap \wh G_a^2|}{\sum_{a=1}^{K}|G_a^1|},\qquad \text{DFNP}=\frac{\sum_{a=1}^{K} |G_a^2\cap \wh G_a^1|}{\sum_{a=1}^{K}|G_a^2|}.
\end{equation}

\begin{figure}[ht!]
	\begin{center}
		\begin{tabular}{cc}
			\\[-20pt]
			$p=400$ &  $p=1000$ \vspace{-0.5mm}\\
			\hskip-15pt
			\includegraphics[width=0.45\textwidth]{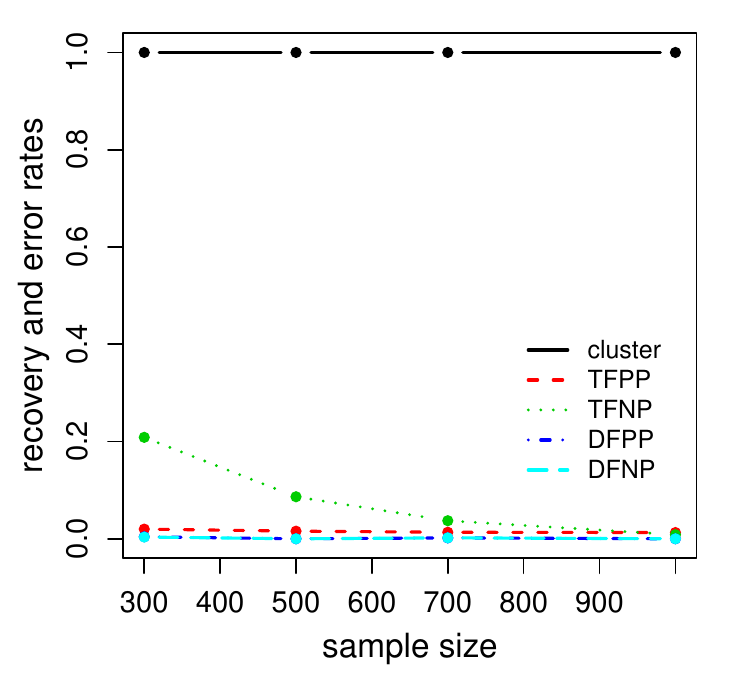} &
			\hskip-15pt
			\includegraphics[width=0.45\textwidth]{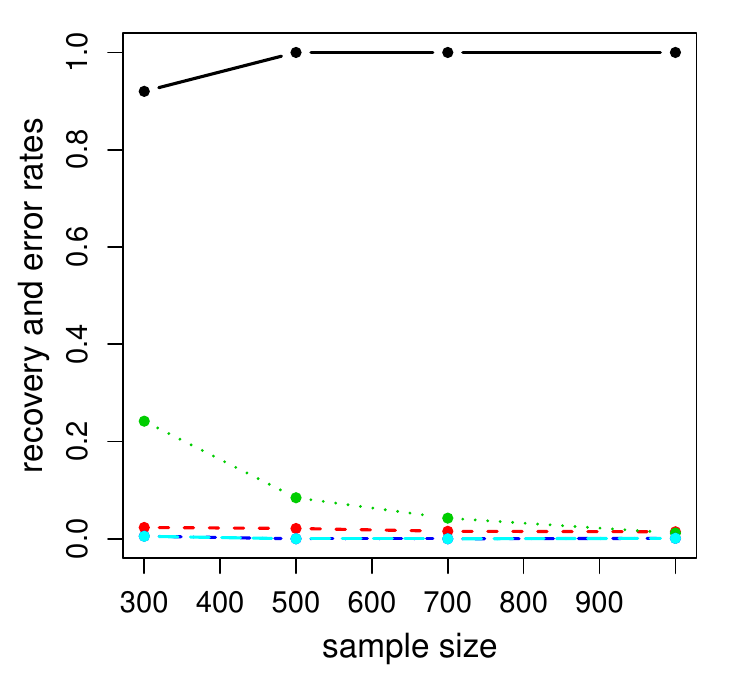}
			\\[-10pt]
		\end{tabular}
	\end{center}
	\caption{Percentage of exact recovery of number of clusters $K$ (cluster), total false positive proportion (TFPP), total false negative proportion (TFNP), direction false positive proportion (DFPP) and direction false negative proportion (DFNP) for LOVE.} \label{fig2}
	\vspace{-3mm}
\end{figure}

Figure \ref{fig2} shows the percentage of exact recovery of number of clusters $K$, TFPP, TFNP, DFPP and DFNP of LOVE. Since the last four measures are well defined only if $\textrm{rank}(\wh A)=K$, we can compute them when the number of clusters is correctly identified. We can see that the proposed method correctly selects $K$ and as long as the number of clusters is correctly selected, TFPP, TFNP, DFPP and DFNP of our method are very close to 0, which implies that the sign and sparsity pattern of $A$ can be correctly recovered.  We  present the estimation error of $\wh A$ as measured by the matrix $\ell_1$ norm scaled by $pK$ and the Frobenius norm scaled by $\sqrt{pK}$ in Table \ref{tab1}. 

\begin{table}[ht!]
	\caption{The average estimation error of $\wh A$ as measured by the matrix $\ell_1$ norm ($\ell_1$) (divided by $pK$) and the Frobenius norm ($\ell_2$) (divided by $\sqrt{pK}$). Numbers
		in parentheses are the simulation standard errors.}
	\begin{center}
		\begin{tabular}{cccccccccc}
			\toprule 
			\multirow{2}{*}{$p$} &\multicolumn{2}{c}{$n=300$} &\multicolumn{2}{c}{$n=500$}
			&\multicolumn{2}{c}{$n=700$} &\multicolumn{2}{c}{$n=1000$}\\
			\cmidrule(r){2-3}\cmidrule(r){4-5} \cmidrule(r){6-7}\cmidrule(r){8-9}
			& $\ell_1$ &  $\ell_2$ & $\ell_1$ &  $\ell_2$ & $\ell_1$ & $\ell_2$ & $\ell_1$ &  $\ell_2$ \\
			\midrule
			200 & 0.018 & 0.062 & 0.015 & 0.053 & 0.013 & 0.048 & 0.012 & 0.041 \\ 
			& (0.001) & (0.005) & (0.001) & (0.003) & (0.001) & (0.008) & (0.001) & (0.002) \\ 
			400 & 0.026 & 0.075 & 0.023 & 0.064 & 0.021 & 0.059 & 0.018 & 0.051 \\ 
			& (0.002) & (0.007) & (0.001) & (0.003) & (0.001) & (0.006) & (0.001) & (0.003) \\ 
			600 & 0.029 & 0.079 & 0.025 & 0.067 & 0.023 & 0.063 & 0.020 & 0.055 \\ 
			& (0.002) & (0.006) & (0.001) & (0.003) & (0.001) & (0.003) & (0.001) & (0.003) \\ 
			800 & 0.031 & 0.083 & 0.026 & 0.068 & 0.024 & 0.064 & 0.022 & 0.057 \\ 
			& (0.002) & (0.006) & (0.001) & (0.004) & (0.001) & (0.004) & (0.001) & (0.004) \\ 
			1000 & 0.032 & 0.083 & 0.027 & 0.069 & 0.025 & 0.065 & 0.022 & 0.057 \\ 
			& (0.002) & (0.006) & (0.001) & (0.003) & (0.001) & (0.004) & (0.001) & (0.004) \\ 
			\bottomrule
		\end{tabular}
	\end{center}\label{tab1}
\end{table}

As expected, the estimation error decreases when the sample size increases from 300 to 1000, which is in line with our theoretical results. The simulations are conducted on an macOS Sierra system version 10.12.6 with 2.2 GHz Intel Core i7 CPU and 16 GB memory. Even with $p=1000$ and $n=1000$,   the computing time of our method for each simulation is around 1 minute.

Moreover, we evaluated the performance of the LOVE procedure  for $K$ varying in a wide range, from 3 to 30,  and when $A_J$ contains   many very small entries. The results are consistent with what we observed in this section and deliver the same message.
The GFPP and GFNP are similar as TFPP and TFNP and the performance of the hard thresholding estimator $\wt A$, defined in Remark \ref{rem3} of Section \ref{stat_guarantee_group},  is similar to $\wh A$. To save space, we have omitted those results. 

We also compared the performance of LOVE  with other off-the-shelf algorithms for overlapping clustering, and  tested LOVE for non-overlapping clustering. We included these results in Sections C.2 and C.3 of the supplementary material. 

\section{Application}\label{genes}

\begin{figure}[H]
	\begin{center}
		\begin{tabular}{cc}
			\includegraphics[width=0.47\textwidth]{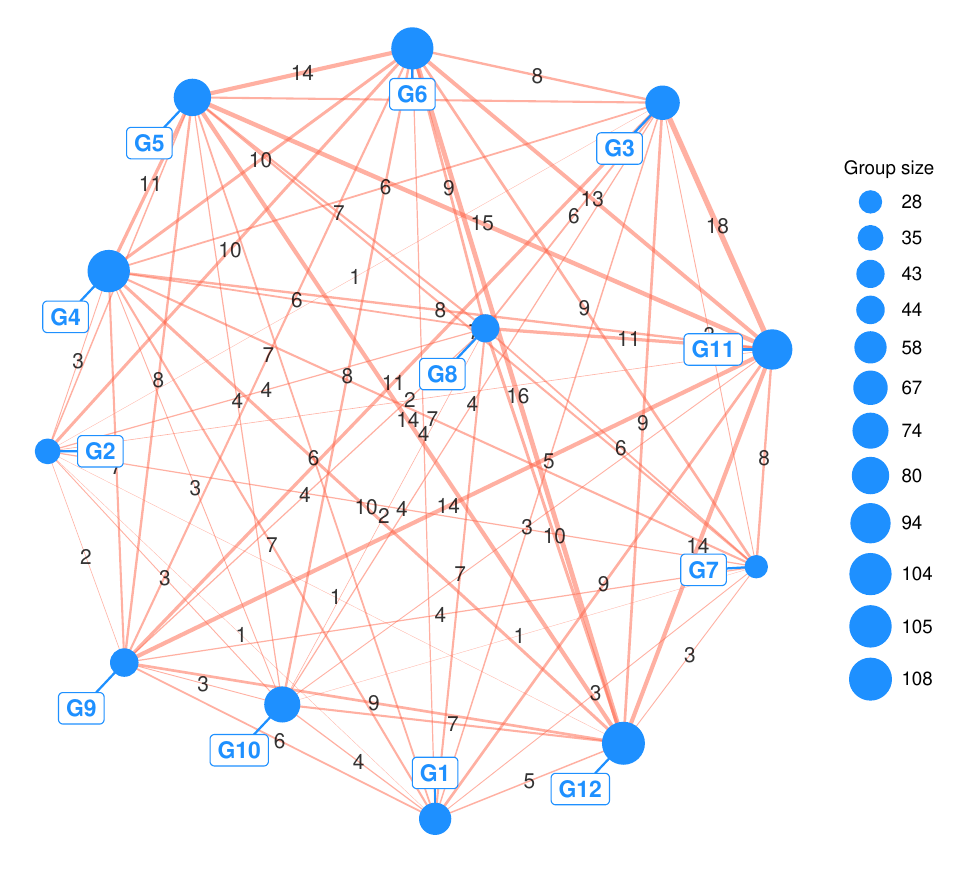} &
			\hskip-10pt
			\includegraphics[width=0.49\textwidth,height = 0.27\textheight]{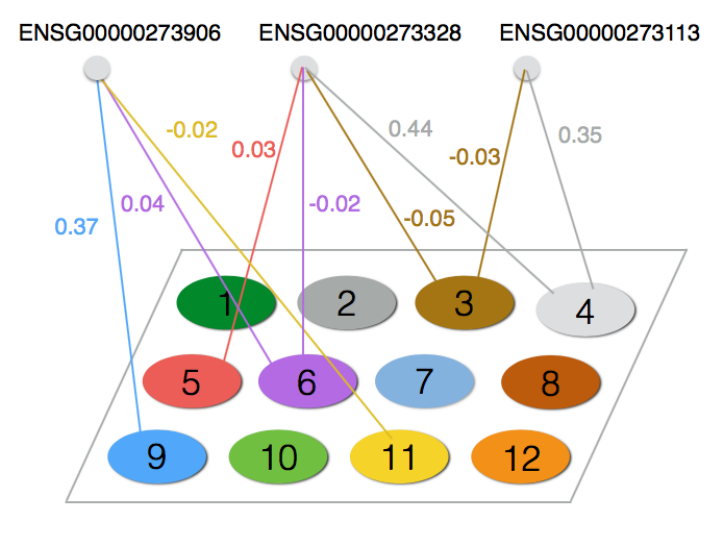}
			\\[-10pt]
		\end{tabular}
	\end{center}
	\caption{Left panel: Number of genes overlapped in different groups. The nodes represent 12 groups with the same labels and sizes as those in Table  \ref{tabdata}. The number shown on the edge between two nodes represents the number of genes shared by the two groups, which corresponds to the width of that edge.  Right panel: Illustration of three genes ENSG00000273906, ENSG00000273328 and ENSG00000273113 and their allocation matrix relative to 12 groups. For instance, the $j$th gene ENSG00000273906 belongs to groups 6, 9 and 11 with $\wh A_{j6}=0.04, \wh A_{j9} = 0.37,  \wh A_{j11}=-0.02$.}
	\label{figdata}
	\vspace{-1mm}
\end{figure}

To benchmark LOVE, we used a publicly available RNA-seq dataset of 285 blood platelet samples from patients with different malignant tumors \citep{best2015rna}. We extracted a small subset of 500 Ensembl genes to test the method. The goal of the benchmarking was to test whether (i) clusters corresponded to biological knowledge, specifically Gene Ontology (GO) functional annotation of the genes \citep{ashburner2000gene}, (ii) overlapping clusters corresponded to pleiotropic gene function. LOVE produced twelve overlapping clusters (Table \ref{tabdata}) which aligned well with a-priori expectation. Table \ref{tabdata} lists the number of pure genes and the total number of genes in twelve overlapping clusters. Figure \ref{figdata} shows that each cluster overlaps with the other and also gives us a clear picture on how two clusters possibly overlap. For example, 18 genes belong to both cluster 3 and cluster 11, whereas cluster 2 and cluster 3 have only one common gene. The genes with the same GO biological process, molecular function or cellular component terms tended to be assigned to the same cluster. For example, ENSG00000273906 and ENSG00000273328 are both RNA genes. They were both assigned to the same cluster (cluster 6, Figure \ref{figdata}). However, they were also assigned to other clusters, suggesting they have pleiotropic functions. This suggests that the latent variables used for clustering are likely to have biological significance and can potentially be used for functional discovery for genes with under-explored functions. We found  308 genes with zero expression across all samples. None of them  were assigned to any of the 12 estimated clusters,  as desired. 
Indeed, our model not only allows for  the existence of pure noise  variables $X_j = E_j$, but variables with structural zero values as well, as $\Gamma_{jj}=\text{Var}(E_j) = 0$ is permitted.
Formally we   place them in the pure noise cluster $G_0$, for further scientific scrutiny.

\begin{table}[H]
	\vspace{-3mm}
	\caption{Number of pure genes and total number of genes in each group.}
	\centering
	\resizebox{\textwidth}{!}{
		\begin{tabular}{ccccccccccccc}
			\\[-15pt]
			\toprule 
			& G1 &  G2 &  G3 &  G4 &  G5 &  G6 &  G7 &  G8 &  G9 &  G10 & G11 & G12\\
			\midrule
			Number of pure genes  & 2&   2&   2&  4&  2& 10& 2 &2 & 2 & 4 & 2 &15\\
			Total number of genes & 58 &35 &67&105 &80& 104  &28 &43 &44 &74 &94&108\\
			\bottomrule
		\end{tabular}
		\label{tabdata}}
	\vspace{-5mm}
\end{table}

\section*{Acknowledgements}
We   thank the referees for their many insightful and helpful suggestions. 
We are grateful to Jishnu Das for help with the interpretation of our data analysis results.
Bunea and Wegkamp were supported in part by NSF grant DMS 1712709. Bing was supported in part by NSF grant DMS-1407600. 

\begin{supplement}
	\sname{Supplement to ``Adaptive Estimation in  Structured  Factor Models 
		with Applications to Overlapping Clustering''}\label{suppA}
	\slink[doi]{COMPLETED BY THE TYPESETTER}
	\sdescription{The supplementary document includes the proofs and additional numerical results.}
\end{supplement}

\appendix
\section{Appendix}
\subsection{Proofs of the results from Section \ref{id}} 
We begin by stating and proving two lemmata that are crucial for the main results of  this section. All results are proved under the condition that model \ref{mod} and (i) - (iii) hold.

\begin{lemma}\label{lem1a} For any $a\in[K]$ and $i\in I_{a}$, we have
	\begin{itemize}
		\item[(a)]  $|\Sigma_{ij} | = C_{aa}$ for all $j\in I_a$,
		\item[(b)]  $|\Sigma_{ij} | < C_{aa}$ for all $j\not\in I_a$.
	\end{itemize}
\end{lemma}
\begin{proof} 
	For given $i\in[p]$, we define the set $s(i) := \{ 1\le a \le K: A_{ia}\ne 0\}$. 
	For any $i\in I_a$ and $j\ne i$, we have
	\begin{eqnarray*}\label{onepure}
		|\Sigma_{ij}| &=&
		\left|\sum_{a\in s(i)}A_{ia}\Bigl(\sum_{b\in s(j)}A_{jb}C_{ab}\Bigr)\right|  \\
		&=& \Biggl|\sum_{b\in s(j)}A_{jb}C_{ab}\Biggr| \text{ from the definition of $ I_a$}\\
		&\le&  \sum_{b\in s(j)} |A_{jb}|\cdot\max_{b\in s(i)} |C_{ab}|\\
		&\le & C_{aa} \text{  using conditions (i) and (iii)}.
	\end{eqnarray*}
	Furthermore, using conditions (i) and (iii), we observe that we have equality in the above display for $j\in I_a$, and strict inequality for $j\not\in I_a$, which proves the lemma.
\end{proof}

\begin{lemma}\label{lem1}
	We have
	\begin{itemize}
		\item[(a)]  $S_i \cap I \ne \emptyset$, for any $i\in[p]$,
		\item[(b)]   $S_i\cup \{i\} = I_a$ and
		$ M_i = C_{aa}$, for any $i\in I_a$ and $a\in[K]$,
	\end{itemize}
	where $M_i$ and $S_i$ are defined in (\ref{mi}) and (\ref{si}), respectively.
\end{lemma}
\begin{proof}
	Lemma \ref{lem1a} implies
	that, for any $i\in I_a$, 
	$M_i= C_{aa}$ and $S_i= I_a\setminus \{i\}$, which proves part (b).\\
	From the result of part (b), it remains  to show $S_i \cap I \ne \emptyset$  for any $i\notin I$.
	Let $i \notin I$ be fixed. We have
	\begin{eqnarray}\label{display2}
	M_i &=&  \max_{j\ne i} | \Sigma_{ij} | \ =\  \max_{j\ne i} 
	\left|\sum_{b\in s(j)}A_{jb}\Biggl(\sum_{a\in s(i)}A_{ia}C_{ab}\Biggr)\right|
	\\ 
	&\le& \max_{j\ne i} 
	\max_{    b\in s(j)}  \left| \sum_{a\in s(i)}A_{ia}C_{ab} \right|\nonumber
	\ =\	 \max_{j\ne i} 	\Biggl|\sum_{a\in s(i)}A_{ia}C_{ab^*}\Biggr| \nonumber
	\end{eqnarray}
	for some $b^*\in[K]$.
	A direct computation yields $|\Sigma_{ij} |=  |\sum_{a\in s(i)}A_{ia}C_{ab^*}|$
	for any $j \in I_{b^*}$, that is, the maximum $M_i$ of $|\Sigma_{ij}|$ is achieved at all $j\in I_{b^*}$. Since $I_{b^*}\not=\emptyset$ by condition (ii),  this completes the proof of claim  (a).
\end{proof}

\noindent {\bf  Proof of Theorem \ref{I}.} 
We have all the necessary ingredients to proceed with the proof of the main result of this section. \\

\noindent {\it Proof of ({\bf a}).}  We first show the sufficiency part.	Consider any  $i\in [p]$  with  $M_i = M_j$ for all $j\in S_i$. 
Part (a) of Lemma \ref{lem1} states that  there exists a $j\in I_a\cap S_i$ for some $a\in[K]$. For  this $j\in I_a$, 
we have $M_j=C_{aa}$ from part (b) of Lemma \ref{lem1}.
Invoking our premise  $M_j= M_i$ as $j\in S_i$, we conclude  that $M_i=C_{aa}$, that is, $\max_{k\ne i} |\Sigma_{ik}|=C_{aa}$. By   Lemma \ref{lem1a}, the maximum is achieved for any pair $i,k\in I_a$. However, if $i\not\in I_a$, we have that $|\Sigma_{ik}|< C_{aa}$ for all $k\ne i$. Hence $i\in I_a$ and this concludes the proof of the sufficiency part. 

It remains to  prove
the necessity part.
Let $i\in I_a$ for some $a\in[K]$ and $j\in S_i$.  Lemma \ref{lem1} implies that $j\in I_a$ and $M_i=C_{aa}$.
Since $j\in S_i$, we have $|\Sigma_{ij}|=M_i=C_{aa}$, while  $j\in I_a$ yields $|\Sigma_{jk}| \le C_{aa}$ for all $k\ne j$,  and $|\Sigma_{jk}| = C_{aa}$ for $k\in I_a$, as a result of Lemma \ref{lem1a}. Hence, $M_j =\max_{k\ne j} |\Sigma_{jk} |= C_{aa}= M_i$ for any $j\in S_i$, which proves our claim. \\

\noindent {\it Proof of ({\bf b}).}  We start with the following constructive approach. Let $N = [p]$ be the set of all variable indices and $O=\emptyset$. Let $M_i$ and $S_i$ be defined in (\ref{mi}) and (\ref{si}), respectively.
\begin{itemize}
	\item[(1)] Choose  $i\in N$ and calculate $S_i$ and $M_i$.
	\begin{itemize}
		\item[(a)] If $M_i = M_j$, 
		for all $j\in S_i$, set $  I^{(i)} := S_i\cup \{i\}$,  $O=O \cup  \{i\}$ and   $N=N\bl   I^{(i)}$. 
		\item[(b)]Otherwise, replace $N$ by $N\bl\{i\}$.
	\end{itemize}
	\item[(2)] Repeat step (1) until $N = \emptyset$.
\end{itemize}
We show that 
$ \{  I^{(i)}:\ {i\in O} \}= \I.$
Let $i\in O$ be arbitrary fixed. By ({\bf a}), we have $i\in I$. Thus, there exists $a\in [K]$ such that $i \in I_a$.  By Lemma \ref{lem1},    $i \in I_a$ implies  $I_a = S_i \cup \{i\} =   I^{(i)}$. On the other hand, let $a\in [K]$ be arbitrary fixed. By condition (ii), there exists at least one $j\in I_a$. Once again, by part (b) of Lemma \ref{lem1}, if $j\in I_a$, then $S_j\cup\{j\} = I_a$, that is,  $I^{(j)} = I_a$.\qed

\subsection*{Proof of Theorem \ref{ident}}

Theorem \ref{I} shows that $\Sigma$ uniquely defines $I$ and its partition $\mathcal I$, up to permutation of labels.
Given $I$ and its partition $\I=\{ I_1,\ldots,I_K\}$, for any $i\in I$, there exists a unique $1\leq a\leq K$ such that $i\in I_a$. Then we set $|A_{i\cdot}|=e_a$, the canonical basis vector in $\RR^K$ that contains 1 in position $a$ and is zero otherwise.  Thus,  the $|I| \times K$  matrix $A_{I}$ with rows $A_{i\cdot}$ is uniquely defined up to multiplication with a signed permutation matrix $P$. 

We show below that $A_J$ is also identifiable up to a signed permutation matrix. We begin by observing that, for each $i\in I_k$, for some $k\in [K]$, and any $j\in J$, Model \ref{mod} implies
$$
\Sigma_{ij} = \sum_{a\in s(i)}\sum_{ b\in s(j)}A_{ia}A_{jb}C_{kb} = A_{ik}\sum_{b\in s(j)}A_{jb}C_{kb}
$$
and since $A_{ik}^2=1$, we obtain 
$$   A_{ik}\Sigma_{ij} = C_{k\cdot}^TA_{j\cdot}
$$
and, after averaging over all $i\in I_k$,
$$
C_{k\cdot}^TA_{j\cdot} = \frac{1}{|I_k|}\sum_{i\in I_k}A_{ik}\Sigma_{ij}.
$$
Repeating this for every $k\in[K]$, we obtain
the formula
\begin{eqnarray*}
	CA_{j\cdot} & =	& \left(\frac{1}{|I_1|}\sum_{i\in I_1}A_{i1}\Sigma_{ij},\ldots,\frac{1}{|I_K|}\sum_{i\in I_K}A_{iK}\Sigma_{ij}
	\right)^T := \theta^j.
\end{eqnarray*}
The covariance matrix $C$  can be uniquely constructed from $\Sigma$  via  
\begin{eqnarray*}
	C_{aa} &=& \frac{1}{|I_a|(|I_a|-1)}\sum_{i, j\in I_a, i\ne j}|\Sigma_{ij}|
\end{eqnarray*}
for any $a\in [K]$, and
\begin{eqnarray*}
	C_{ab} &=& \frac{1}{|I_a||I_b|}\sum_{i\in I_a, j\in I_b} A_{ia}A_{jb}\Sigma_{ij}
\end{eqnarray*}
for $a, b\in[K]$ with $a\ne b$.  Notice that $\min_{a \in [K]} |I_a| \geq 2$, which is part of our model requirement (ii),  is needed for the construction of $C_{aa}$.  Since   the covariance matrix $C$ is assumed to be positive definite, 
$A_{j\cdot} = C^{-1}\theta^j$, for each $j\in J$, which shows that $A_J$ can be determined uniquely from $\Sigma$ up to a signed permutation. Therefore, $A_J$ is identifiable which concludes the proof. 
\qed

\subsection{Proofs of the  results from Section \ref{whI} }

The proof of Theorem \ref{consistI}  will repeatedly use Lemma \ref{lem3}, stated and proved below.  Let 
\begin{equation}\label{mihat}
\wh M_i := \max_{j\in[p]\bl\{i\}}|\wh \Sigma_{ij}|.
\end{equation}

\begin{lemma}\label{lem3}
	Under the conditions in Theorem \ref{consistI}, for any $i\in I_a$ with some $a\in [K]$, the following inequalities hold on the event $\E$:
	\begin{eqnarray}
	\label{ineq1}
	&\Bigl||\wh \Sigma_{ij}| - |\wh \Sigma_{ik}|\Bigr| \le 2\delta,& \quad\text{ for all }j, k\in I_a\bl\{i\} \text{ and }j\ne k;\\
	\label{ineq2}
	&|\wh \Sigma_{ij}| - |\wh \Sigma_{ik}| > 2\delta ,& \quad\text{ for all }j\in I_a\bl\{i\},\ k\notin (I_a\cup J_1^a);\\
	\label{ineq5}
	&|\wh \Sigma_{ij}| - |\wh \Sigma_{ik}|< 2\delta,& \quad\text{ for all }j\in J_1^a \text{ and }k\in I_a\bl\{i\}.
	\end{eqnarray}
	For any $i\in J_1^a$, we have
	\begin{equation}\label{ineq4}
	\wh M_i - |\wh \Sigma_{ij}| \le 2\delta,\qquad \text{for any }j\in I_a.
	\end{equation} 
\end{lemma}
\begin{proof}[Proof of Lemma \ref{lem3}]
	For the entire proof, we work on the event $\E$ defined in (\ref{delta}).
	To prove (\ref{ineq1}), we observe that, for any $i, j,k \in I_a$, $\Sigma_{ij}=\Sigma_{ik}=C_{aa}$ by Lemma \ref{lem1a}, whence
	\begin{eqnarray*}
		\Bigl| |\wh \Sigma_{ij}| - |\wh \Sigma_{ik}|\Bigr|
		&\le& \bigl||\Sigma_{ij}|-|\Sigma_{ik}|\bigr|+2\delta = 2\delta.
	\end{eqnarray*}  
	To prove (\ref{ineq2}), we first observe that, for any $j\in I_a$,	$|\Sigma_{ij}|=C_{aa}$ by Lemma \ref{lem1a}, whence 
	\begin{equation}\label{helpeq1}
	|\wh \Sigma_{ij}| \overset{\E}{\ge}  C_{aa} -\delta.
	\end{equation}
	Next, we notice that, for any $\ell\in [p]$, 
	\begin{align}\nonumber
	|\Sigma_{i\ell}| &\ =\ \Biggl|\sum_{b=1}^KA_{\ell b}C_{ab}\Biggr| = \Biggl|A_{\ell a}C_{aa} + \sum_{b\ne a}A_{\ell b}C_{ab}\Biggr|\\
	&\overset{(iii)}{\le}\  |A_{\ell a}|C_{aa} + (1-|A_{\ell a}|)(C_{aa}-\nu)\ =\ C_{aa} - (1-|A_{\ell a}|)\nu. \label{display7}
	\end{align}
	For any $j\in I_a$ and $k\in [p]\setminus (I_a \cup J_1^a)$, the definition of $J_1$ implies $|A_{ka}| \le 4\delta/\nu$, hence
	\begin{equation*}\label{display8}
	|\wh \Sigma_{ik}| \overset{\E}{\le} |\Sigma_{ik}|+\delta \overset{(\ref{display7})}{\le}C_{aa}-(1-|A_{ka}|)\nu +\delta \le  C_{aa}-\nu+5\delta,
	\end{equation*}	
	so that
	\begin{eqnarray*}
		|\wh\Sigma_{ij}| - |\wh \Sigma_{ik}| \overset{\E}{\ge}  |\Sigma_{ij}|-\delta - |\wh \Sigma_{ik}| \ge |\Sigma_{ij}|-C_{aa}+\nu -6\delta > 2\delta,
	\end{eqnarray*}
	by using $\nu > 8 \delta\cdot(\| C \|_\infty/\nu) \ge 8\delta $. To prove (\ref{ineq5}), observe that, for any $j\in J_1^a$ and $k\in I_a\setminus \{i\}$, 
	\begin{equation*}
	|\wh \Sigma_{ij}| \overset{(\ref{display7})}{\le} C_{aa} - (1-|A_{ja}|)\nu + \delta < C_{aa}+\delta = | \Sigma_{ik}| +\delta\overset{\E}{\le} |\wh \Sigma_{ik}|+2\delta.
	\end{equation*}
	So far, we have proved (\ref{ineq1}) - (\ref{ineq5}) and it remains to
	show (\ref{ineq4}). For any $i\in J_1^a$, we have, for some $c\in[K]$,
	\begin{align*}
	\wh M_i &\overset{\E}{\le}  \max_{k\in[p]\bl{i}}|\Sigma_{ik}|+\delta\overset{(\ref{display2})}{ = }\Biggl|\sum_{b=1}^KA_{ib}C_{bc}\Biggr|+\delta\\
	&\overset{(\ast)}{\le} \Biggl|\sum_{b=1}^KA_{ib}C_{ba}\Biggr|+\delta = |\Sigma_{ij}|+\delta \overset{\E}{\le}|\wh \Sigma_{ij}|+2\delta.
	\end{align*}
	It remains to show that inequality $(\ast)$ holds, 
	for any $c\ne a$. On the one hand, we have
	\begin{equation*}
	\Biggl|\sum_{b=1}^KA_{ib}C_{bc}\Biggr| \le |A_{ia}||C_{ac}|+(1-|A_{ia}|)C_{cc}\overset{({iii})}{\le} |A_{ia}|(C_{aa}-\nu)+(1-|A_{ia}|)C_{cc},
	\end{equation*} while
	on the other hand, we find
	\begin{equation*}
	\Biggl|\sum_{b=1}^KA_{ib}C_{ab}\Biggr| \overset{({iii})}{\ge} |A_{ia}||C_{aa}|-(1-|A_{ia}|)(C_{aa}-\nu).
	\end{equation*}
	Combining the preceding two display yields
	\begin{equation*}
	\Biggl|\sum_{b=1}^KA_{ib}C_{ab}\Biggr| - \Biggl|\sum_{b=1}^KA_{ib}C_{bc}\Biggr| \ge \nu - (1-|A_{ia}|)(C_{aa}+C_{cc})  .	\end{equation*}
	The term on the right is positive, since condition (\ref{nu}) guarantees that
	\begin{equation*}
	\nu > \frac{4\delta}{\nu}(C_{aa}+C_{cc})\ge (1-|A_{ia}|)(C_{aa}+C_{cc}),
	\end{equation*}
	where the last inequality is due to the definition of $J_1$. This concludes the proof.  
\end{proof}
Lemma \ref{lem3} remains valid under the conditions of Remark \ref{rem1} in which case $J_1 = \emptyset$
and we only need $\nu > 4\delta$ to prove (\ref{ineq2}).\\

\noindent{\bf Proof of Theorem \ref{consistI}.}
We work on the event $\E$ throughout the proof.
Without loss of generality, we assume that the label permutation $\pi$ is the identity. We start by pointing out that the following three claims are sufficient to prove (a) - (c).  Let $\wh I^{(i)}$ be defined in step 4 of Algorithm \ref{alg1}.
\begin{enumerate}
	\item[(1)] For any $i\in J\setminus J_1$, we have $Pure(i) = False$.
	\item[(2)] For any $i \in I_a$ and $a\in [K]$, we have $Pure(i) = True$, $I_a\subseteq \wh I^{(i)}$ and $\wh I^{(i)}\bl I_a \subseteq J_1^a$.
	\item[(3)] For any $i\in J_1^a$ and $a\in [K]$, 
	we have $I_a \subseteq \wh I^{(i)}$.
\end{enumerate}
If we can prove these claims, then (1) implies that none of variables in $J\setminus J_1$ will be selected in any set of $\wh \I$ via $i\in J\setminus J_1$. 
(2) implies that for any $a\in [K]$, there exists $\wh I_a$ such that $I_a\subseteq \wh I_{a}$ and $\wh I_{a} \bl I_a \subseteq J_1^a$. Moreover, this together with $\textsc{Merge}$ in Algorithm \ref{alg1} prevents $\wh I_a$ from selecting any variable from $[p]\setminus (I_a\cup J_1^a)$. Finally, (3) guarantees that none of pure variables will be excluded by any $i\in J_1$ in the \textsc{Merge} step. Thus, $\wh K = K$ and $\wh \I = \{\wh I_1,\ldots,\wh I_K\}$ is the desired partition. Therefore, in the following we proceed to prove (1) - (3). 

To prove (1), let $i\in J\setminus J_1$ be fixed. We first prove that $Pure(i) = False$ when $\wh I^{(i)} \cap I \ne \emptyset$. It suffices to show that, there exists $j\in \wh I^{(i)}$ such that the following does not hold
\begin{equation}\label{keyeq2}
\wh M_j - |\wh \Sigma_{ij}|\le 2\delta.
\end{equation}
Let $\wh I^{(i)} \cap I \ne \emptyset$, so there exists $j\in I_b\cap \wh I^{(i)}$ for some $b\in [K]$. For such $j$, we have $|\Sigma_{ij}|=|\sum_{a=1}^K A_{ia} C_{ab}|$ and
\begin{equation}\label{left}
|\wh \Sigma_{ij}|\overset{\E}{\le}
\Biggl|\sum_{a=1}^KA_{ia}C_{ab}\Biggr|+ \delta \overset{({iii})}{\le} |A_{ib}|C_{bb}+(1-|A_{ib}|)(C_{bb}-\nu)+\delta < C_{bb}-3\delta,
\end{equation}
using the definition of $J_1$ to justify
the last inequality.	On the other hand, since $j\in I_b$, part (b) of Lemma \ref{lem1} implies
\begin{equation}\label{right}
\wh M_j = \max_{k\in [p]\bl\{i\}}|\wh \Sigma_{jk}| \overset{\E}{\ge} \max_{k\in [p]\bl\{i\}} |\Sigma_{jk}| -\delta = C_{bb} - \delta.
\end{equation}
Combining (\ref{left}) with (\ref{right}) gives
$
\wh M_j - |\wh\Sigma_{ij}|> 2\delta.
$
This shows that for any $i\in J\setminus J_1$, if $\wh I^{(i)}\cap I \ne \emptyset$, then $Pure(i) = False$. Therefore, to complete the proof of (1), we show $\wh I^{(i)} \cap I = \emptyset$ is impossible when $i\in J\setminus J_1$ under our assumptions.
If  $\wh I^{(i)}\cap I = \emptyset$, then there exists some $j\in J\cap \wh I^{(i)}$ and
\begin{equation*}
|\Sigma_{ij}|  = \left|\sum_{b=1}^K \sum_{a=1}^KA_{ia}A_{jb}C_{ab}\right|
\le  \max_{1\le b\le K} \left|\sum_{a=1}^KA_{ia}C_{ab}\right|  = \left|\sum_{a=1}^KA_{ia}C_{ab^*}\right|= |\Sigma_{ik}|
\end{equation*}
for some $b^*\in [K]$ and any $k\in I_{b^*}$ (the set  $I_{b^*}$ is non-empty  by condition (ii)). 
Therefore,  
\begin{eqnarray*}
	|\wh \Sigma_{ij}| - |\wh \Sigma_{ik }|  & \overset{\E}{ \le } & |\Sigma_{ij}| - |\Sigma_{ik}| +2\delta\quad \le \quad 2\delta 
\end{eqnarray*}
However, since $\wh I^{(i)} \cap I=\emptyset$ and $k \in I_{b^*}$, we know $k\notin \wh I^{(i)}$, which implies 
\begin{eqnarray*}
	|\wh \Sigma_{ij}| - |\wh \Sigma_{ik}| > 2\delta,
\end{eqnarray*}
from Step 4 of Algorithm \ref{alg1}. The last two displays contradict each other, and
we conclude that, for any $i\in J\setminus J_1$, $\wh I^{(i)} \cap I \ne \emptyset$. 

To prove (2), let $i\in I_a$ be arbitrarily fixed with some $a\in [K]$. We first show that $Pure(i) = True$. From steps 7 - 8 of Algorithm \ref{alg1}, it suffices to show that, for any $j\in \wh I^{(i)}$,  (\ref{keyeq2}) holds.
From (\ref{ineq2}) in Lemma \ref{lem3}, given Step 4 of Algorithm \ref{alg1}, we know that, for any $j\in \wh I^{(i)}$, $j\in I_a\cup J_1^a$. Thus, we write	$\wh I^{(i)} = (\wh I^{(i)} \cap I_a) \cup (\wh I^{(i)}\cap J_1^a)$. For any $j \in \wh I^{(i)} \cap I_a$, by the same reasoning, $\wh M_j$ is achieved by some element in either $I_a$ or $J_1^a$. For both cases, since $i,j\in I_a$ and $i\ne j$, (\ref{ineq1}) and (\ref{ineq5}) in Lemma \ref{lem3} guarantee that (\ref{keyeq2}) holds. On the other hand, for any $j\in \wh I^{(i)} \cap J_1^a$, (\ref{ineq4}) in Lemma \ref{lem3} implies that (\ref{keyeq2}) still holds. Thus, we have shown that, for any $i\in I_a$, $Pure(i) = True$. To show $I_a\subseteq \wh I^{(i)}$, let any $j\in I_a\bl\{i\}$ and observe that  $\wh M_i$ can only be achieved by indices in $I_a\cup J_1^a$. In both cases, (\ref{ineq1}) and (\ref{ineq5}) imply $j\in \wh I^{(i)}$. Thus, $I_a\subseteq \wh I^{(i)}$. Finally, $\wh I^{(i)}\bl I_a\subseteq J_1^a$ follows immediately from (\ref{ineq2}).

We conclude the proof by noting that (3) immediately follows from (\ref{ineq4}). \qed

\subsection{Proofs of the results from Section \ref{stat_guarantee_A}} 
We divide the proof of Theorem \ref{prop_rates} into three steps: \\
\noindent  {\bf Step  1. } We show that there exists a signed permutation $\wh P$ such that the columns of $\wh A_I$ aligns with those of $A_I$ in terms of label and sign, as detailed in Lemma \ref{lemP};\\
\noindent {\bf Step 2.}   We write $\bar A =  A\wh P$,  and prove first the error bounds  for  $\wh A_{\wh I}-\bar A_{\wh I}$;\\
\noindent  {\bf Step 3.}  We prove the error bounds for $\wh A - \bar A = \wh A - A\wh P$, with the same $\wh P$, which further implies that $\wh P$ aligns the columns of $\wh A$ and $A$.

\begin{lemma}\label{lemP}
	Under conditions of Theorem \ref{prop_rates}, there exists a signed permutation matrix $Q$ such that $\bar A = AQ$ satisfies that $\sgn(\bar A_{ia}) = \sgn(\wh A_{ia})$ for any $i\in \wh I_a$ with each $a\in[K]$.
\end{lemma}
\begin{proof}[Proof of Lemma \ref{lemP}]
	Theorem \ref{consistI} guarantees $\wh K=K$, $I \subseteq \wh I\subseteq I\cup J_1$ and $I_{\pi(a)} \subseteq \wh I_a \subseteq I_{\pi(a)} \cup J_1^{\pi(a)}$, with high probability, for any $a\in [K]$ and some label permutation $\pi$. 
	Let us write $Q = Q_1Q_2$, with  the unsigned permutation matrix $Q_1$ which relabels the columns of $A_I$ according to those of $\wh A_I$, and with  $Q_2 = \rm{diag}(q_1,\ldots,q_K)$ with $q_a\in \{+1,-1\}$ for each $a\in [K]$. 
	
	Denoting $\wc A = AQ_1$,  we proceed to show that, for each $a\in [K]$, $\sgn(\wh A_{ia}) = \sgn(\wc A_{ia})\cdot q_a$ holds for any $i\in \wh I_a$, in which case each $q_a$ can be uniquely constructed. 
	Since $\wh I_a \subseteq I_{\pi(a)}\cup J_1^{\pi(a)}$, it suffices to prove that, for any $a\in[K]$,
	\begin{equation}\label{goallemP}
	\frac{\sgn(\wh A_{ia})}{\sgn(\wc A_{ia})} = \frac{\sgn(\wh A_{ja})}{\sgn(\wc A_{ja})},\quad \text{ for any } i,j\in I_{\pi(a)} \text{ or } i,j\in J_1^{\pi(a)}\text{ with }i\ne j.
	\end{equation}
	From the definition of $A_I$ and the way we construct $\wh A_I$, for any $i,j\in I_{\pi(a)}$ or $i,j\in J_1^{\pi(a)}$, we consider the following two cases:
	
	If $\sgn({A_{i{\pi(a)}}}) = \sgn(A_{j{\pi(a)}})$, this implies $\sgn(\wc A_{ia}) = \sgn(\wc A_{ja})$. To show $\wh A_{ia} = \wh A_{ja}$, from (\ref{subgroup}), we need to show $i,j\in \wh I_a^1$ or $i,j\in \wh I_a^2$ which is equivalent to show $\wh \Sigma_{ij} >0$. For any $i, j \in I_{\pi(a)}$ or $i, j \in J_1^{\pi(a)}$ with $i\ne j$, display (\ref{J1}) gives $|A_{k{\pi(a)}}|\ge 1-4\delta/\nu$ and $\sum_{b\ne {\pi(a)}}|A_{kb}|\le 4\delta/\nu$, for $k=i,j$. Thus, using $\sgn(A_{i{\pi(a)}})=\sgn(A_{j{\pi(a)}})$, we have 
	\begin{align*}
	\Sigma_{ij} &\ =\ A_{i{\pi(a)}}A_{j{\pi(a)}}C_{\pi(a)\pi(a)} + A_{i\pi(a)}\sum_{c\ne a }A_{jc}C_{\pi(a)c}+A_{j\pi(a)}\sum_{b\ne a}A_{ib}C_{\pi(a)b}\\
	&\qquad+\sum_{b,c\ne {\pi(a)}}A_{ib}A_{jc}C_{bc}\\
	&\overset{(iii)}{\ge} \ A_{i{\pi(a)}}A_{j{\pi(a)}}C_{\pi(a)\pi(a)}- |A_{i\pi(a)}|(1-|A_{j\pi(a)}|)(C_{\pi(a)\pi(a)}-\nu)\\
	&\qquad-|A_{j\pi(a)}|(1-|A_{i\pi(a)}|)(C_{\pi(a)\pi(a)}-\nu) -\sum_{b,c\ne {\pi(a)}}A_{ib}A_{jc}C_{bc}\\
	&\ \ge\ \left(1-\frac{4\delta}{\nu}\right)^2C_{\pi(a)\pi(a)} -\frac{8\delta}{\nu}\cdot\left(1-\frac{4\delta}{\nu}\right) C_{\pi(a)\pi(a)}-\frac{16\delta^2}{\nu^2}C_{b^*b^*}+8\delta\\
	&\ \ge\ \left(1+\frac{48\delta^2}{\nu^2}-\frac{16\delta}{\nu}\right)C_{\pi(a)\pi(a)} -\frac{16\delta^2}{\nu^2}C_{b^*b^*}+8\delta,
	\end{align*}
	for some $b^*\ne \pi(a)$.
	Since (\ref{nu}) implies $8\delta C_{b^*b^*}< \nu^2$ and $\nu > 8\delta$, on the event $\E$, we have 
	$\wh\Sigma_{ij} \ge \Sigma_{ij}-\delta > 3\delta>0$.
	
	If $\sgn(A_{i{\pi(a)}}) \ne  \sgn(A_{j{\pi(a)}})$, this gives $\sgn(\wc A_{ia}) \ne \sgn(\wc A_{ja})$. Similarly, to show $\wh A_{ia} \ne \wh A_{ja}$, we prove $\wh \Sigma_{ij}<0$. Using the same arguments yields 
	\begin{eqnarray*}
		\wh \Sigma_{kl}\overset{\E}{\le} \Sigma_{kl} +\delta < -3\delta< 0.
	\end{eqnarray*}
	Therefore, given $\wh \I = \{\wh I_a\}_{a\in[K]}$, we can construct the signed permutation $\wh P = Q$ which alligns the columns of $A_I$ with those of $\wh A_I$.
\end{proof}
\vspace*{2mm}

{\em 	For ease of notation and without loss of generality, we make the blanket assumption that the signed permutation $\wh P$ is the identity so that $\bar A = A$ for the remainder of the proof. We note that the signed permutation $\wh P$ will be the same when estimating each row $A_{j\cdot}$ for $j\in J$.}\\

\noindent{\text{\bf Proof of step 2}:} 
From the construction of $\wh A_{\wh I}$ and 
parts (a) - (c) in Theorem \ref{consistI},  we can write, for each $a\in [K]$, $\wh I_a = I_{a} \cup L_{a}$ with $L_{a} := \wh I_{a} \cap J_1^{a}$. For any $i\in \wh I_a$, the definitions of $I$ and $J_1^{a}$ imply 
$|A_{i{a}}| \ge 1-4\delta/\nu$. Since Lemma \ref{lemP} guarantees that $\sgn(A_{ia}) = \sgn(\wh A_{ia})$, we have
\begin{eqnarray*}
	\|\wh A_{\wh I} - {A}_{\wh I}\|_\i = \max_{i\in \wh I}\|\wh A_{i.} -  A_{i.}\|_\i \le \frac{4}{\nu} \delta. \end{eqnarray*}
Let $s_i = \| A_{i\cdot}\|_0$ for $i\in [p]$. Then, for any $i\in \wh I$, we have 
\begin{eqnarray*}
	\bigl\|\wh A_{j\cdot} - \bar A_{j\cdot}\bigr\|_q \le \frac{4}{\nu}s_i^{1/q}\delta,\qquad 1\le q\le \infty.
\end{eqnarray*}\qed
\medskip

\noindent For {\bf Step 3} of the proof of Theorem \ref{prop_rates},  we will make use of the results of Lemmas \ref{lem4} and \ref{lemzero}, stated here first and proved at the end of this section, in order to preserve the flow of the presentation.

\begin{lemma}\label{lem4}
	Under the conditions of Theorem \ref{prop_rates}, on the event $\E$, we have
	\begin{equation}\label{sup}
	\|  \wh C - C\|_\i \leq 2\delta',\qquad
	\max_{j\in \wh J}\|  \wh \theta^j - \theta^j  \|_\infty \leq  \delta',
	\end{equation}
	where $\delta'$ is given in (\ref{delta'}).
\end{lemma}

\begin{lemma}\label{lemzero} Under the conditions of Theorem \ref{prop_rates}, on the event $\E$, we have $\beta_a^j = 0$ implies $\wh \beta_a^j=0$, for any $j\in\wh J$ and $a\in [\wh K]$.
\end{lemma}
\medskip

\noindent {\bf Proof of Step 3}.   	For each $j\in \wh J$, recall that $\beta^j = C^{-1}\theta^j= \O\theta^j $ since $C$ is invertible. Also recall that $\bar\beta^j = \wh \O\wh \theta^j$. We first show $\|\bar \beta^j - \beta^j\|_\i\le 5\|\O\|_\r \delta'$. For notational convenience, we remove all the super indices.
From Lemma \ref{lem4}, the following event 
\[
\E' = \left\{ 
\|  \wh C - C\|_\i \leq 2\delta',\ 
\max_{j\in \wh J}\|  \wh \theta^j - \theta^j  \|_\infty \leq  \delta'
\right\},
\]
is implied by the event $\E=\E_\delta$. On the event $\E'$, the true $\Omega:=C^{-1}$ satisfies the constraint since
\begin{equation*}
\| \Omega\wh C-I\|_{\infty} \ =\ \|\Omega (\wh C-C)\|_{\infty} \ \le\ \|\wh C-C\|_{\infty} \|\Omega\|_{\infty,1}\ \le\ 2\delta' \|\Omega\|_{\infty,1}.
\end{equation*}	
Then the pair $(\|\O\|_\r,\O)$ of $(t,\O)$ is feasible. 	Consequently, the optimality and feasibility of $(\wh t,\wh \Omega)$ imply
\begin{equation}\label{display3}
\|\wh \Omega\|_\r\le \wh t\le  \|\O\|_\r, \quad\| \wh \Omega\wh C-I\|_{\infty} \le 2\delta' \wh t \le 2\delta' \|\Omega\|_{\infty,1}.
\end{equation}
Then, on the event $\E'$, we obtain
\begin{eqnarray*}		\|\bar \beta - \beta\|_\infty
	&=& \|\wh \Omega\wh \theta- \wh \Omega \theta + \wh\Omega\theta- \beta\|_\infty\\
	&\le & \|\wh \Omega\|_{\infty,1} \|\wh \theta-\theta\|_\infty+ \|\wh \Omega \theta-\beta\|_\i\\
	& \le & \delta'\|\wh \Omega\|_{\infty,1}+ \|\wh \Omega C\beta - \beta\|_\infty\\
	&\le& \delta'\|\Omega\|_{\infty,1}+ \|\wh \Omega C  -I  \|_\infty \| \beta\|_1\\
	&\le & \delta'\|\Omega\|_{\infty,1}+ \|\wh \Omega \wh C -I\|_\infty+ \|\wh \Omega \wh C - \wh \Omega C\|_\infty\quad (\text{since }\|\beta\|_1 \le 1)\\
	&\le & 3\delta'\|\Omega\|_{\infty,1}+ \|\wh \Omega\|_{\infty,1}\|\wh C-C\|_\i\\
	&\le & 5\delta' \| \Omega\|_{\infty,1}.
\end{eqnarray*}
The feasibility of $\wh\beta^j$ implies that $\|\wh \beta^j - \bar{\beta}^j\|_\i \le \mu$. By the  triangle inequality, we obtain
\begin{eqnarray*}
	\|\wh\beta^j - \beta^j\|_\i &\le& \|\wh\beta^j - \bar\beta^j\|_\i+\|\bar\beta^j - \beta^j\|_\i \ \le\ 2\mu,
\end{eqnarray*} since $\mu=5\delta' \| \Omega\|_{\infty,1}$. Then following from Lemma \ref{lemzero} and using $\wh K=K$ on the event $\E$ gives
\[
\| \wh A_{j\cdot}-A_{j\cdot}\|_q = \left(\sum_{a=1}^{  K}| \wh \beta^j_a-\beta^j_a|^q\right)^{1/q} = \left(\sum_{a\in s_j}| \wh \beta^j_a-\beta^j_a|^q\right)^{1/q} \le\ 2 s_j^{1/q}\mu,
\]
for any $1\le q\le \infty$. This completes the proof of the last step and of Theorem \ref{prop_rates}. 
\qed \\

\noindent To conclude this  section we give below the proofs of the intermediary results used in the proof. \\

\noindent {\it Proof of Lemma \ref{lem4}.} \ 
On the event $\E$, we showed that $\wh K = K$. Then,  from the definition of $\wh C_{aa}$, we have
\begin{align*}
\max_{1\leq a\leq K} |  \wh C_{aa} - C_{aa} | &\ \le\ 
\max_{1\leq a\leq K}\frac{1}{|\wh I_a|(|\wh I_a|-1)}\sum_{i,j\in \wh I_a, i\ne j}\bigl||\wh\Sigma_{ij}|-C_{aa}\bigr|\\
&\ \overset{\E}{\le}\ \delta + \frac{1}{|\wh I_a|(|\wh I_a|-1)}\sum_{i, j\in \wh I_a, i\ne j}\bigl||\Sigma_{ij}|-C_{aa}\bigr|.
\end{align*}
Theorem \ref{consistI} states that, on the event $\E$, 
$\wh I_a = I_a \cup L_a$ where $L_a = \wh I_a\cap J_1^a$, for any $a\in[K]$. Therefore, we consider the following three cases:\\
(1) For any $i, j\in I_a$ and $i\ne j$,  Lemma \ref{lem1a}  implies $\bigl||\Sigma_{ij}|-C_{aa}\bigr|= 0$.\\
(2) For any $i\in I_a$ and $j\in L_a$, the definition of $J_1^a$ gives
\begin{eqnarray*}
	\bigl||\Sigma_{ij}|-C_{aa}\bigr| \le \bigl(1-|A_{ja}|\bigr)(2C_{aa}-\nu)\le \frac{8\delta}{\nu}\| C \|_\infty  - 4\delta.
\end{eqnarray*}
(3) For any $i, j\in L_a$ and $i\ne j$, since $i, j\in J_1^a$, we know $|A_{ka}| \ge 1-4\delta/\nu$ and $\sum_{b\ne a}|A_{kb}|\le 4\delta/\nu$, for $k=i,j$. Thus,
\begin{align*}
\bigl||\Sigma_{ij}|-C_{aa}\bigr| &\le\bigl(1-|A_{ia}||A_{ja}|\bigr)C_{aa} + |A_{ia}|\sum_{c\ne a}|A_{jc}||C_{ac}|+ |A_{ja}|\sum_{b\ne a}|A_{ib}||C_{ab}|\\
&\quad+\sum_{b,c\ne a}|A_{ib}A_{jc}|C_{bc}\\
&\le\bigl(1-|A_{ia}||A_{ja}|\bigr)C_{aa} + |A_{ia}|(1-|A_{ja}|)(C_{aa}-\nu)\\
&\quad + |A_{ja}|(1-|A_{ib}|)(C_{aa}-\nu)+(1-|A_{ib}|)(1-|A_{jc}|)C_{b^*b^*}\\
&\le \left[1-\left(1-\frac{4\delta}{\nu}\right)^2\right]C_{aa}+ \frac{8\delta}{\nu}(C_{aa}-\nu)+\frac{16\delta^2}{\nu^2}C_{b^*b^*}\\
&\le \frac{16\delta}{\nu}\| C \|_\infty-8\delta,\qquad (\text{by }(\ref{nu})).
\end{align*}
for some $b^*\ne a$, where we use the definition of $J_1$ in the third inequality. Therefore, by combining cases (1) - (3), we have
\begin{align*}
\max_{1\leq a\leq K} |  \wh C_{aa} - C_{aa} | &\ \le\ \delta +  \frac{|I_a||L_a| +|L_a|(|L_a|-1)}{|\wh I_a|(|\wh I_a|-1)}\cdot\left(\frac{16\delta \| C \|_\infty}{\nu}-8\delta\right)\\		
&\ \le\ \left(\frac{16}{\nu}\| C \|_\infty-7\right)\delta.
\end{align*}
where the last inequality comes from that $|L_a|+|I_a| = |\wh I_a|$. For the off-diagonal entries, since $\sgn(\wh A_{ia}) = \sgn(A_{ia})$, for any $i\in \wh I$ and $a\in [K]$, we have
\begin{eqnarray*}
	\max_{1\leq a, b\leq K, a\ne b} |\wh C_{ab} - C_{ab}| &\le& \delta + \frac{1}{|\wh I_a||\wh I_b|}\sum_{i\in \wh I_a, j\in \wh I_b}\bigl||\Sigma_{ij}|-|C_{ab}|\bigr|,
\end{eqnarray*}
we consider the following three cases:\\
(1) For any $i\in I_a, j\in I_b$, we have $|\Sigma_{ij}| - |C_{ab}| = 0$.\\
(2) For any $i\in I_a$, $j\in J_1^b$, we have
\begin{eqnarray*}
	\bigl||\Sigma_{ij}|-|C_{ab}|\bigr| \le (1-|A_{jb}|)|C_{ab}|+\sum_{c\ne b}|A_{jc}||C_{ac}|\le \frac{8\delta}{\nu}\| C \|_\infty - 4\delta.
\end{eqnarray*}
(3) For any $i\in J_1^a$, $j\in J_1^b$, we obtain
\begin{eqnarray*}
	\Sigma_{ij} &=& A_{ia}A_{jb}C_{ab}+A_{ia}\sum_{d\ne b}A_{jd}C_{ad} + \sum_{c\ne a}A_{ic}\sum_{d\in s(j)}A_{jd}C_{cd}.
\end{eqnarray*}
Thus, 
\begin{align*}
\bigl||\Sigma_{ij}|-|C_{ab}|\bigr| &\le (1-|A_{ia}||A_{jb}|)|C_{ab}| + |A_{ia}|(1-|A_{jb}|)\| C \|_\infty + (1-|A_{ia}|)\| C \|_\infty\\
&\le \left(\frac{8\delta}{\nu}-\frac{16\delta^2}{\nu^2}\right)(C_{aa}-\nu)+\left(\frac{8\delta}{\nu}-\frac{16\delta^2}{\nu^2}\right)\| C \|_\infty\\
&\le \frac{16\delta}{\nu}\|C \|_\infty - 8\delta. \qquad (\text{by } \nu < \|C\|_\i)
\end{align*}
Therefore, combining the three cases gives
\begin{align*}
\max_{1\leq a, b\leq K, a\ne b} |\wh C_{ab} - C_{ab}|& \le \delta + \frac{|I_a||L_b| + |L_a||I_b| + 2|L_a||L_b|}{2|\wh I_a||\wh I_b|}\cdot \left(\frac{16\delta\| C \|_\infty}{\nu}-8\delta\right)\\
&\le \left(\frac{16}{\nu}\| C \|_\infty-7\right)\delta. 
\end{align*}
Combining the diagonal and off-diagonal cases yields
\[
\|\wh C-C\|_\i \ \le\ \left(\frac{16}{\nu}\| C \|_\infty-7\right)\delta\ \le\ 2\delta'
\]
We now proceed to bound  $\max_{j\in \wh J}\|  \wh \theta^j - \theta^j  \|_\infty$. From $\sgn (A_{ia}) = \sgn(\wh A_{ia})$ for any $i\in \wh I$, we obtain
\begin{eqnarray*}
	\max_{j\in \wh J}\|  \wh \theta^j - \theta^j  \|_\infty &\le& \delta + \max_{a\in[K],j\in \wh J} \frac{1}{|\wh I_a|}\sum_{i\in \wh I_a}\Bigl|\Sigma_{ij} - \sum_{b\in s(j)}A_{jb}C_{ab}\Bigr|. 
\end{eqnarray*}
Since for any $i\in I_a$ and any $j\in J$, $\Sigma_{ij} = \sum_{b\in s(j)}A_{jb}C_{ab}$, we focus on the case when $i\in L_a$. For any $i\in L_a$ and $j\in J$, (\ref{display2}) yields
\begin{eqnarray*}
	\sum_{b\in s(j)}A_{jb}C_{ab} - \Sigma_{ij} &=& (1-A_{ia})\sum_{b\in s(j)}A_{jb}C_{ab} - \sum_{c\ne a }A_{ic}\sum_{b\in s(j)}A_{jb}C_{bc},
\end{eqnarray*}
which, by the definition of $J_1$, implies
\begin{align*}
&\left| \sum_{b\in s(j)}A_{jb}C_{ab} - \Sigma_{ij}\right|\\ &\qquad\le
(1-|A_{ia}|)|A_{jd}||C_{ad}| + (1-|A_{ia}|)|A_{jd'}||C_{ad'}|\quad (\text{for some }d,d'\in [K])\\
&\qquad\le  \frac{4\delta}{\nu}(2\|C\|_\i-\nu)\ \le\  \frac{8\delta}{\nu}\| C \|_\infty-4\delta.
\end{align*}
Since we have $\wh J\subseteq J$, we have
\begin{align*}
\max_{j\in \wh J}\|  \wh \theta^j - \theta^j  \|_\infty &\le \delta + \max_{a}\frac{|L_a|}{|\wh I_a|}\cdot\left( \frac{8}{\nu}\| C \|_\infty-4\right)\delta \le \left(\frac{8}{\nu}\| C \|_\infty-3\right)\delta = \delta',
\end{align*}
which concludes the proof of Lemma \ref{lem4}. \qed \\

\noindent {\it Proof of Lemma \ref{lemzero}.} Let $j\in\wh J$ be arbitrarily fixed and $\wh \beta^j$ be the optimal solution of (\ref{obj_beta}) with $\mu = 5\|\O\|_\r\delta'$. For simplicity, we remove the super indices. Starting with the following Karush-Kuhn-Tucker  condition:
\begin{eqnarray}\label{kkt1}
\sgn(\wh \beta_a) + \lambda_a\sgn (\wh \beta_a - \bar \beta_a)=0,
\end{eqnarray}
subject to 
\begin{equation}\label{kkt2}
\lambda_a(|\wh \beta_a-\bar \beta_a|-\mu) =0,\quad
\lambda_a \ge 0,\quad \text{for }a=\{1,\ldots,K\},
\end{equation}
we obtain 
\begin{equation}\label{kkt3}
0= \sgn(\wh \beta_a)\left(\wh \beta_a - \bar \beta_a\right) + \lambda_a\left|\wh \beta_a - \bar\beta_a\right| \overset{(\ref{kkt2})}{=} \sgn(\wh \beta_a)\left(\wh \beta_a - \bar \beta_a\right)+\lambda_a\mu,
\end{equation}
by multiplying both sides of (\ref{kkt1}) by $\wh\beta_a - \bar \beta_a$. In what follows we   prove that  if  $\beta_a=0$, for some $a$,  then $\wh \beta_a=0$. Since this is true when $\lambda_a=0$ from (\ref{kkt1}), we only consider when $\lambda_a\ne 0$.  Note this implies $|\wh \beta_a - \bar \beta_a| =\mu$ from (\ref{kkt2}). If  we assume $\wh\beta_a>0$, then    (\ref{kkt3}) gives
\[
\bar \beta_a - \wh \beta_a = \lambda_a\mu.
\]
Since
$
|\wh \beta_a - \bar\beta_a| = \mu,
$
we further obtain $\lambda_a=1$ and 
\begin{equation}\label{dispbetabar}
\bar\beta_a = \mu + \wh \beta_a > \mu.
\end{equation}
Recall that 
$
\|\beta -\bar \beta\|_\i \le \mu.$
This implies $\bar{\beta}_a \le \mu  + |\beta_a| = \mu$, which contradicts (\ref{dispbetabar}), so $\wh\beta_a$ cannot be strictly positive. Similarly, $\wh{\beta}_a <0$ cannot hold based on similar arguments. Thus, $\wh \beta_a = 0$ from which we conclude $\supp(\wh\beta^j)\subseteq\supp(\beta^j)$ for any $j\in\wh J$. \qed\\

\noindent{\bf Proof of Theorem \ref{thm:Dantzig}.}
Estimation of the submatrix $A_I$ is as  in Step 2 of the proof of Theorem \ref{prop_rates}.  
We denote by $\widehat \beta_D^j$, $j\in \wh J$, the minimizer of (\ref{Dantzig}) under the constraint (\ref{Dantzig_cond}).
First, we observe that the true $\beta^j$ satisfies the constraint (\ref{Dantzig_cond}) on the event $\E$. Indeed,
\begin{eqnarray*}
	\| \widehat C \beta^j - \widehat \theta^j\|_\infty &\le& \| \widehat C \beta^j - C \beta^j \|_\infty + \| C\beta^j -\widehat \theta^j\|_\infty\\
	&\le& \| \widehat C - C \|_\infty \| \beta^j \|_1 + \| \theta^j -\widehat \theta^j\|_\infty\\
	&\le& \| \widehat C - C \|_\infty    + \| \theta^j -\widehat \theta^j\|_\infty\\
	&\le & 3\delta'= \lambda',
\end{eqnarray*}
by Lemma \ref{lem4}.
Second,
this implies, on the event $\E$, that $\| \widehat \beta_D^j\|_1\le \|\beta^j\|_1$
and  $\widehat \beta_D^j-\beta^j$ is in the cone $\mathcal{C}_S$ with $S=\supp(\beta^j)$ by a standard argument. 
Finally, by the definition of the $\ell_q$-sensitivity of $C$ and the feasibility of $\wh \beta^j_D$, we get for $\Delta=\widehat \beta_D^j - \beta^j $
\begin{align*}
\| \Delta \|_q & \kappa_q(C,s)\\
&\le \| C \Delta\|_\infty \\
&\le  \| C\widehat \beta_D^j - \widehat \theta^j\|_\infty + \| \widehat \theta^j - \theta^j\|_\infty\quad (\text{since }\theta^j = C\beta^j)\\
&\le \| \widehat C \widehat \beta_D^j - 
\wh \theta^j\|_\infty + \| \widehat C-C\|_\infty \| \wh \beta^j_D\|_1 + \| \widehat \theta^j-\theta^j\|_\infty\\
&\le \| \widehat C \widehat \beta_D^j - 
\wh \theta^j\|_\infty + \| \widehat C-C\|_\infty + \| \widehat \theta^j-\theta^j\|_\infty\quad (\text{since }\| \wh \beta^j_D\|_1 \le 1)\\
&\le 2 \lambda'
\end{align*}
and the conclusion (\ref{uno}) follows. It remains to prove  the second inequality (\ref{duo}).
First, we observe that
$\|v\|_q\le \| v\|_\infty (2s)^{1/q}$ for all $v\in \mathcal{C}_S$ and $s=|S|$ by the following computation:
\begin{eqnarray*} \| v\|_q^q &\le & \| v\|_1 \| v\|_\infty^{q-1}\\
	&\le& 2\| v_S\|_1 \| v\|_\infty^{q-1}\quad  (\text{ since $v\in \mathcal{C}_S$})\\
	&\le& 2s \|v\|_\infty^q.
\end{eqnarray*}
This implies that $\kappa_q(C,s) \ge (2s)^{-1/q} \kappa_\infty(C,s)$, and
clearly $[\kappa_\infty(C,s)]^{-1} \le \| C^{-1} \|_\r$ for all $s\le K$, with equality for $s=K$. Now
(\ref{duo}) follows from (\ref{uno}).
\qed

\bigskip 
\noindent {\bf Proof of Theorem \ref{thm_minimax}.}
Without loss of generality, we assume that $\lambda_1(C)<\infty$, since otherwise the lower bound is trivially zero.\\
First we construct a set of ``hypotheses'' of $A$. Let 
\[
\M := \{v\in \{0,1\}^{K}: d_H(0,v) = s\}
\]
where 
$d_H(\cdot)$ denotes the Hamming distance between two binary vectors. Following Lemma A.3 in \cite{rigollet2011} when $s\le 4K/5$, there exists $\M'\subset \M$ such that, for any $w^{(i)}\ne w^{(j)} \in M'$, 
\begin{equation}\label{dist_w}
d_H\left(w^{(i)}, w^{(j)}\right) > s/16,
\end{equation}
and
\begin{equation}\label{cardM}
\log |\M'| \ge c_0s\log(K/s),
\end{equation}
for some constant $c_0>0$. We let $w^{(0)} = (0,\ldots, 0)\in \RR^K$. Then, we choose
\begin{equation}\label{constrA}
A^{(j)} = \begin{bmatrix}
B \\ \eta\left(w^{(j)}\right)^T
\end{bmatrix} \in\RR^{p\times K}, \quad \text{for each $j=0,1,\ldots, |\M'|$,}
\end{equation}
where 
\begin{equation}\label{constrA2}
B = \begin{bmatrix}
B_1 \\ B_2 \\ \vdots \\ B_K
\end{bmatrix}\in \RR^{(p-1)\times K}, \quad
B_ k=
\begin{bmatrix}	e_k^T\\ e_k^T\\	 \vdots \\ e_k^T	\end{bmatrix} 
\in \RR^{|I_k| \times K}, ~\text{ for } k\in[K],
\end{equation}
and
\begin{equation}\label{def_eta}
\eta = \sqrt{c_0\sigma^2\over 8\lambda_1(C)}\sqrt{\frac{\log(K/s)}{n}}.
\end{equation}
We use $e_k$ to denote the canonical basis of $K$ dimensional space and $\bm{0}$ to denote the zero vector. Note that, for each $B_k$, the only non-zero values are at the $k$th column. By specifying as above, we choose $\sum_{k=1}^K |I_k| = p-1$ and consider the $A^{(j)}$ with only one non-pure row.
It is easy to verify that $A^{(j)} \in \A_s$ for each $j = 0, 1, \ldots, |\M'|$ under (\ref{ass_rate}).

We denote by $\KL(\PP,\QQ)$ the Kullback-Leibler divergence between two probability distributions $\PP$ and $\QQ$. Since we particularize into one choice of $C$, we write $\PP_{A}:=\PP_{A, C}$ for simplicity.
In order to apply Theorem 2.5 in \cite{np_sasha} to prove (\ref{minimax}), for fixed $\alpha \in (0,1/8)$, we need to check the following three conditions:
\begin{itemize}
	\itemsep0.5em
	\item[(a)]  $\KL(\PP_{A^{(i)}},\PP_{A^{(0)}}) \le \alpha\log|\M'|$, for each $i= 1,\ldots, |\M'|$.
	\item[(b)] For any $0\le i<j\le |\M'|$, with some constant $c'>0$, $$L_q\left(A^{(i)},A^{(j)}\right) \ge c's^{1/q}\sqrt{\frac{\log(K/s)}{n}}.$$
	\item[(c)] $L_q(\ \cdot\ )$ satisfies the triangle inequality. 
\end{itemize}

To show (a), since $X\sim N(0, ACA^T+\sigma^2\bI_p)$,
invoking Lemma \ref{lem_kl} gives
\begin{equation}\label{kl_eq2}
\KL\left(\PP_{A^{(i)}}, \PP_{A^{(0)}}\right) \le \lambda_1(C) {n\eta^2s \over 2\sigma^2}\le \frac{1}{16}\log|M'|,\quad \forall\ i = 1,\ldots, |M'|, 
\end{equation}
by using (\ref{cardM}) and (\ref{def_eta}). 

To prove (b), for any $i=1,\ldots,|\M'|$, observe that
\begin{equation*}
L_q\left(A^{(i)}, A^{(0)}\right) = \eta\|w^{(i)}\|_q = s^{1/q}\eta
\end{equation*}
and, for any $i\ne j$ different from $0$, 
\begin{equation*}
L_q\left(A^{(i)}, A^{(j)}\right) = \eta\|w^{(i)}-w^{(j)}\|_q \ge (s / 16)^{1/q}\eta\ge (s^{1/q}\eta)/16,
\end{equation*}
by using (\ref{dist_w}). Combining these two and using the expression of $\eta$ yield
\begin{eqnarray}\label{dist_eq}
L_q\left(A^{(i)}, A^{(j)}\right) \ge c' s^{1/q}\sqrt{\sigma^2 \over  \lambda_1(C)}\sqrt{\frac{\log(K/s)}{n}},
\end{eqnarray}
for $0\le i<j\le |\M'|$. 

Finally, we verify (c) by showing that $L_q(\cdot)$ satisfies the triangle inequality. Consider $(A, \wt A, \wh A)$ and observe that
\begin{eqnarray*}
	L_q(A,\wt A) &=& \min_{P\in \H_K}\|AP-\wt A\|_\q\\
	& = &\min_{P,Q\in \H_K}\|AP-\wt AQ\|_\q \\
	&\le &  \min_{P,Q\in \H_K}\left(\|AP-\wh A\|_\q+\|\wh A-\wt AQ\|_\q\right)\\
	& = & \min_{P\in \H_K}\|AP-\wh A\|_\q + \min_{Q\in \H_K}\|\wh A-\wt AQ\|_\q\\
	& = & L_q(A,\wh A)  + L_q(\wt A,\wh A).
\end{eqnarray*}
Therefore, we conclude the proof of (\ref{minimax})  by invoking the Theorem 2.5 in \cite{np_sasha}.
\qed \\

\begin{lemma}\label{lem_kl}
	Assume model (\ref{mod}) and $X\sim N_p(\0, ACA^T+\sigma^2\bI_p)$. Let $A^{(0)}$ and $A^{(i)}$ be constructed as (\ref{constrA}) and (\ref{constrA2}), for any $1\le i \le M'$ with $M'$ satisfying (\ref{cardM}). Let $\PP_{A^{(0)}}$ and $\PP_{A^{(i)}}$ be the probability densities of $X$ parametrized by $A^{(0)}$ and $A^{(i)}$, respectively.  Then we have
	\begin{equation}\label{kl_eq}
	\KL\left(\PP_{A^{(i)}},\PP_{A^{(0)}}\right) \le \lambda_1(C){ n\eta^2s \over 2\sigma^2}.
	\end{equation}
\end{lemma} 

\begin{proof}[Proof of Lemma \ref{lem_kl}]
	From the property of Kullback-Leibler divergence, we only need to verify the case when $n=1$. We consider arbitrary $A^{(i)}$ constructed as (\ref{constrA}) and (\ref{constrA2}) for some $0\le i\le M'$. For notational simplicity, we write
	$A = A^{(i)} = (B^T, \xi)^T$ where $\xi = \eta w^{(i)}\in \RR^{K}$.
	For this $A\in \RR^{p\times K}$, from (\ref{constrA}) and (\ref{constrA2}), we observe that
	\[
	\Sigma = ACA^T+\Gamma = \begin{bmatrix}
	BCB^T + \sigma^2\bI_{p-1} &  BC\xi \\
	\xi^TCB^T & \xi^TC\xi + \sigma^2
	\end{bmatrix} := \begin{bmatrix}
	\ll & \lr \\
	\rl & \rr\\
	\end{bmatrix}.
	\]
	Similarly, for any $\wt A \ne A$ constructed in the same way, we have
	\[
	\wt\Sigma = \wt AC\wt A^T+\sigma^2\bI_{p-1} = \begin{bmatrix}
	BCB^T + \Gamma_B &  BC\wt\xi \\
	\wt \xi^TCB^T & \wt\xi^TC\wt \xi  + \sigma^2
	\end{bmatrix} := 
	\begin{bmatrix}
	\ll & \wt\Sigma_{12} \\
	\wt\Sigma_{21}  & \wt\Sigma_{22} \\
	\end{bmatrix}
	\]
	Recall that the Kullback-Leibler divergence between two $p$-dimensional multivariate normal distributions $\mathcal{N}_0 := N_p(\0, \Sigma)$ and $\mathcal{N}_1 :=N_{p}(\0, \wt \Sigma)$ is given by
	\begin{equation}\label{kl_eq1}
	\KL(\PP_{\wt A},\PP_{A})={1 \over 2}\left[\mathrm {tr} \left(\Sigma^{-1}\wt \Sigma\right)-p+\log \left({\det \Sigma} \over \det \wt\Sigma\right)\right].
	\end{equation}
	By using the formula of the inverse of a block matrix
	\[
	{\begin{bmatrix} {A} & {B} \\ {C} & {D} \end{bmatrix}}^{-1}
	\!\!\!={\begin{bmatrix} {A} ^{-1}+ {A} ^{-1} {B} ( {D} - {CA} ^{-1} {B} )^{-1} {CA} ^{-1}&- {A} ^{-1} {B} ( {D} - {CA} ^{-1} {B} )^{-1}\\-( {D} - {CA} ^{-1} {B} )^{-1} {CA} ^{-1}&( {D} - {CA} ^{-1} {B} )^{-1}\end{bmatrix}},
	\]
	for square matrices $A$ and $D$ and non-singular matrices $A$ and $D-CA^{-1}B$, we have
	\[
	\Sigma^{-1} = \begin{bmatrix}
	\ll^{-1} + \ll^{-1}\lr\rrl^{-1}\rl\ll^{-1} & -\ll^{-1}\lr\rrl^{-1}\\
	-\rrl^{-1}\rl\ll^{-1} & \rrl^{-1}
	\end{bmatrix} := \begin{bmatrix}
	\Omega_{11} & \Omega_{12} \\
	\Omega_{21} & \Omega_{22}
	\end{bmatrix},
	\]
	with $\rrl = \rr - \rl\ll^{-1}\lr$. This gives
	\[
	\tr\left(\Sigma^{-1}\wt\Sigma\right) = \underbrace{\tr\left(\Omega_{11}\ll + \Omega_{12} (BC\wt\xi)^T \right)}_{T_1}+ \underbrace{\Omega_{21}BC\wt{\xi} + \Omega_{22}(\wt\xi^TC\wt \xi + \sigma^2)}_{T_2}.
	\]
	We first calculate $T_1$ by observing that
	\begin{eqnarray}\label{eq_disp1}\nonumber
	T_1 &=& \tr\left(\bm{I}_{p-1}+ \ll^{-1}\lr\rrl^{-1}\rl -\ll^{-1}\lr\rrl^{-1}\wt \xi^TCB^T\right)\\\nonumber
	&=& p-1+ \tr\left(\ll^{-1}\lr\rrl^{-1}\Delta^TCB^T\right) \qquad \left(\rl = \xi^TCB^T\right)\\
	&=&  p-1+ \rrl^{-1}\Delta^TCB^T\ll^{-1}BC\xi
	\end{eqnarray}
	where $\Delta := \xi - \wt{\xi} \in\RR^{K}$. On the other hand, we have 
	\begin{eqnarray}\label{eq_disp2}
	T_2 = \rrl^{-1}\left(\wt\xi^TC\wt \xi + \sigma^2-\xi^TCB^T\ll^{-1}BC\wt \xi \right).
	\end{eqnarray}
	Since our specification of $A = A^{(0)}$ and $\wt A = A^{(i)}$ in (\ref{constrA}) and (\ref{constrA2}) gives $\xi = \0$ and $\wt \xi = \eta w^{(i)}$, it implies $\|\wt \xi\|^2 = \eta^2s$ and 
	\begin{align}\label{disp_rrl}
	\rrl &= \xi^TC\xi + \sigma^2 - \xi^TCB^T\ll^{-1}BC\xi=\sigma^2,\\\nonumber
	\wt \Sigma_{22\cdot1} &= \sigma^2 + \wt \xi^T\left(C - CB^T\ll^{-1}BC\right)\wt \xi.
	\end{align}
	Hence combining (\ref{eq_disp1}) with (\ref{eq_disp2}) yields
	\begin{equation}\label{disp_trace}
	\tr\left(\Sigma^{-1}\wt{\Sigma} \right)= p+ {\wt \xi^TC\wt \xi \over \sigma^2}\le p + {\eta^2s\over \sigma^2}\lambda_1(C).
	\end{equation}
	To calculate the determinant of $\Sigma$ and $\wt \Sigma$, recall that the inverse formula of a block matrix is
	\[
	\det {\begin{pmatrix}A&B\\C&D\end{pmatrix}}=\det(A)\det(D-CA^{-1}B)
	\]
	for any invertible matrix $A$. We thus obtain
	\[
	\det\Sigma = \det\ll\cdot \rrl, \quad \det\wt \Sigma = \det\ll\cdot\wt{\Sigma}_{22\cdot1} ,
	\]
	from which, the display (\ref{disp_rrl}) further gives
	\begin{equation*}
	\log\left({\det \Sigma \over \det \wt \Sigma}\right) = \log \rrl - \log \wt\Sigma_{22\cdot1} = \log \sigma^2 - \log \left(\sigma^2 + \xi^TM\xi\right)
	\end{equation*}
	with $M := C - CB^T\ll^{-1}BC$. It is easy to see that $M$ is positive definite. Indeed, since
	\[
	\left\| C^{1/2}B^T\ll^{-1}BC^{1/2}\right\|_{op} =  \left\|(BCB^T+\sigma^2\bI_{p-1})^{-1}BCB^T\right\|_{op} < 1,
	\]
	$\lambda_{\min}(M)>0$ follows from 
	\[
	M = C^{1/2}\left(\bI_p-C^{1/2}B^T\ll^{-1}BC^{1/2}\right)C^{1/2}
	\]
	and an application of Weyl's inequality. This implies 
	\begin{equation}\label{eq_disp3}
	\log\left({\det \Sigma \over \det \wt \Sigma}\right) < 0.
	\end{equation}
	Finally, plugging (\ref{disp_trace}) and (\ref{eq_disp3}) into (\ref{kl_eq1}) concludes the proof of  Lemma \ref{lem_kl}.
\end{proof}

\subsection{Proofs for the results from Section \ref{stat_guarantee_group}} 
We first prove the three statements of Theorem \ref{propgroup}, then present the proofs of Remark \ref{rem3}. Without loss of generality, we assume that the signed permutation $P$ is identity.
\vspace*{2mm}
\\
\noindent{\bf Proof of Theorem \ref{propgroup}.} We first give the proof for part (a). Then, for ease of the presentation, we prove part (c) first and then part (b).
\vspace{2mm}

\noindent{\it Proof of part (a).} Recall that Lemma \ref{lemzero} immediately implies $\supp(\wh A_{\wh J})\subseteq \supp(A_{\wh J})$. In addition, Theorem \ref{consistI} yields $\wh I_a \subseteq I_a\cup J_1^a$, for any $a\in \wh K$. From the way we construct $\wh A_{\wh I}$, we have $\supp(\wh A_{\wh I})\subseteq \supp(A_{\wh I})$. Therefore, we have proved $\supp(\wh A)\subseteq \supp(A)$. 

On the other hand, for any $(j,a)\in \supp(A_{J_2})$, we know 
$
|\beta^j_a| > 2\mu.
$ This and the fact that
$\|\wh\beta^j-\beta^j\|_\i \le 2\mu$, immediately gives
\[
|\wh \beta^j_a| \ge |\beta^j_a| - \|\wh\beta^j-\beta^j\|_\i > 0,
\]
which implies $\supp(A_{J_2})\subseteq \supp(\wh A_{J_2})$.

To show $\sgn(\wh A_{\wh S}) = \sgn(A_{\wh S})$, since Lemma \ref{lemP} guarantees $\sgn(\wh A_{ia}) = \sgn(A_{ia})$ for any $(i,a) \in \wh S$ and $i\in \wh I$, we focus on any fixed $(j, a)\in \wh S$ and $j\in \wh J$. First,  we consider the case $\wh A_{ja} = \wh \beta_a^j> 0$. Removing super indices, if $\wh \beta_a>0$, (\ref{dispbetabar}) gives
$\bar{\beta}_a > \mu$. 
Thus, 
$ \beta_a \ge \bar \beta_a - \|\beta-\bar{\beta}\|_\i > 0$ by recalling $\|\beta-\bar{\beta}\|_\i \le \mu$.  So far, we have shown that, for any $\wh A_{ja}>0$, $(j,a)\in \wh S$ and $j\in\wh J$, we have $A_{ja}>0$. Since the same argument holds for any $\wh A_{ja}<0$, the proof of $\sgn(\wh A_{\wh S}) = \sgn(A_{\wh S})$ is completed.

\medskip

\noindent{\it Proof of part (c).}
Recall that, for any $i\in[p]$ and $a\in [K]$, 
$$i \in G_a \iff A_{ia}\ne 0,\qquad i\in \wh G_a \iff \wh A_{ia}\ne 0.$$
We start our proof by rewriting the equivalent expression of TFPP and TFNP:
\begin{eqnarray*}\label{fprfnr}
	\textrm{TFPP}&=&\frac{\sum_{i\in [p],a\in[K]} 1{\{A_{ia}=0,\wh A_{ia}\ne0\}}}{\sum_{i\in[p],a\in[K]} 1{\{A_{ia}=0\}}},\\ 
	\textrm{TFNP}&=&\frac{\sum_{i\in [p],a\in[K]} 1{\{A_{ia}\ne0,\wh A_{ia}=0\}}}{\sum_{i\in [p],a\in[K]} 1{\{A_{ia}\ne0\}}}.
\end{eqnarray*}

We first show $\text{TFPP} = 0$. From the result of part (a), we know $\supp(\wh A)\subseteq \supp(A)$. Thus,
\begin{eqnarray*}
	\sum_{i\in [p],a\in [K]} 1{\{A_{ia} = 0, \wh A_{ia} \ne 0\}} =0,
\end{eqnarray*}
which implies $\text{TFPP}=0$. 

In order to prove the result of $\text{TFNP}$, observe
\begin{equation}\label{disp1}
\sum_{i\in [p],a\in [K]} 1{\{A_{ia} \ne 0\}} = |I| +\sum_{i\in J}s_i.
\end{equation}
with $s_i = \|A_{i.}\|_0$ for each $j\in J$. For given $\wh I$, we partition $[p] = I\cup J_1 \cup J_2 \cup J_3= I\cup (L_1 \cup L_2) \cup J_2\cup J_3$ with $L_1 = \wh I \cap J_1$ and $L_2 = J_1\bl L_1$.
Let us consider the set $I\cup L_1$ first. Theorem \ref{consistI} implies $I\cup L_1 = \wh I$ and $\wh I_a \bl I_a\subseteq J_1^a$. 
From the way we construct $\wh A_{\wh I}$, we have
\begin{equation*}
\sum_{i\in I\cup L_1,a\in [K]} 1{\{A_{ia} \ne 0, \wh A_{ia} = 0\}} = \sum_{i\in L_1,a\in [K]} 1{\{A_{ia} \ne 0, \wh A_{ia} = 0\}}.
\end{equation*}
Since the definition of $J_1$ implies that, for any $j\in J_1^a$ and $a\in [K]$, $|A_{ja}| \ge 1- 4\delta/\nu$ and $|A_{jb}| \le 4\delta/\nu$, for any $b\ne a$, this implies 
$$
\sum_{a\in [K]} 1{\{A_{ia} \ne 0, \wh A_{ia} = 0\}} =   \sum_{b\ne a} 1\{A_{ib} \ne 0\} = t_i,
$$
for any $i \in J_1^a \cap L_1$ and $a\in [K]$. Thus, we have
\begin{equation}\label{disp2}
\sum_{i\in I\cup L_1,a\in [K]} 1{\{A_{ia} \ne 0, \wh A_{ia} = 0\}} = \sum_{i\in L_1}t_i.
\end{equation}
Next we consider the set $L_2$. On the event $\E$,  for any $i \in J_1^a \cap L_2$, we have
$$|\wh A_{ia}| \ge |A_{ia}| - \| \wh A-A\|_\i \ge 1-\frac{4\delta}{\nu}-2\mu > 0.$$
Thus, $\wh A_{ia}\ne 0$, which implies
\begin{equation}\label{disp3}
\sum_{i\in L_2,a\in [K]} 1{\{A_{ia} \ne 0, \wh A_{ia} = 0\}} \le \sum_{i\in L_2}t_i.
\end{equation}
Then we consider the set $J_2$. Part (a) gives $\supp(A_{J_2}) = \supp(\wh A_{J_2})$ which yields
\begin{equation}\label{disp4}
\sum_{i\in J_2, a\in [K]} 1{\{A_{ia} \ne 0, \wh A_{ia} = 0\}}  = 0.
\end{equation}
Finally, we consider the set $J_3$. By examining the proof of Part (a), it is easy to verify that
$\wh A_{ja} \ne 0$ if $|A_{ja}| \ge (2\mu) \vee (4\delta/\nu)$, for any $j\in J_3$ and $a\in [K]$. Thus,
\begin{equation}\label{disp5}
\sum_{i\in J_3, a\in [K]} 1{\{A_{ia} \ne 0, \wh A_{ia} = 0\}} \le \sum_{i\in J_3}t_i.
\end{equation}
At last, combining (\ref{disp1}) - (\ref{disp5}) gives
\begin{align*}
\textrm{TFNP}=\frac{\sum_{i\in [p],a\in[K]} 1{\{A_{ia}\ne0,\wh A_{ia}=0\}}}{\sum_{i\in [p],a\in[K]} 1{\{A_{ia}\ne0\}}}&\le \frac{\sum_{j\in J_1\cup J_3}t_j}{|I| + \sum_{j\in J}s_j}.
\end{align*}

\medskip

\noindent{\it Proof of part (b).}
Similarly, we can express $\rm{GFPP}(\wh G_a)$ and $\rm{GFNP}(\wh G_a)$ by the following:
\begin{eqnarray*}
	\textrm{GFPP}(\wh G_a)&=&\frac{\sum_{i\in [p]} 1{\{A_{ia}=0,\wh A_{ia}\ne0\}}}{\sum_{i\in[p]} 1{\{A_{ia}=0\}}},\\ 
	\textrm{GFNP}(\wh G_a)&=&\frac{\sum_{i\in [p]} 1{\{A_{ia}\ne0,\wh A_{ia}=0\}}}{\sum_{i\in [p]} 1{\{A_{ia}\ne0\}}}.
\end{eqnarray*}
For any given $a\in [\wh K]$, $\rm{GFPP}(\wh G_a) = 0$ follows immediately by noting that 
$$0\ =\ \rm{TFPP}\ \ge\ \frac{|(G_a)^c\cap \wh G_a|}{\sum_{b=1}^{K}|(G_b)^c|}\ =\ \frac{|(G_a)^c|}{\sum_{b=1}^{K}|(G_b)^c|}\rm{GFPP}(\wh G_a),$$
with the convention $\rm{GFPP}(\wh G_a) = 0$ if $(G_a)^c = 0$.
To show the expression of $\rm{GFNP}(\wh G_a)$, by the definition of $I$ and Theorem \ref{consistI}, we obtain
\[
\sum_{i\in I} 1{\{A_{ia}\ne0,\wh A_{ia}=0\}} = 0, \qquad \sum_{i\in I} 1\{A_{ia}\ne0\} = |I_a|.
\]
The latter immediately implies  
$$ \sum_{i\in [p]} 1{\{A_{ia}\ne0\}}  = |I_a| + \sum_{i\in J}s_i^a$$
In addition, following the same arguments in the proof of part (b), we have
\[
\sum_{i\in J_2} 1{\{A_{ia}\ne0,\wh A_{ia}=0\}} = 0,\quad \sum_{i\in J_1\cup J_3} 1{\{A_{ia}\ne0,\wh A_{ia}=0\}} = \sum_{i\in J_1\cup J_3\setminus J_1^a}t_i^a.
\]
Combining these two concludes the proof.	
\qed\\

\noindent{\bf Proofs of Remark \ref{rem3}.}
We briefly verify the first claim. It suffices to verify 
$\|A_{j\cdot}\|_1 \le 1$ which is equivalent with $\|\wh \beta^j\|_1 \le 1$, for any $j\in \wh J$. Recall (\ref{obj_beta}), since $\beta^j$ is feasible, the optimality of $\wh \beta^j$ immediately gives $\|\wh \beta^j\|_1 \le 1$.\qed\\

To verify the expression of TFNP in the second claim, we assume $t_j = t$ and $s_j=s$, for $j\in J$, and $|J_1|+|J_3| = \alpha(|I|+|J_3|)$. Note that $|I|+|J_1|+|J_2|+|J_3| = p$ implies $|J_1|+|J_3| = \alpha p/(1+\alpha)$. We therefore obtain
\begin{eqnarray*}
	\rm{TFNP} &\le& \frac{t(|J_1|+|J_3|)}{s(|J_1|+|J_3|)+s|J_2|+|I|}\ =\  \frac{t}{s+\frac{s|J_2|+|I|}{\alpha}\cdot\frac{1+\alpha}{p}}\\
	& = & t\bigg/\left(s+\frac{1}{\alpha}\cdot\frac{|I|+s|J_2|}{|I|+|J_2|}\right)\qquad \bigl(\text{using }(1+\alpha)(|I|+|J_2|) = p\bigr),
\end{eqnarray*}
as desired.\qed\\

We verify the third claim. On the event $\E$, when $J_2 = J$, Remark \ref{rem1} yields $\wh I=I$, $\wh\I =\I$ and $\wh J = J$. After careful examination of the proof of Lemma \ref{lem4}, we  derive that 
$\|\wh C-C\|_\i \le \delta$ and $ \max_{j\in J}\|\wh \theta^j - \theta\|_\i \le \delta$, on the event $\E$.
Therefore, choosing $\lambda = \delta$ and $\mu = 3\|\O\|_\r\delta$ proves the claim, following the proof of Theorems \ref{prop_rates} and \ref{propgroup} step by step.\qed\\

Finally,  we verify the fourth claim on the hard-threshold estimator $\wt\beta^j$ for any $j\in J$. For simplicity, we remove the super indices. Recall that, $\wt\beta$ is defined coordinate-wisely by $\bar \beta_a 1\{|\bar \beta_a| > \mu\}$ with $\mu = 5\|\O\|_\r \delta'$.

First, we show $\|\wt \beta - \beta\|_\i \le 2\mu$.  For any $a\in [K]$ such that $|\bar{\beta}_a|\le \mu$, we have
\[
|\wt\beta_a-\beta_a|=|\beta_a| \le \|\bar \beta-\beta\|_\i + |\bar{\beta}_a|\le 2\mu,
\] while the same bound is obtained above for the case $|\bar{\beta}_a|> \mu$. This proves $\|\wt A-A\|_\i \le 2\mu$ where $\wt A$ combines $\wh A_{\wh I}$ and $\wt \beta^j$ for each $j\in \wh J$. To prove the same rate in Theorem \ref{prop_rates} for $\wt A$, it suffices to show that Lemma \ref{lemzero} still holds for $\wt\beta^j$. Recall that,
on the event $\E$, we have $\|\bar\beta - \beta\|_\i \le \mu$. For any $\beta_a = 0$, we thus have $|\bar \beta_a| \le \|\bar{\beta}-\beta\|_\i \le \mu$, which implies $\wt\beta_a=0$. This concludes the proof of Theorem \ref{prop_rates} for $\wt A$.

To show part (a) of Theorem \ref{propgroup}, let $\wh S$ denote the support of $\wt A$ and we write $(i, a) \in \wh S$ if $|\wt A_{ia}|\ne 0$. Let $(i,a)\in \wh S$ be arbitrary fixed and consider the following two cases:
\begin{itemize}
	\item[-] If $i\in \wh I$, from Theorem \ref{consistI} and the way we construct $\wh A_{\wh I}$, we have $|\wt A_{ia}| = 1$. Thus, $|A_{ia}| \ge  |\wt A_{ia}| - \|\wt A-A\|_\i \ge 1-2\mu > 0$.
	\item[-] If $i\in \wh J$, then $|\wt A_{ia}| = |\wt \beta^i_a| = |\bar \beta^i_a| > \mu$. Therefore, 
	$|A_{ia}| = |\beta^i_a| \ge |\bar \beta^i_a| - \|\bar{\beta}^i-\beta^i\|_\i > 0$.
\end{itemize}
Thus, we have proved that $\supp(\wt A) \subseteq \supp(A)$. To show $\supp(A_{J_2}) \subseteq \supp(\wt A)$, for any $(i,a)\in \supp(A_{J_2})$, by the definition of $J_2$, $|A_{ia}| > 2\mu$. Thus, $|\wt A_{ia}| \ge |A_{ia}|-\|\wt A-A\|_\i \ge 0$. Therefore, $(i,a)\in \supp(\wt A)$.

To show $\sgn(\wt A_{\wh S}) = \sgn(A_{\wh S})$, since Lemma \ref{lemP} guarantees $\sgn(\wh A_{ia}) = \sgn(A_{ia})$ for any $(i,a) \in \wh S$ and $i\in \wh I$, we focus on each $(i, a)\in \wh S$ and $i\in \wh J$. Assuming $\wt A_{ia} = \wt \beta^i_a> 0$, we know $ \wt \beta^i_a = \bar \beta^i_a> \mu$. Since 
$
A_{ia} = \beta^i_a \ge \bar \beta^i_a - \|\bar \beta^i-\beta^i\|_\i > 0,
$
we have proved that $A_{ia}>0$ for any $\wt A_{ia}>0$ with $(i,a)\in \wh S$ and $i\in\wh J$. Since the same argument holds for any $\wt A_{ia}<0$, we conclude the proof of $\sgn(\wt A_{\wh S}) = \sgn(A_{\wh S})$.

The same conclusion in part (b) and (c) of Theorem \ref{propgroup} holds for GFPP, GFNP, TFPP and TFNP based on the hard-threshold estimator $\wt A$, as it shares the same property in part (a).\qed 
\\

\section{Cross-validation illustration}\label{appendix:B}
We consider a simple case, when $C$ is diagonal and the signed permutation matrix $P$ is $I$, to illustrate our cross-validation method.\\

\noindent\textbf{Example 1.}
Let $C = \textrm{diag}(\tau,\tau,\tau)$, $\I = \bigl\{\{1,2\},\{3,4\},\{5,6\}\bigr\}$ and
\begin{eqnarray*}
	A = \begin{bmatrix}
		1 & 0 & 0 \\
		-1 & 0 & 0\\
		0 & 1 & 0\\
		0 & 1  & 0\\
		0 & 0 & -1\\
		0 & 0 & -1\\
		0.4 & 0.6 & 0\\
		-0.5 & 0 & 0.4
	\end{bmatrix},\quad A_ICA_I^T = \begin{bmatrix}
		* & \tau & 0 & 0 & 0& 0\\
		\tau & * & 0 & 0 & 0& 0\\
		0 & 0 & * & \tau & 0 & 0\\
		0 & 0 & \tau & * & 0& 0\\
		0 & 0 & 0& 0& * & \tau\\
		0 & 0 & 0& 0 & \tau & * \\
	\end{bmatrix},
\end{eqnarray*}
where we use $*$ to reflect the fact that our algorithm  ignores the diagonal elements. For the true $I$ and $\I$, we have $\wh A_{I} = A_I$,
\begin{eqnarray*}
	\left\|\wh \Sigma_{II}^{(1)}-A_I \wh CA_I^T\right\|_{\textrm{F-off}} &\le&
	\left\|\wh \Sigma_{II}^{(1)}-\Sigma_{II}\right\|_{\textrm{F-off}}+\left\|A_I \wh CA_I^T-\Sigma_{II}\right\|_{\textrm{F-off}}\\
	& \le &\left\|\wh \Sigma_{II}^{(1)}-\Sigma_{II}\right\|_{\textrm{F-off}} + \sqrt{|I|(|I|-1)}\cdot\|\wh C-C\|_\i.
\end{eqnarray*}
For
$$\epsilon = \left(\max_{i\ne j}\left|\wh\Sigma^{(1)}_{ij}-\Sigma_{ij}\right|\right) \vee \left(\max_{i\ne j}\left|\wh\Sigma^{(2)}_{ij}-\Sigma_{ij}\right|\right),$$ 
we obtain
\begin{eqnarray*}
	CV(\I) = \frac{1}{\sqrt{| I|\bigl(|I|-1\bigr)}}\left\|\wh \Sigma_{II}^{(1)}-A_I \wh CA_I^T\right\|_{\textrm{F-off}}\le 2\epsilon.
\end{eqnarray*} 
Suppose that  $\wh \I = \bigl\{\{1,2\},\{3,5\},\{4,6\}\bigr\}$, so $\wh I=I$, yet $\wh\I\ne\I$, we would have 
\begin{equation*}
\wh A_{\wh I}\wh C\wh A_{\wh I}^T = \begin{bmatrix}
* & \wh \tau_1 & 0 & 0 & 0& 0\\
\wh \tau_1 & * & 0 & 0 & 0& 0\\
0 & 0 & * & 0 & \wh \tau_2 & 0\\
0 & 0 & 0 & * & 0 & \wh \tau_3\\
0 & 0 & \wh \tau_2&  0& * & 0\\
0 & 0 & 0& \wh \tau_3 & 0 & * \\
\end{bmatrix},
\end{equation*}
\begin{equation*}
\wh A_{\wh I}\wh C\wh A_{\wh I}^T - \Sigma_{\wh I\wh I} = 
\begin{bmatrix}
* &  \Delta \tau_1 & 0 & 0 & 0& 0\\
\Delta \tau_1 & * & 0 & 0 & 0& 0\\
0 & 0 & * & \bm{-\tau} & \bm{\wh \tau_2} & 0\\
0 & 0 & \bm{-\tau} & * & 0& \bm{\wh \tau_3}\\
0 & 0 &\bm{ \wh \tau_2}& 0& * & \bm{-\tau}\\
0 & 0 & 0& \bm{\wh \tau_3} & \bm{-\tau}& * \\
\end{bmatrix}.
\end{equation*}
Here $\Delta\tau_a = \wh\tau_a - \tau_a$, using estimates $\wh \tau_a$ defined in lieu of $\wh C_{aa}$ from (\ref{Chat}) for each $a\in [\wh K]$. Thus, the cross-validation criterion  in (\ref{cvcrit}) would satisfy 
\begin{align*}
CV(\wh \I) &\ge  \frac{\left\|\wh A_{\wh I}\wh C\wh A_{\wh I}^T - \Sigma_{\wh I\wh I}\right\|_{\rm{F-off}}-\left\|\wh \Sigma_{\wh I\wh I}^{(1)}-\Sigma_{\wh I\wh I}\right\|_{\textrm{F-off}}}{\sqrt{| \wh I|\bigl(|\wh I|-1\bigr)}}\ge  \sqrt{\frac{4\tau^2 + 2\wh\tau_2^2+2\wh \tau_3^2}{|\wh I|\bigl(|\wh I|-1\bigr)}}-2\epsilon.
\end{align*}
From noting that $|\wh \tau_a-\tau|\le \epsilon$, for $a = 2,3$, it gives
\begin{eqnarray*}
	CV(\wh \I) \ge \sqrt{\frac{4\tau^2-4\tau\epsilon+2\epsilon^2}{15}}-2\epsilon > 2\epsilon \ge CV(\I),
\end{eqnarray*}
for $\tau \ge 9\epsilon$. We conclude in this example, with $\wh I=I$, incorrectly specifying $  \I$ will induce a large loss. It is easily   verified that this is also the case when $\wh I = I$ but $\wh K \ne K$ and $\wh\I \ne \I$. 

On the other hand, suppose we mistakenly  included    some non-pure variable in $\wh I$. For instance, suppose we found $\wh \I = \bigl\{\{1,2\},\{3,4\},\{5,6,7\}\bigr\}$. Then we would  have
\begin{equation*}
\Sigma_{\wh I'\wh I'} = \begin{bmatrix}
* & \tau & 0 & 0 & 0& 0 & 0.4\tau\\
\tau & * & 0 & 0 & 0& 0 & -0.4\tau\\
0 & 0 & * & \tau & 0 & 0 & 0.6\tau\\
0 & 0 & \tau & * & 0& 0 & 0.6\tau\\
0 & 0 & 0& 0& * & \tau & 0\\
0 & 0 & 0& 0 & \tau & * & 0\\
0.4\tau & -0.4\tau & 0.6\tau & 0.6\tau & 0 & 0 & *
\end{bmatrix},
\end{equation*}
and
\begin{equation*}
\wh A_{\wh I'}\wh C\wh A_{\wh I'}^T =  
\begin{bmatrix}
* & \wh\tau_1 & 0 & 0 & 0& 0 & 0\\
\wh\tau_1 & * & 0 & 0 & 0& 0 & 0\\
0 & 0 & * & \wh \tau_2 & 0 & 0 & 0\\
0 & 0 & \wh \tau_2 & * & 0& 0 & 0\\
0 & 0 & 0& 0& * & \wh\tau_3 & \wh\tau_3\\
0 & 0 & 0& 0 & \wh\tau_3 & * & \wh\tau_3\\
0 & 0& 0 & 0& \wh\tau_3 & \wh\tau_3 & *
\end{bmatrix}.
\end{equation*}
We thus have 
\begin{eqnarray*}
	\wh A_{\wh I'}\wh C\wh A_{\wh I'}^T - \Sigma_{\wh I'\wh I'} = 
	\begin{bmatrix}
		* &  \Delta \tau_1 & 0 & 0 & 0& 0 & \bm{-0.4\tau}\\
		\Delta \tau_1 & * & 0 & 0 & 0& 0 & \bm{0.4\tau}\\
		0 & 0 & * & \Delta \tau_2& 0 & 0 & \bm{-0.6\tau}\\
		0 & 0 & \Delta \tau_2 & * & 0& 0 &\bm{-0.6\tau}\\
		0 & 0 & 0& 0 & * & \Delta \tau_3 & \bm{\wh\tau_3}\\
		0 & 0 & 0& 0 & \Delta \tau_3 & * & \bm{\wh\tau_3}\\
		\bm{-0.4\tau} & \bm{0.4\tau} & \bm{-0.6\tau} & \bm{-0.6\tau} & \bm{\wh\tau_3} & \bm{\wh\tau_3} & *
	\end{bmatrix}
\end{eqnarray*}
and, by similar arguments, for $\tau \ge 12\epsilon$, we find
\begin{eqnarray*}
	CV(\wh \I') \ge 
	\sqrt{\frac{4\wh \tau_3^2 + 4\times 0.36\tau^2+4\times 0.16\tau^2}{42}}-2\epsilon > 2\epsilon.
\end{eqnarray*}
Thus, the cross-validation loss in this example will be large even if  only one non-pure variable is mistakenly classified as pure variable. 
In rare cases, the cross-validation criterion might miss a very small subset of $I$ but this can be rectified in our later estimation of $A_J$.\\

\section{Additional Simulation Results}
\subsection{Related work on the estimation of $A$}\label{moreBai} 

As we explained in Section \ref{sec_related_works}, the existing procedures for estimating $A$ in (\ref{mod}) are developed  for models satisfying identifiability conditions different than our (i)-(iii).  Specifically, \cite{bai2012}  propose to first optimize, via EM, a  quasi-likelihood objective under the identifiability conditions  (a)  $C=\bI_K$ and (b) $A^T \Gamma ^{-1} A$ is diagonal. 
The major advantage of this setting is that the computationally demanding EM algorithm only needs to determine $A$ and $\Gamma$ as $C=\bI_K$ is given. The EM algorithm, however, is only guaranteed to  find  
stationary point $\wh B$ with the property that $\wh B^T \wh B$ is diagonal. In the context of this problem, as the authors note, the EM algorithm requires a delicate initialization and is  computationally demanding,  even if only one of $K$, $n$ and $p$ is moderately large.  Next, the authors propose to link this estimator with an estimator of a model  no longer satisfying (a) and (b) as identifiability conditions,  but satisfying instead (1) $C$ is an arbitrary positive definite matrix; (2) There exists a {\it known} set $S$ of $K$ pure variables, with only one pure variable per latent factor allowed. No further sparsity conditions on $A$ are imposed.  To estimate $A$ under (2), they suggest to solve for $A$ and $C$ the equation $ACA^T= \wh B\wh B^T$. This yields the estimator  $\wt A= \wh B \wh B_S^{-1}$ of  $A$. However, when $K$ is relatively large,  $\wh B_S$ may not be invertible, and the estimator may not exist. 
Finally, although $\wt A_S= \wh B_S \wh B_S^{-1}= \bI_K$, the submatrix  $\wt A_{S^c}$ is not sparse in general. One possibility is to threshold $\wt A$,  but it is unclear how  to choose the correct threshold level, for the following reason. Although the authors establish the asymptotic limit of the MLE of $A$ under (1) and (2), the estimator of $A$ explained above is not guaranteed to be the MLE in this model:  if it exists, it is a transformation of a  stationary point that estimates parameters under the  model specifications (a) and (b), different from (1) and (2).  The immediate practical implication  is that the variation of $\wh A$ around $A$  under (1) and (2)  is not known, which makes the thresholding level of $\wh A$ difficult to assess. For all these reasons,    we cannot compare numerically our estimation procedure with the procedure proposed in \cite{bai2012}, even in the  (unrealistic) case when the pure variable set is known.

\subsection{LOVE for non-overlapping cluster estimation}

In applications, one may not have prior information on whether the clusters may overlap or not. Thus, one would prefer a clustering method that works well in both overlapping and non-overlapping scenarios. In the previous section, we have demonstrated that LOVE outperforms the existing clustering methods if data are generated from a model that yields variable clusters with overlaps.   In this section, we study the numerical performance of the proposed method under  non-overlapping data generating schemes.

To generate data with non-overlapping clusters, we set the number of variables in each cluster to be $20$. We generate the diagonal elements of $C$ from the uniform distribution in $[1,2]$ and use the same method as in  Section \ref{sec_sim_love} to generate the off-diagonal elements. The variance $\sigma_j^2$ of the error $E_j$ is generated from the uniform distribution in $[3,4]$. In Table \ref{tab2}, we compare the sensitivity and specificity of the proposed method with the CORD estimator \citep{bunea2015minimax} under non-overlapping scenarios, where the sensitivity and specificity are defined in  (\ref{eqspsn}). The CORD estimator can be viewed as a benchmark method for variable clustering without overlaps and is shown to outperform K-means and hierarchical clustering, via an extensive numerical study presented in \cite{bunea2015minimax}. For this reason,  we only focus on the comparison between LOVE and  CORD. From Table \ref{tab2}, we see that for small $p$ (i.e., $p=100$) the performance of LOVE is only slightly worse than CORD. As $p$ increases, the specificity of LOVE and that of CORD remain close to $1$, but LOVE yields in fact higher sensitivity than CORD when $n=300$. This confirms that the performance of the proposed method is comparable to the benchmark method under non-overlapping scenarios. Of course, LOVE is much more flexible as it can detect possible overlaps.

\begin{table}[ht!]
	\begin{center}
		\caption{Sensitivity (SN) and specificity (SP) of the proposed method (LOVE) and CORD under non-overlapping scenarios. Numbers in parentheses are the simulation standard errors.}
		\label{tab2}
		\begin{tabular}{ccccccccc}
			\toprule 
			$p$ &\multicolumn{4}{c}{$n=300$} &\multicolumn{4}{c}{$n=500$}\\
			\cmidrule(r){2-5}\cmidrule(r){6-9} 
			&\multicolumn{2}{c}{LOVE} &\multicolumn{2}{c}{CORD}&\multicolumn{2}{c}{LOVE} &\multicolumn{2}{c}{CORD}\\
			\cmidrule(r){2-5}\cmidrule(r){6-9} 
			& SN & SP & SN &  SP & SN & SP & SN & SP \\
			\midrule
			100 & 0.87 & 0.90 & 0.92 & 0.98 & 0.93 & 0.97 & 0.98 & 1.00 \\
			& (0.09) & (0.10) &(0.05) & (0.02) & (0.05) & (0.03) & (0.02) &(0.01) \\
			500 & 0.86 & 0.98 & 0.82 & 0.98 & 0.87 & 0.99 & 0.94 & 1.00 \\
			& (0.05) & (0.01) &(0.03) & (0.01) & (0.04) & (0.00) & (0.02) &(0.01) \\
			1000 & 0.84 & 0.97 & 0.78 & 0.97 & 0.87 & 1.00 & 0.90 & 1.00 \\
			& (0.05) & (0.02) &(0.03) & (0.01) & (0.04) & (0.00) & (0.02) &(0.01) \\
			\bottomrule
		\end{tabular}
	\end{center}
\end{table}

\subsection{Comparison with other overlapping clustering algorithms}\label{other}

We adopt the same data generating procedure except that we set $K=10$ and the negative entries of $A$ are replaced by their absolute values, since existing  overlapping clustering algorithms typically return an estimator of $A$ with positive entries. 
We compare the proposed method with the following overlapping clustering algorithms:  fuzzy K-means, and fuzzy K-medoids \citep{krishnapuram2001low}, the latter being more robust to noise and outliers.  We describe the methods  briefly in what follows.  Both of them  aim to estimate a degree of membership matrix $M\in \RR^{p\times K}$ by minimizing the average within-cluster $L_2$ or $L_1$ distances \citep{bezdek2013pattern}. 
Specifically, denote $\wt X_j=(X_{1j},...,X_{nj})$, and $\wt X=\{\wt X_1,...,\wt X_p\}$. Let $W=\{w_1,...,w_K\}$, where $w_k\in \mathbb{R}^n$, be a subset of $\wt X$ with $K$ elements. The fuzzy algorithms aim to find the set $W$ such that $J(W)$ defined as 
$$
J(W)=\sum_{j=1}^p\sum_{k=1}^K M_{jk} r(\wt X_j,w_k),
$$
is minimized. Here, $M_{jk}>0$ can be interpreted as the degree of membership matrix which is a known function of $r(\wt X_j,w_k)$. Some commonly used expressions of $M_{jk}$ are shown by \cite{krishnapuram2001low}. In addition, $r(\wt X_j,w_k)$ is a measure of dissimilarity between $\wt X_j$ and $w_k$. For instance, if $r(x,\theta)=\|x-\theta\|_2^2$, this corresponds to the fuzzy K-means. Similarly, the fuzzy K-medoids is given by $r(x,\theta)=\|x-\theta\|_1$. Since searching over all possible subsets of $\wt X$ is computationally infeasible, an approximate algorithm for minimizing $J(W)$ is proposed by \cite{krishnapuram2001low}, we refer to their original paper for further details.


Their degree of membership matrix $M$ plays the same role as our allocation matrix $A$, but is typically non-sparse. In order to construct overlapping clusters based on $M$ one needs to specify a cut-off value $v$ and assign variable $j$  to cluster $k$ if $M_{jk}>v$. Moreover, the number of clusters $K$ is a required input of the algorithm. In the simulations presented in this section we set $K = 10$ for these two methods, which  have been implemented by the functions {\verb FKM } and {\verb FKM.med } in R. \\
\indent We compare their performance with  our proposed method LOVE. We emphasize that our method does not require 
the specification of $K$ and that the tuning parameters are chosen in a data adaptive fashion, as explained in the previous sections.  We follow the pairwise approach of  \cite{wiwie2015comparing} for this comparison. Recall that $\G=(G_1,...,G_K)$ denotes the true overlapping clusters. For notational simplicity, we use $\wh \G=(\wh G_1,...,\wh G_{\bar K})$ to denote clusters computed from an algorithm. Since LOVE estimates the number of clusters, we allow $\wh K$ to be different from $K$. For any pair $1\leq j<k\leq p$, define 
\begin{align*}
TP_{jk} &=\1\left\{ \textrm{if $j,k\in G_a$ and $j,k\in \wh G_b$ for some $1\leq a\leq K$ and $1\leq b\leq \wh K$}\right\},
\\
TN_{jk} &=\1\left\{ \textrm{if $j,k\notin G_a$ and $j,k\notin \wh G_b$ for any $1\leq a\leq K$ and $1\leq b\leq \wh K$}\right\},
\\
FP_{jk}&=\1\left\{\textrm{if $j,k\notin G_a$ for any $1\leq a\leq K$ and $j,k\in \wh G_b$ for some  $1\leq b\leq \wh K$}\right\},
\\
FN_{jk} &=\1\left\{\textrm{if $j,k\in G_a$ for some $1\leq a\leq K$ and $j,k\notin \wh G_b$ for any $1\leq b\leq \wh K$}\right\}.
\end{align*}
and we define 
\begin{align*}
TP&=\sum_{1\leq j<k\leq p} TP_{jk},~TN=\sum_{1\leq j<k\leq p} TN_{jk},\\
FP&=\sum_{1\leq j<k\leq p} FP_{jk},~FN=\sum_{1\leq j<k\leq p} FN_{jk}.
\end{align*}
We use sensitivity (SN) and specificity (SP) to evaluate the performance of different methods, where
\begin{equation}\label{eqspsn}
SP=\frac{TN}{TN+FP}, ~\textrm{and}~SN=\frac{TP}{TP+FN}.
\end{equation}
Recall that for the fuzzy methods, variable $j$ belongs to cluster $k$ if the estimated membership matrix $M_{jk}$ is beyond a cut-off $v$, i.e., $M_{jk}>v$. We search for the optimal cut-off $v$ in a grid $\{0.01, 0.1, ..., 0.3\}$ such that $SP + SN$ is maximized. The corresponding sensitivity and specificity for LOVE, fuzzy K-means (F-Kmeans) and fuzzy K-medoids (F-Kmed) are shown in Figure \ref{fig1}. To save space, we only present the results for $p=500$ since the other scenarios illustrate the same patterns. The following findings are observed. First, the F-Kmeans is superior to F-Kmed in most scenarios in terms of both sensitivity and specificity. Second, LOVE clearly outperforms these two existing methods and its specificity and sensitivity are very close to 1, which implies that our method leads to very few false positives and false negatives. The conclusions hold with $n$ from $300$ to $1000$. Moreover, we reiterate  that the true value $K = 10$ is used as input in the competing methods, whereas it is estimated from the data in LOVE. This illustrates the net advantage of the proposed method over the existing overlapping clustering methods, for data generated from Model (\ref{mod}). 
{
	\begin{figure}[htp!]
		\begin{center}
			\begin{tabular}{cc}
				\\[-10pt]
				\vspace{-0in} \hskip-15pt
				\includegraphics[width=2.5in]{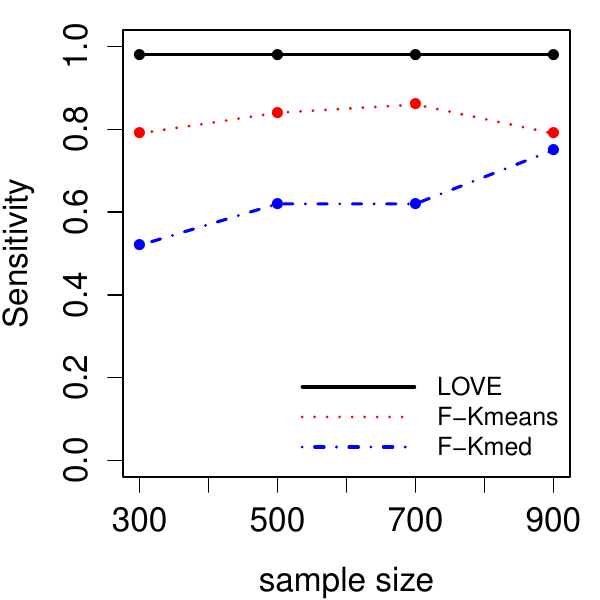} &
				\hskip-15pt
				\includegraphics[width=2.5in]{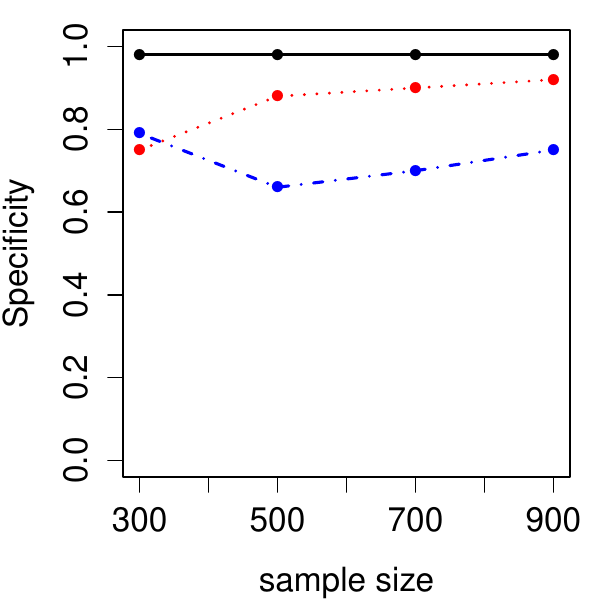}
			\end{tabular}
		\end{center}
		\caption{Plot of specificity and sensitivity for LOVE, fuzzy K-means (F-Kmeans), and fuzzy K-medoids (F-Kmed) when $p=500$. } \label{fig1}
	\end{figure}
}

\vspace{.5cm}
\bibliographystyle{imsart-nameyear}
\bibliography{ref}

\begin{thebibliography}{52}

\bibitem[\protect\citeauthoryear{Anderson}{2003}]{anderson_book}
\begin{bbook}[author]
\bauthor{\bsnm{Anderson},~\bfnm{T.~W.}\binits{T.~W.}}
(\byear{2003}).
\btitle{An Introduction to Multivariate Statistical Analysis}.
\bseries{Wiley Series in Probability and Statistics}.
\bpublisher{Wiley}.
\end{bbook}
\endbibitem

\bibitem[\protect\citeauthoryear{Anderson and Amemiya}{1988}]{anderson1988}
\begin{barticle}[author]
\bauthor{\bsnm{Anderson},~\bfnm{T.~W.}\binits{T.~W.}} \AND
  \bauthor{\bsnm{Amemiya},~\bfnm{Yasuo}\binits{Y.}}
(\byear{1988}).
\btitle{The Asymptotic Normal Distribution of Estimators in Factor Analysis
  under General Conditions}.
\bjournal{Ann. Statist.}
\bvolume{16}
\bpages{759--771}.
\bdoi{10.1214/aos/1176350834}
\end{barticle}
\endbibitem

\bibitem[\protect\citeauthoryear{Anderson and Rubin}{1956}]{anderson1956}
\begin{binproceedings}[author]
\bauthor{\bsnm{Anderson},~\bfnm{T.~W.}\binits{T.~W.}} \AND
  \bauthor{\bsnm{Rubin},~\bfnm{Herman}\binits{H.}}
(\byear{1956}).
\btitle{Statistical Inference in Factor Analysis}.
In \bbooktitle{Proceedings of the Third Berkeley Symposium on Mathematical
  Statistics and Probability, Volume 5: Contributions to Econometrics,
  Industrial Research, and Psychometry}
\bpages{111--150}.
\bpublisher{University of California Press}, \baddress{Berkeley, Calif.}
\end{binproceedings}
\endbibitem

\bibitem[\protect\citeauthoryear{Arora et~al.}{2013}]{arora2013practical}
\begin{binproceedings}[author]
\bauthor{\bsnm{Arora},~\bfnm{Sanjeev}\binits{S.}},
  \bauthor{\bsnm{Ge},~\bfnm{Rong}\binits{R.}},
  \bauthor{\bsnm{Halpern},~\bfnm{Yonatan}\binits{Y.}},
  \bauthor{\bsnm{Mimno},~\bfnm{David~M}\binits{D.~M.}},
  \bauthor{\bsnm{Moitra},~\bfnm{Ankur}\binits{A.}},
  \bauthor{\bsnm{Sontag},~\bfnm{David}\binits{D.}},
  \bauthor{\bsnm{Wu},~\bfnm{Yichen}\binits{Y.}} \AND
  \bauthor{\bsnm{Zhu},~\bfnm{Michael}\binits{M.}}
(\byear{2013}).
\btitle{A Practical Algorithm for Topic Modeling with Provable Guarantees.}
In \bbooktitle{ICML (2)}
\bpages{280--288}.
\end{binproceedings}
\endbibitem

\bibitem[\protect\citeauthoryear{Ashburner et~al.}{2000}]{ashburner2000gene}
\begin{barticle}[author]
\bauthor{\bsnm{Ashburner},~\bfnm{Michael}\binits{M.}},
  \bauthor{\bsnm{Ball},~\bfnm{Catherine~A}\binits{C.~A.}},
  \bauthor{\bsnm{Blake},~\bfnm{Judith~A}\binits{J.~A.}},
  \bauthor{\bsnm{Botstein},~\bfnm{David}\binits{D.}},
  \bauthor{\bsnm{Butler},~\bfnm{Heather}\binits{H.}},
  \bauthor{\bsnm{Cherry},~\bfnm{J~Michael}\binits{J.~M.}},
  \bauthor{\bsnm{Davis},~\bfnm{Allan~P}\binits{A.~P.}},
  \bauthor{\bsnm{Dolinski},~\bfnm{Kara}\binits{K.}},
  \bauthor{\bsnm{Dwight},~\bfnm{Selina~S}\binits{S.~S.}} \AND
  \bauthor{\bsnm{Eppig},~\bfnm{Janan~T}\binits{J.~T.}}
(\byear{2000}).
\btitle{Gene Ontology: tool for the unification of biology}.
\bjournal{Nature genetics}
\bvolume{25}
\bpages{25--29}.
\end{barticle}
\endbibitem

\bibitem[\protect\citeauthoryear{Bai and Li}{2012}]{bai2012}
\begin{barticle}[author]
\bauthor{\bsnm{Bai},~\bfnm{Jushan}\binits{J.}} \AND
  \bauthor{\bsnm{Li},~\bfnm{Kunpeng}\binits{K.}}
(\byear{2012}).
\btitle{Statistical analysis of factor models of high dimension}.
\bjournal{Ann. Statist.}
\bvolume{40}
\bpages{436--465}.
\bdoi{10.1214/11-AOS966}
\end{barticle}
\endbibitem

\bibitem[\protect\citeauthoryear{Bai and Ng}{2002}]{BaiNg2002}
\begin{barticle}[author]
\bauthor{\bsnm{Bai},~\bfnm{Jushan}\binits{J.}} \AND
  \bauthor{\bsnm{Ng},~\bfnm{Serena}\binits{S.}}
(\byear{2002}).
\btitle{Determining the Number of Factors in Approximate Factor Models}.
\bjournal{Econometrica}
\bvolume{70}
\bpages{191--221}.
\bdoi{10.1111/1468-0262.00273}
\end{barticle}
\endbibitem

\bibitem[\protect\citeauthoryear{Bekker and ten Berge}{1997}]{Bekker1997}
\begin{barticle}[author]
\bauthor{\bsnm{Bekker},~\bfnm{Paul~A.}\binits{P.~A.}} \AND
  \bauthor{\bparticle{ten} \bsnm{Berge},~\bfnm{Jos M.~F.}\binits{J.~M.~F.}}
(\byear{1997}).
\btitle{Generic global identification in factor analysis}.
\bjournal{Linear Algebra and its Applications}
\bvolume{264}
\bpages{255 - 263}.
\bnote{Sixth Special Issue on Linear Algebra and Statistics}.
\bdoi{https://doi.org/10.1016/S0024-3795(96)00363-1}
\end{barticle}
\endbibitem

\bibitem[\protect\citeauthoryear{Belloni, Rosenbaum and
  Tsybakov}{2017}]{belloni2016}
\begin{barticle}[author]
\bauthor{\bsnm{Belloni},~\bfnm{Alexandre}\binits{A.}},
  \bauthor{\bsnm{Rosenbaum},~\bfnm{Mathieu}\binits{M.}} \AND
  \bauthor{\bsnm{Tsybakov},~\bfnm{Alexandre~B.}\binits{A.~B.}}
(\byear{2017}).
\btitle{Linear and conic programming estimators in high dimensional
  errors-in-variables models}.
\bjournal{Journal of the Royal Statistical Society: Series B (Statistical
  Methodology)}
\bvolume{79}
\bpages{939--956}.
\bdoi{10.1111/rssb.12196}
\end{barticle}
\endbibitem

\bibitem[\protect\citeauthoryear{Best et~al.}{2015}]{best2015rna}
\begin{barticle}[author]
\bauthor{\bsnm{Best},~\bfnm{Myron~G}\binits{M.~G.}},
  \bauthor{\bsnm{Sol},~\bfnm{Nik}\binits{N.}},
  \bauthor{\bsnm{Kooi},~\bfnm{Irsan}\binits{I.}},
  \bauthor{\bsnm{Tannous},~\bfnm{Jihane}\binits{J.}},
  \bauthor{\bsnm{Westerman},~\bfnm{Bart~A}\binits{B.~A.}},
  \bauthor{\bsnm{Rustenburg},~\bfnm{Fran{\c{c}}ois}\binits{F.}},
  \bauthor{\bsnm{Schellen},~\bfnm{Pepijn}\binits{P.}},
  \bauthor{\bsnm{Verschueren},~\bfnm{Heleen}\binits{H.}},
  \bauthor{\bsnm{Post},~\bfnm{Edward}\binits{E.}},
  \bauthor{\bsnm{Koster},~\bfnm{Jan}\binits{J.}} \betal{et~al.}
(\byear{2015}).
\btitle{RNA-Seq of tumor-educated platelets enables blood-based pan-cancer,
  multiclass, and molecular pathway cancer diagnostics}.
\bjournal{Cancer cell}
\bvolume{28}
\bpages{666--676}.
\end{barticle}
\endbibitem

\bibitem[\protect\citeauthoryear{Bezdek}{2013}]{bezdek2013pattern}
\begin{bbook}[author]
\bauthor{\bsnm{Bezdek},~\bfnm{James~C}\binits{J.~C.}}
(\byear{2013}).
\btitle{Pattern recognition with fuzzy objective function algorithms}.
\bpublisher{Springer Science \& Business Media}.
\end{bbook}
\endbibitem

\bibitem[\protect\citeauthoryear{Bhattacharya and Dunson}{2011}]{dunson2011}
\begin{barticle}[author]
\bauthor{\bsnm{Bhattacharya},~\bfnm{A.}\binits{A.}} \AND
  \bauthor{\bsnm{Dunson},~\bfnm{D.~B.}\binits{D.~B.}}
(\byear{2011}).
\btitle{Sparse Bayesian infinite factor models}.
\bjournal{Biometrika}
\bvolume{98}
\bpages{291--306}.
\bdoi{10.1093/biomet/asr013}
\end{barticle}
\endbibitem

\bibitem[\protect\citeauthoryear{Bien, Bunea and Xiao}{2016}]{bien2016convex}
\begin{barticle}[author]
\bauthor{\bsnm{Bien},~\bfnm{Jacob}\binits{J.}},
  \bauthor{\bsnm{Bunea},~\bfnm{Florentina}\binits{F.}} \AND
  \bauthor{\bsnm{Xiao},~\bfnm{Luo}\binits{L.}}
(\byear{2016}).
\btitle{Convex banding of the covariance matrix}.
\bjournal{Journal of the American Statistical Association}
\bvolume{111}
\bpages{834--845}.
\end{barticle}
\endbibitem

\bibitem[\protect\citeauthoryear{Bing, Bunea and Wegkamp}{2018}]{topic18}
\begin{barticle}[author]
\bauthor{\bsnm{Bing},~\bfnm{Xin}\binits{X.}},
  \bauthor{\bsnm{Bunea},~\bfnm{Florentina}\binits{F.}} \AND
  \bauthor{\bsnm{Wegkamp},~\bfnm{Marten~H.}\binits{M.~H.}}
(\byear{2018}).
\btitle{A fast algorithm with minimax optimal guarantees for topic models with
  an unknown number of topics}.
\bjournal{ArXiv e-prints:1805.06837}.
\end{barticle}
\endbibitem

\bibitem[\protect\citeauthoryear{Bing and Wegkamp}{2018}]{bing2018}
\begin{barticle}[author]
\bauthor{\bsnm{Bing},~\bfnm{Xin}\binits{X.}} \AND
  \bauthor{\bsnm{Wegkamp},~\bfnm{Marten~H.}\binits{M.~H.}}
(\byear{2018}).
\btitle{Adaptive estimation of the rank of the coefficient matrix in high
  dimensional multivariate response regression models}.
\bjournal{ArXiv}
\banumber{1704.02381}.
\end{barticle}
\endbibitem

\bibitem[\protect\citeauthoryear{Bittorf et~al.}{2012}]{bittorfNMF}
\begin{barticle}[author]
\bauthor{\bsnm{Bittorf},~\bfnm{Victor}\binits{V.}},
  \bauthor{\bsnm{Recht},~\bfnm{Benjamin}\binits{B.}},
  \bauthor{\bsnm{Re},~\bfnm{Christopher}\binits{C.}} \AND
  \bauthor{\bsnm{Tropp},~\bfnm{Joel~A}\binits{J.~A.}}
(\byear{2012}).
\btitle{Factoring nonnegative matrices with linear programs}.
\bjournal{arXiv:1206.1270}.
\end{barticle}
\endbibitem

\bibitem[\protect\citeauthoryear{Bollen}{1989}]{bollen1989}
\begin{bbook}[author]
\bauthor{\bsnm{Bollen},~\bfnm{Kenneth~A.}\binits{K.~A.}}
(\byear{1989}).
\btitle{Structural Equations with Latent Variables}.
\bpublisher{Wiley}.
\end{bbook}
\endbibitem

\bibitem[\protect\citeauthoryear{Bunea, Giraud and
  Luo}{2016a}]{bunea2015minimax}
\begin{barticle}[author]
\bauthor{\bsnm{Bunea},~\bfnm{Florentina}\binits{F.}},
  \bauthor{\bsnm{Giraud},~\bfnm{Christophe}\binits{C.}} \AND
  \bauthor{\bsnm{Luo},~\bfnm{Xi}\binits{X.}}
(\byear{2016}a).
\btitle{Minimax Optimal Variable Clustering in G-models via Cord}.
\bjournal{arXiv preprint arXiv:1508.01939}.
\end{barticle}
\endbibitem

\bibitem[\protect\citeauthoryear{Bunea, She and Wegkamp}{2011}]{bunea2011}
\begin{barticle}[author]
\bauthor{\bsnm{Bunea},~\bfnm{Florentina}\binits{F.}},
  \bauthor{\bsnm{She},~\bfnm{Yiyuan}\binits{Y.}} \AND
  \bauthor{\bsnm{Wegkamp},~\bfnm{Marten~H.}\binits{M.~H.}}
(\byear{2011}).
\btitle{Optimal selection of reduced rank estimators of high-dimensional
  matrices}.
\bjournal{Ann. Statist.}
\bvolume{39}
\bpages{1282--1309}.
\bdoi{10.1214/11-AOS876}
\end{barticle}
\endbibitem

\bibitem[\protect\citeauthoryear{Bunea et~al.}{2016b}]{bunea2016pecok}
\begin{barticle}[author]
\bauthor{\bsnm{Bunea},~\bfnm{Florentina}\binits{F.}},
  \bauthor{\bsnm{Giraud},~\bfnm{Christophe}\binits{C.}},
  \bauthor{\bsnm{Royer},~\bfnm{Martin}\binits{M.}} \AND
  \bauthor{\bsnm{Verzelen},~\bfnm{Nicolas}\binits{N.}}
(\byear{2016}b).
\btitle{PECOK: a convex optimization approach to variable clustering}.
\bjournal{arXiv preprint arXiv:1606.05100}.
\end{barticle}
\endbibitem

\bibitem[\protect\citeauthoryear{Bunea et~al.}{2018}]{bunea_cord_pecok}
\begin{barticle}[author]
\bauthor{\bsnm{Bunea},~\bfnm{Florentina}\binits{F.}},
  \bauthor{\bsnm{Christophe},~\bfnm{Giraudm}\binits{G.}},
  \bauthor{\bsnm{Luo},~\bfnm{Xi}\binits{X.}},
  \bauthor{\bsnm{Royer},~\bfnm{Martin}\binits{M.}} \AND
  \bauthor{\bsnm{Verzelen},~\bfnm{Nicolas}\binits{N.}}
(\byear{2018}).
\btitle{Model Assisted Variable Clustering: Minimax-optimal Recovery and
  Algorithms}.
\bjournal{ArXiv e-prints:1508.01939}.
\end{barticle}
\endbibitem

\bibitem[\protect\citeauthoryear{Cai, Liu and Luo}{2011}]{cai_CLIME}
\begin{barticle}[author]
\bauthor{\bsnm{Cai},~\bfnm{Tony}\binits{T.}},
  \bauthor{\bsnm{Liu},~\bfnm{Weidong}\binits{W.}} \AND
  \bauthor{\bsnm{Luo},~\bfnm{Xi}\binits{X.}}
(\byear{2011}).
\btitle{A Constrained ℓ1 Minimization Approach to Sparse Precision Matrix
  Estimation}.
\bjournal{Journal of the American Statistical Association}
\bvolume{106}
\bpages{594-607}.
\bdoi{10.1198/jasa.2011.tm10155}
\end{barticle}
\endbibitem

\bibitem[\protect\citeauthoryear{Cai, Liu and Zhou}{2016}]{cai2016}
\begin{barticle}[author]
\bauthor{\bsnm{Cai},~\bfnm{T.~Tony}\binits{T.~T.}},
  \bauthor{\bsnm{Liu},~\bfnm{Weidong}\binits{W.}} \AND
  \bauthor{\bsnm{Zhou},~\bfnm{Harrison~H.}\binits{H.~H.}}
(\byear{2016}).
\btitle{Estimating sparse precision matrix: Optimal rates of convergence and
  adaptive estimation}.
\bjournal{Annals of Statistics}
\bvolume{44}
\bpages{455--488}.
\bdoi{10.1214/13-AOS1171}
\end{barticle}
\endbibitem

\bibitem[\protect\citeauthoryear{Cand{\`e}s et~al.}{2011}]{2011robust}
\begin{barticle}[author]
\bauthor{\bsnm{Cand{\`e}s},~\bfnm{Emmanuel~J}\binits{E.~J.}},
  \bauthor{\bsnm{Li},~\bfnm{Xiaodong}\binits{X.}},
  \bauthor{\bsnm{Ma},~\bfnm{Yi}\binits{Y.}} \AND
  \bauthor{\bsnm{Wright},~\bfnm{John}\binits{J.}}
(\byear{2011}).
\btitle{Robust principal component analysis?}
\bjournal{Journal of the ACM (JACM)}
\bvolume{58}
\bpages{11}.
\end{barticle}
\endbibitem

\bibitem[\protect\citeauthoryear{Carvalho et~al.}{2008}]{Carvalho2008}
\begin{barticle}[author]
\bauthor{\bsnm{Carvalho},~\bfnm{C.~M.}\binits{C.~M.}},
  \bauthor{\bsnm{Chang},~\bfnm{J.}\binits{J.}},
  \bauthor{\bsnm{Lucas},~\bfnm{J.~E.}\binits{J.~E.}},
  \bauthor{\bsnm{Nevins},~\bfnm{J.~R.}\binits{J.~R.}},
  \bauthor{\bsnm{Wang},~\bfnm{Q.}\binits{Q.}} \AND
  \bauthor{\bsnm{West},~\bfnm{M.}\binits{M.}}
(\byear{2008}).
\btitle{High-Dimensional Sparse Factor Modeling: Applications in Gene
  Expression Genomics}.
\bjournal{Journal of the American Statistical Association}
\bvolume{103}
\bpages{1438–1456}.
\end{barticle}
\endbibitem

\bibitem[\protect\citeauthoryear{Chandrasekaran, Parrilo and
  Willsky}{2012}]{ch12a}
\begin{barticle}[author]
\bauthor{\bsnm{Chandrasekaran},~\bfnm{Venkat}\binits{V.}},
  \bauthor{\bsnm{Parrilo},~\bfnm{Pablo~A.}\binits{P.~A.}} \AND
  \bauthor{\bsnm{Willsky},~\bfnm{Alan~S.}\binits{A.~S.}}
(\byear{2012}).
\btitle{Latent variable graphical model selection via convex optimization}.
\bjournal{Ann. Statist.}
\bvolume{40}
\bpages{1935-1967}.
\end{barticle}
\endbibitem

\bibitem[\protect\citeauthoryear{Chandrasekaran et~al.}{2011}]{ch11}
\begin{barticle}[author]
\bauthor{\bsnm{Chandrasekaran},~\bfnm{Venkat}\binits{V.}},
  \bauthor{\bsnm{Sanghavi},~\bfnm{Sujay}\binits{S.}},
  \bauthor{\bsnm{Parrilo},~\bfnm{Pablo~A.}\binits{P.~A.}} \AND
  \bauthor{\bsnm{Willsky},~\bfnm{Alan~S.}\binits{A.~S.}}
(\byear{2011}).
\btitle{Rank-Sparsity Incoherence for Matrix Decomposition}.
\bjournal{{SIAM} J. Optim.}
\bvolume{21}
\bpages{572-596}.
\end{barticle}
\endbibitem

\bibitem[\protect\citeauthoryear{Craddock et~al.}{2012}]{Craddock2}
\begin{barticle}[author]
\bauthor{\bsnm{Craddock},~\bfnm{R~Cameron}\binits{R.~C.}},
  \bauthor{\bsnm{James},~\bfnm{G~Andrew}\binits{G.~A.}},
  \bauthor{\bsnm{Holtzheimer},~\bfnm{Paul~E}\binits{P.~E.}},
  \bauthor{\bsnm{Hu},~\bfnm{Xiaoping~P}\binits{X.~P.}} \AND
  \bauthor{\bsnm{Mayberg},~\bfnm{Helen~S}\binits{H.~S.}}
(\byear{2012}).
\btitle{A whole brain fMRI atlas generated via spatially constrained spectral
  clustering}.
\bjournal{Human brain mapping}
\bvolume{33}
\bpages{1914--1928}.
\end{barticle}
\endbibitem

\bibitem[\protect\citeauthoryear{Craddock et~al.}{2013}]{Craddock}
\begin{barticle}[author]
\bauthor{\bsnm{Craddock},~\bfnm{R~Cameron}\binits{R.~C.}},
  \bauthor{\bsnm{Jbabdi},~\bfnm{Saad}\binits{S.}},
  \bauthor{\bsnm{Yan},~\bfnm{Chao-Gan}\binits{C.-G.}},
  \bauthor{\bsnm{Vogelstein},~\bfnm{Joshua~T}\binits{J.~T.}},
  \bauthor{\bsnm{Castellanos},~\bfnm{F~Xavier}\binits{F.~X.}},
  \bauthor{\bsnm{Di~Martino},~\bfnm{Adriana}\binits{A.}},
  \bauthor{\bsnm{Kelly},~\bfnm{Clare}\binits{C.}},
  \bauthor{\bsnm{Heberlein},~\bfnm{Keith}\binits{K.}},
  \bauthor{\bsnm{Colcombe},~\bfnm{Stan}\binits{S.}} \AND
  \bauthor{\bsnm{Milham},~\bfnm{Michael~P}\binits{M.~P.}}
(\byear{2013}).
\btitle{Imaging human connectomes at the macroscale}.
\bjournal{Nature methods}
\bvolume{10}
\bpages{524--539}.
\end{barticle}
\endbibitem

\bibitem[\protect\citeauthoryear{Donoho and Stodden}{2004}]{donohoNMF}
\begin{bincollection}[author]
\bauthor{\bsnm{Donoho},~\bfnm{David}\binits{D.}} \AND
  \bauthor{\bsnm{Stodden},~\bfnm{Victoria}\binits{V.}}
(\byear{2004}).
\btitle{When Does Non-Negative Matrix Factorization Give a Correct
  Decomposition into Parts?}
In \bbooktitle{Advances in Neural Information Processing Systems 16}
(\beditor{\bfnm{S.}\binits{S.}~\bsnm{Thrun}},
  \beditor{\bfnm{L.~K.}\binits{L.~K.}~\bsnm{Saul}} \AND
  \beditor{\bfnm{P.~B.}\binits{P.~B.}~\bsnm{Sch\"{o}lkopf}}, eds.)
\bpages{1141--1148}.
\bpublisher{MIT Press}.
\end{bincollection}
\endbibitem

\bibitem[\protect\citeauthoryear{Everitt}{1984}]{everitt1984}
\begin{bbook}[author]
\bauthor{\bsnm{Everitt},~\bfnm{B.~S.}\binits{B.~S.}}
(\byear{1984}).
\btitle{An Introduction to Latent Variable Models}.
\bpublisher{Monographs on Statistics and Applied Probability. Springer}.
\end{bbook}
\endbibitem

\bibitem[\protect\citeauthoryear{Fan, Liao and Mincheva}{2013}]{fan2013large}
\begin{barticle}[author]
\bauthor{\bsnm{Fan},~\bfnm{Jianqing}\binits{J.}},
  \bauthor{\bsnm{Liao},~\bfnm{Yuan}\binits{Y.}} \AND
  \bauthor{\bsnm{Mincheva},~\bfnm{Martina}\binits{M.}}
(\byear{2013}).
\btitle{Large covariance estimation by thresholding principal orthogonal
  complements}.
\bjournal{Journal of the Royal Statistical Society: Series B (Statistical
  Methodology)}
\bvolume{75}
\bpages{603--680}.
\end{barticle}
\endbibitem

\bibitem[\protect\citeauthoryear{Friedman, Hastie and
  Tibshirani}{2008}]{friedman2008sparse}
\begin{barticle}[author]
\bauthor{\bsnm{Friedman},~\bfnm{Jerome}\binits{J.}},
  \bauthor{\bsnm{Hastie},~\bfnm{Trevor}\binits{T.}} \AND
  \bauthor{\bsnm{Tibshirani},~\bfnm{Robert}\binits{R.}}
(\byear{2008}).
\btitle{Sparse inverse covariance estimation with the graphical lasso}.
\bjournal{Biostatistics}
\bvolume{9}
\bpages{432--441}.
\end{barticle}
\endbibitem

\bibitem[\protect\citeauthoryear{Gautier and Tsybakov}{2011}]{gt2011}
\begin{barticle}[author]
\bauthor{\bsnm{Gautier},~\bfnm{Eric}\binits{E.}} \AND
  \bauthor{\bsnm{Tsybakov},~\bfnm{Alexandre~B.}\binits{A.~B.}}
(\byear{2011}).
\btitle{High-dimensional instrumental variables regression and confidence
  sets}.
\bjournal{arXiv preprint arXiv:1105.2454v4}
\bvolume{1105.2454}.
\end{barticle}
\endbibitem

\bibitem[\protect\citeauthoryear{Geweke and Zhou}{1996}]{geweke1996}
\begin{barticle}[author]
\bauthor{\bsnm{Geweke},~\bfnm{J.}\binits{J.}} \AND
  \bauthor{\bsnm{Zhou},~\bfnm{G.}\binits{G.}}
(\byear{1996}).
\btitle{Measuring the pricing error of the arbitrage pricing theory}.
\bjournal{The review of financial studies}
\bvolume{9}
\bpages{557--587}.
\end{barticle}
\endbibitem

\bibitem[\protect\citeauthoryear{Hsu, Kakade and Zhang}{2011}]{Hsu11}
\begin{barticle}[author]
\bauthor{\bsnm{Hsu},~\bfnm{Daniel}\binits{D.}},
  \bauthor{\bsnm{Kakade},~\bfnm{Sham~M.}\binits{S.~M.}} \AND
  \bauthor{\bsnm{Zhang},~\bfnm{Tong}\binits{T.}}
(\byear{2011}).
\btitle{Robust Matrix Decomposition with Sparse Corruptions}.
\bjournal{{IEEE} Trans. Inform. Theory}
\bvolume{57}
\bpages{7221-7234}.
\end{barticle}
\endbibitem

\bibitem[\protect\citeauthoryear{Izenman}{2008}]{izenman}
\begin{bbook}[author]
\bauthor{\bsnm{Izenman},~\bfnm{Alan~Julian}\binits{A.~J.}}
(\byear{2008}).
\btitle{Modern Multivariate Statistical Techniques: Regression, Classification,
  and Manifold Learning}.
\bpublisher{Series: Springer Texts in Statistics}.
\end{bbook}
\endbibitem

\bibitem[\protect\citeauthoryear{Jiang, Tang and
  Zhang}{2004}]{GeneExpressionCluster}
\begin{barticle}[author]
\bauthor{\bsnm{Jiang},~\bfnm{Daxin}\binits{D.}},
  \bauthor{\bsnm{Tang},~\bfnm{Chun}\binits{C.}} \AND
  \bauthor{\bsnm{Zhang},~\bfnm{Aidong}\binits{A.}}
(\byear{2004}).
\btitle{Cluster analysis for gene expression data: A survey}.
\bjournal{IEEE Transactions on knowledge and data engineering}
\bvolume{16}
\bpages{1370--1386}.
\end{barticle}
\endbibitem

\bibitem[\protect\citeauthoryear{Koopmans and Reiersol}{1950}]{koopmans1950}
\begin{barticle}[author]
\bauthor{\bsnm{Koopmans},~\bfnm{T.~C.}\binits{T.~C.}} \AND
  \bauthor{\bsnm{Reiersol},~\bfnm{O.}\binits{O.}}
(\byear{1950}).
\btitle{The Identification of Structural Characteristics}.
\bjournal{Ann. Math. Statist.}
\bvolume{21}
\bpages{165--181}.
\bdoi{10.1214/aoms/1177729837}
\end{barticle}
\endbibitem

\bibitem[\protect\citeauthoryear{Krishnapuram
  et~al.}{2001}]{krishnapuram2001low}
\begin{barticle}[author]
\bauthor{\bsnm{Krishnapuram},~\bfnm{Raghu}\binits{R.}},
  \bauthor{\bsnm{Joshi},~\bfnm{Anupam}\binits{A.}},
  \bauthor{\bsnm{Nasraoui},~\bfnm{Olfa}\binits{O.}} \AND
  \bauthor{\bsnm{Yi},~\bfnm{Liyu}\binits{L.}}
(\byear{2001}).
\btitle{Low-complexity fuzzy relational clustering algorithms for web mining}.
\bjournal{IEEE transactions on Fuzzy Systems}
\bvolume{9}
\bpages{595--607}.
\end{barticle}
\endbibitem

\bibitem[\protect\citeauthoryear{Lawley and Maxwell}{1971}]{lawyley1971}
\begin{bbook}[author]
\bauthor{\bsnm{Lawley},~\bfnm{D.~N.}\binits{D.~N.}} \AND
  \bauthor{\bsnm{Maxwell},~\bfnm{A.~E.}\binits{A.~E.}}
(\byear{1971}).
\btitle{Factor analysis as a statistical method}, \bedition{Second} ed.
\bpublisher{American Elsevier Publishing Co., Inc., New York}.
\bmrnumber{0343471}
\end{bbook}
\endbibitem

\bibitem[\protect\citeauthoryear{Ledermann}{1937}]{Ledermann1937}
\begin{barticle}[author]
\bauthor{\bsnm{Ledermann},~\bfnm{Walter}\binits{W.}}
(\byear{1937}).
\btitle{On the rank of the reduced correlational matrix in multiple-factor
  analysis}.
\bjournal{Psychometrika}
\bvolume{2}
\bpages{85--93}.
\end{barticle}
\endbibitem

\bibitem[\protect\citeauthoryear{McDonald}{1999}]{mcdonaldbook}
\begin{bbook}[author]
\bauthor{\bsnm{McDonald},~\bfnm{Roderick~P.}\binits{R.~P.}}
(\byear{1999}).
\btitle{Test theory: a unified treatment}.
\bpublisher{Taylor and Francis}.
\end{bbook}
\endbibitem

\bibitem[\protect\citeauthoryear{Meinshausen and
  B{\"u}hlmann}{2006}]{meinshausen2006high}
\begin{barticle}[author]
\bauthor{\bsnm{Meinshausen},~\bfnm{Nicolai}\binits{N.}} \AND
  \bauthor{\bsnm{B{\"u}hlmann},~\bfnm{Peter}\binits{P.}}
(\byear{2006}).
\btitle{High-dimensional graphs and variable selection with the lasso}.
\bjournal{Annals of Statistics}
\bpages{1436--1462}.
\end{barticle}
\endbibitem

\bibitem[\protect\citeauthoryear{Rigollet and Tsybakov}{2011}]{rigollet2011}
\begin{barticle}[author]
\bauthor{\bsnm{Rigollet},~\bfnm{Philippe}\binits{P.}} \AND
  \bauthor{\bsnm{Tsybakov},~\bfnm{Alexandre}\binits{A.}}
(\byear{2011}).
\btitle{Exponential Screening and optimal rates of sparse estimation}.
\bjournal{Ann. Statist.}
\bvolume{39}
\bpages{731--771}.
\bdoi{10.1214/10-AOS854}
\end{barticle}
\endbibitem

\bibitem[\protect\citeauthoryear{Rubin and Thayer}{1982}]{Rubin1982}
\begin{barticle}[author]
\bauthor{\bsnm{Rubin},~\bfnm{Donald~B.}\binits{D.~B.}} \AND
  \bauthor{\bsnm{Thayer},~\bfnm{Dorothy~T.}\binits{D.~T.}}
(\byear{1982}).
\btitle{EM algorithms for ML factor analysis}.
\bjournal{Psychometrika}
\bvolume{47}
\bpages{69--76}.
\end{barticle}
\endbibitem

\bibitem[\protect\citeauthoryear{Shapiro}{1982}]{Shapiro1982}
\begin{barticle}[author]
\bauthor{\bsnm{Shapiro},~\bfnm{Alexander}\binits{A.}}
(\byear{1982}).
\btitle{Rank-reducibility of a symmetric matrix and sampling theory of minimum
  trace factor analysis}.
\bjournal{Psychometrika}
\bvolume{47}
\bpages{187--199}.
\end{barticle}
\endbibitem

\bibitem[\protect\citeauthoryear{Shapiro}{1985}]{Shapiro1985}
\begin{barticle}[author]
\bauthor{\bsnm{Shapiro},~\bfnm{A.}\binits{A.}}
(\byear{1985}).
\btitle{Identifiability of factor analysis: some results and open problems}.
\bjournal{Linear Algebra and Its Applications}
\bvolume{70}
\bpages{1-7}.
\end{barticle}
\endbibitem

\bibitem[\protect\citeauthoryear{Tsybakov}{2009}]{np_sasha}
\begin{bbook}[author]
\bauthor{\bsnm{Tsybakov},~\bfnm{Alexandre~B.}\binits{A.~B.}}
(\byear{2009}).
\btitle{Introduction to Nonparametric Estimation}.
\bpublisher{Springer, New York}.
\bdoi{10.1007/b13794}
\end{bbook}
\endbibitem

\bibitem[\protect\citeauthoryear{Wegkamp and Zhao}{2016}]{wegkamp2016}
\begin{barticle}[author]
\bauthor{\bsnm{Wegkamp},~\bfnm{Marten}\binits{M.}} \AND
  \bauthor{\bsnm{Zhao},~\bfnm{Yue}\binits{Y.}}
(\byear{2016}).
\btitle{Adaptive estimation of the copula correlation matrix for semiparametric
  elliptical copulas}.
\bjournal{Bernoulli}
\bvolume{22}
\bpages{1184--1226}.
\bdoi{10.3150/14-BEJ690}
\end{barticle}
\endbibitem

\bibitem[\protect\citeauthoryear{Wiwie, Baumbach and
  R{\"o}ttger}{2015}]{wiwie2015comparing}
\begin{barticle}[author]
\bauthor{\bsnm{Wiwie},~\bfnm{Christian}\binits{C.}},
  \bauthor{\bsnm{Baumbach},~\bfnm{Jan}\binits{J.}} \AND
  \bauthor{\bsnm{R{\"o}ttger},~\bfnm{Richard}\binits{R.}}
(\byear{2015}).
\btitle{Comparing the performance of biomedical clustering methods}.
\bjournal{Nature methods}
\bvolume{12}
\bpages{1033--1038}.
\end{barticle}
\endbibitem

\bibitem[\protect\citeauthoryear{Yuan and Lin}{2007}]{yuan2007model}
\begin{barticle}[author]
\bauthor{\bsnm{Yuan},~\bfnm{Ming}\binits{M.}} \AND
  \bauthor{\bsnm{Lin},~\bfnm{Yi}\binits{Y.}}
(\byear{2007}).
\btitle{Model selection and estimation in the Gaussian graphical model}.
\bjournal{Biometrika}
\bvolume{94}
\bpages{19--35}.
\end{barticle}
\endbibitem

\end{thebibliography}
\end{document}